\documentclass[journal,onecolumn]{IEEEtran}

\makeatletter
\def\ps@headings{%
\def\@oddhead{\mbox{}\scriptsize\rightmark \hfil \thepage}%
\def\@evenhead{\scriptsize\thepage \hfil \leftmark\mbox{}}%
\def\@oddfoot{}%
\def\@evenfoot{}}
\makeatother

\usepackage{graphics}
\usepackage{latexsym}
\usepackage[dvips]{epsfig}
\usepackage{amsmath,amsfonts,amssymb,amscd,amsthm,xspace}
\usepackage{bookmark,hyperref}
\usepackage{rotating}
\usepackage{multirow}
\usepackage{bm}
\usepackage{bbm}
\usepackage{mathrsfs}

\newtheorem{definition}{Definition}
\newtheorem{theorem}{Theorem}
\newtheorem{proposition}{Proposition}

\newtheorem{lemma}{Lemma}
\newtheorem{claim}{Claim}

\theoremstyle{remark}
\newtheorem{remark}{Remark}

\newcommand{\entropy}[1]{H\left( #1 \right)}
\newcommand{\condentropy}[2]{H\left( {#1} | {#2} \right)}
\newcommand{\mutual}[2]{I\left( {#1} ; {#2} \right)}
\newcommand{\condmutual}[3]{I\left( {#1} ; {#2} | {#3} \right)}
\newcommand{\variation}[2]{\mathbb{V}\left({#1}, {#2}\right)}
\newcommand{\KL}[2]{D\left({#1} \parallel {#2}\right)}
\newcommand{\wt}[1]{\mathrm{wt}_H ( #1 )}

\newcommand{\Oh}[1]{\mathcal{O}\left( #1 \right)}
\newcommand{\dist}{\mathrm{d}_H}

\newcommand{\prob}{\Pr}

\newcommand{\code}{\mathcal{C}}

\newcommand{\tranrv}{\mathbf{T}}

\newcommand{\Exp}{\mathbb{E}}
\newcommand{\Var}{\mathbb{V}ar}

\newcommand{\thr}{r}
\newcommand{\rate}{R}

\newcommand{\vM}{\mathbf{M}}
\newcommand{\msg}{m}
\newcommand{\priran}{m_s}
\newcommand{\messrv}{\vM}
\newcommand{\vx}{\vec{x}}
\newcommand{\vy}{\vec{y}}

\newcommand{\vX}{\vec{\mathbf{X}}}
\newcommand{\vY}{\vec{\mathbf{Y}}}
\newcommand{\vZ}{\vec{\mathbf{Z}}}

\newcommand{\bS}{\mathbf{S}}

\newcommand{\vS}{\bS}

\newcommand{\fx}{f_{1*}}

\newcommand{\fyb}{f_{b,*1}}
\newcommand{\fyw}{f_{w,*1}}
\newcommand{\fzzb}{f_{b,00}}
\newcommand{\fzob}{f_{b,01}}
\newcommand{\fozb}{f_{b,10}}
\newcommand{\foob}{f_{b,11}}
\newcommand{\fzzw}{f_{w,00}}
\newcommand{\fzow}{f_{w,01}}
\newcommand{\fozw}{f_{w,10}}
\newcommand{\foow}{f_{w,11}}

\newcommand{\Pb}{\mathbf{P}_b}
\newcommand{\Pw}{\mathbf{P}_w}
\newcommand{\pb}{p_b}
\newcommand{\pw}{p_w}

\newcommand{\bobtypyz}{\mathcal{A}_0(\vY_b)}
\newcommand{\bobtypyo}{\mathcal{A}_1(\vY_b)}
\newcommand{\bobtypxgyb}{\mathcal{A}_1(\vX|\vy_b)}
\newcommand{\bobtypxgybrv}{\mathcal{A}_1(\vX|\vY_b)}
\newcommand{\bobtypexgyb}{\mathcal{T}_{\vX|\vy_b, \tranrv = 1}(\fozb, \foob)}
\newcommand{\typyw}{\mathcal{A}_1(\vY_w)}
\newcommand{\typxgyw}{\mathcal{A}_1(\vX|\vy_w)}
\newcommand{\typexgyw}{\mathcal{T}_{\vX | \vy_w, \tranrv = 1}(\fozw, \foow)}
\newcommand{\typFw}{\mathcal{F}_w}
\newcommand{\typFb}{\mathcal{F}_b}
\newcommand{\typx}{\mathcal{A}(\vX)}

\newcommand{\hpsyw}{\mathcal{B}(\vY_w)}
\newcommand{\hpsxgyw}{\mathcal{B}(\vX|\vy_w)}
\newcommand{\typeBxgyw}{\mathcal{T}(\vX|\vy_w)(\fozw, \foow)}

\newcommand{\pz}{p_0}
\newcommand{\po}{p_1}
\newcommand{\qz}{q_0}
\newcommand{\qo}{q_1}
\newcommand{\Pz}{p_0}
\newcommand{\Po}{p_1}
\newcommand{\Qz}{q_0}
\newcommand{\Qo}{q_1}

\newcommand{\lw}{L_w}
\newcommand{\lb}{L_b}
\newcommand{\uw}{U_w}
\newcommand{\ub}{U_b}

\newcommand{\thre}{t}

\newcommand{\di}{\mathrm{d}}

\newcommand{\delx}{\Delta_{1*}}

\newcommand{\deln}{\delta_n}

\newcommand{\delybz}{\Delta_{b,*1}^{(0)}}
\newcommand{\delybo}{\Delta_{b,*1}^{(1)}}

\newcommand{\delozb}{\Delta_{b,10}}
\newcommand{\deloob}{\Delta_{b,11}}

\newcommand{\delyw}{\Delta_{w,*1}}

\newcommand{\delozw}{\Delta_{w,10}}
\newcommand{\deloow}{\Delta_{w,11}}

\newcommand{\errb}{\mathcal{E}_b}

\newcommand{\crho}{c_\rho}

\newcommand{\epsreli}{\epsilon_r}
\newcommand{\epsdeni}{\epsilon_d}
\newcommand{\epstyp}{\epsilon_{\mathcal{A}}}

\newcommand{\epn}{\epsilon_n}

\newcommand{\capacity}{C}

\newcommand{\parc}{\rho}

\newcommand{\py}{f_{*1}}
\newcommand{\poz}{f_{10}}
\newcommand{\poo}{f_{11}}
\newcommand{\pzo}{f_{01}}
\newcommand{\pzz}{f_{00}}
\newcommand{\px}{f_{1*}}

\newcommand{\bX}{\mathbf{X}}

\newcommand{\eqspacing}{\hspace{1cm}}

\begin{document}

\title{Reliable Deniable Communication: \\Hiding Messages in Noise}

\author{\IEEEauthorblockN{Pak Hou Che, Mayank Bakshi, Sidharth Jaggi\\}
\IEEEauthorblockA{The Chinese University of Hong Kong}}
\maketitle




\begin{abstract}
A transmitter Alice may wish to {\it reliably} transmit a message to a receiver Bob over a binary symmetric channel (BSC), while simultaneously ensuring that her transmission is {\it deniable} from an eavesdropper Willie. That is, if Willie listening to Alice's transmissions over a {``significantly noisier''} BSC than the one to Bob, he should be unable to estimate even whether Alice is transmitting.  We consider two scenarios. In our first scenario, we assume that the channel transition probability from Alice to Bob and Willie is perfectly known to all parties. Here, even when Alice's (potential) communication scheme is publicly known to Willie (with {\it no} common randomness between Alice and Bob), we prove that over $n$ channel uses Alice can transmit a message of length ${\cal O}(\sqrt{n})$ bits to Bob, deniably from Willie. We also prove information-theoretic order-optimality of this result. In our second scenario, we allow uncertainty in the knowledge of the channel transition probability parameters. In particular, we assume that the channel transition probabilities for both Bob and Willie are uniformly drawn  from a known interval. Here, we show that, in contrast to the previous setting, Alice can communicate ${\cal O}(n)$ bits of message reliably and deniably (again, with no common randomness). We give both an achievability result and a matching converse for this setting. Our work builds upon the work of Bash {\em et al} on AWGN channels (but with common randomness) and differs from other recent works (by Wang {\em et al} and Bloch) in two important ways - firstly our deniability metric is variational distance (as opposed to Kullback-Leibler divergence), and secondly, our techniques are significantly different from these works.

\end{abstract}
\IEEEpeerreviewmaketitle














%

\section{Introduction}
\label{chap:intro}

Consider the following scenario -- Alice, an undercover agent lurking in a foreign country, wishes to send occasional messages to a journalist Bob without attract the attention of the secret agent Willie. In this case, anonymous communication is not an option, since the secret agent Willie is constantly monitoring Alice. On the other hand, information-theoretically or cryptographically secure communication may also be undesirable, since an indecipherable message may be viewed as suspicious by Willie. To facilitate her communication, Alice tries to present the appearance of innocence (silence) to Willie, while attempting to communicate with Bob. The goal for Alice is to communicate with Bob reliably, while ensuring that Willie does not know whether Alice is transmitting or not. That is, Willie's best estimate of Alice's transmission status should be essentially statistically independent of his observations. It may be possible to instantiate such communication due to some ``asymmetry'' in the communication system between Alice and Bob, and that between Alice and Willie.

\subsection{Steganography}

The scenario we consider is a variant of the classical steganography problem. It is broadly defined as ``hiding a undetectable message in a plain sight'' -- brief but colourful historical perspectives on a variety of steganographic models and methods (including various techniques used by Xerxes, Herodotus, Mary Queen of Scots, and Margaret Thatcher, and even one which involves killing dogs...) can be found in \cite{anderson98} and \cite{kahn96}. 

The modern information-theoretic model of steganography started appear in the literature in the 1980's and 1990's. Simmons \cite{sim84} formalized the ``prisoner's problem'', and the connection between steganography and hypothesis testing appears first in Maurer \cite{maurer96} and Cachin \cite{cachin98}. More recently, there are books \cite{fridrich09, ingemar08} that fairly describe the theory of steganography and its applications.

\begin{itemize}
	\item {\bf Shared keys (Steganography):}
	The problem of steganography is usually defined as concealing some information within another file, image, video or article. Some implementations of steganography in physical manner are cataloged in history, for instance, a message could be hidden by writing in invisible ink that could reveal the message when heating the paper.
	
	The problem of steganography is at least relying on one of the following assumptions:
	
	\begin{itemize}
		\item {\bf Non-zero covertext/stegotext:}
		In most of the literature, Alice accesses a {\it covertext} (-- a length-$n$ vector) drawn from some distribution (which is known by Bob and Willie). In this scenario, Alice is able to transmit a slightly perturbed covertext (which is called {\it stegotext}) that only Bob can discover the value of these perturbation. Willie, in this case, is not able to detect the perturbation. One critical point here is that Alice's transmission is always non-zero even if she does not have a stegotext to hide. For example, Alice uploads an image to a website. Willie is difficult to tell whether there is any stegotext to Bob hidden in this image.
		\item {\bf Shared secret key/common randomness:}
		A shared secret key between Alice and Bob is needed in many of the steganography protocols, but the size of the key is usually as large as the message \cite{fridrich09, wang08}. In particular, the key allows Alice and Bob to coordinate a code to use, and keeps Willie in the dark. There are also some steganographic protocols that do not need a key \cite{ker10, ryabko09}.
		\item {\bf Noiseless communication:}
		Some works consider the communication channel between Alice and Bob is noiseless. This has some important consequences –- the optimal throughput can sometimes be scaled by a multiplicative factor of $\log n$ \cite[Chapters 8 and 13]{fridrich09}.
	\end{itemize}
	
	\item {\bf Differential channels (Deniability/Stealth):}
	Alice leverages asymmetries in channel parameters in channels to Bob and to Willie. There are two types of asymmetry -- differential noise and differential network access. We will describe these types of asymmetry in the following. In particular, our focus in this thesis is differential noise.
	\begin{itemize}
		\item {\bf Differential noise (Deniability/Stealth):}
		In this class of models, whether Alice is transmitting or not, is not known by Willie. The goal for Willie is to detect Alice's transmission status, that is, whether Alice is communicating to Bob or not. It is crucial that the noise in Willie's channel is higher than Bob's. For example, Alice may use directional antenna points to Bob. Or, in the ``prisoner's problem'', Bob is locked in the next cell of Alice's, where the warden Willie is not as close as Bob's. We can see that Willie's channel noise is higher than Bob's in both cases. In this scenario, Alice cannot communicate to Bob to loudly. Any ``constant'' (-- non-decaying to zero) noise in fact will trigger Willie's suspicion by simply comparing the noise level. On the other hand, if Alice whispers too soft to Bob, Bob may not be able to know what Alice is trying to communicating to him. Therefore, there is a trade-off between Bob's reliability and Willie's deniability.
		\item {\bf Differential network access (Network deniability):}
		In this class of models, Alice has several links that communicate Bob. Due to the shortage of resource, Willie can only wire-tap some of these links. If Alice trying to communicate to Bob covertly, the covert message can only be find out when one can collect all the information. In this scenario, Alice's goal is to encode her message and transmit it through all the links so that any combinations of Willie's wire-tapped information still looks innocent to him. A more detail description of differential network access can be found in \cite{kadhe14}.
	\end{itemize}
	
\end{itemize}

\subsection{The ``Square Root Law''}
The ``Square Root Law'' (often abbreviated as SRL in the literature) can be perhaps characterized as an observation that in a variety of steganographic models, the throughput (the length of the message that Alice can communicate deniably and reliably with Bob) scales as $\Oh{\sqrt{n}}$ (here $n$ is the number of ``channel uses'' that Alice has access to).

Some recent works (for instance~\cite{ker10}) have begun to theoretically justify the square root law under some (fairly restrictive) assumptions on the class of steganographic protocols. Nonetheless, results in this class should still be taken with a pinch of salt, since they do not offer a universally robust characterization for all models which may be of interest. For instance, in some works (for instance~\cite[Chapters 8 and 13]{fridrich09})) the throughput scales as $\Oh{\sqrt{n}\log n}$.
More drastically, the works of~\cite{wang08} (which gives an information-theoretically optimal characterization of the rate-region of many variants of the steganography problem) and that of~\cite{ryabko09} (which design computationally efficient steganography protocols) both allow throughput that scales linearly in $n$, rather than $\Oh{\sqrt{n}}$ as would be indicated by the SRL. The major difference between the models of~\cite{wang08, ryabko09}, and those that satisfy the SRL, seems to lie in a disagreement as to what comprises ``realistic'' steganographic algorithms.

\subsection{Related Works -- Reliable Deniable Communication}

\begin{itemize}
	\item {\bf Low Probability of Detection (LPD) \cite{BasGT:12J}:} The major difference between our model and that of \cite{BasGT:12J} (and the reason we state that our model is more ``realistic'') is that there is no shared secret key between Alice and Bob that is hidden from Willie in our setting. Hence our codes are ``public''. A setting in \cite{BasGT:12J} requires the secret keys to be significantly longer than her throughput $\Oh{\sqrt{n}}$ to Bob. The reason we are able to achieve such performance is due to a more intricate analysis of random binary codes than is carried out in \cite{BasGT:12J}. This includes a novel and intricate analysis of concentration inequalities.

In our work all channels are discrete (finite input and output alphabets)  which allows using the language of types and type classes. In contrast, the results of \cite{BasGT:12J} are for channels wherein the channel noise is continuous. It is conceivable that our construction of public codes also carries over to the AWGN model of \cite{BasGT:12J}, but significant extensions may be required to translate our techniques from the discrete world over to the continuous version.
	
	\item {\bf Stealth \cite{hou2013effective}:} Hou and Kramer \cite{hou2013effective} first proposed the ``pretend innocence'' model in the context of wiretap channel. The authors define a notion stealth, which is very similar to deniability. The key difference between stealth and deniability is technical, the model in \cite{hou2013effective} allows the innocent distribution to be non-zero, and the stealth measures the difference between the innocent and active distributions in terms of Kullback-Leibler divergence, whereas deniability measures it in terms of variational distance. Under fairly general conditions, Hou and Kramer characterize the reliable deniable and secret communication capacity of channels. The requirement of stealth is similar to the notion of approximation of channel output statistics, which is related to the notion of resolvability \cite{han1993approximation}. 
	
	\item {\bf Covert Communication \cite{Bloch2015}:} The covertness measure in \cite{Bloch2015} between the innocent and active distributions is Kullback-Leibler divergence, which is the same as the stealth measurement \cite{hou2013effective}. Bloch first considers the trade-off between the minimum amount of secret keys required and the asymmetries between Bob and Willie's channel. Under fairly general conditions, Bloch characterize the reliable deniable communication capacity of discrete memoryless channels, and the minimum amount of secret keys required when Bob's channel is noisier than Willie's channel. Quantitiaively, a key difference between our work and~\cite{Bloch2015} is that the error probability (with respect to the random codebook construction) in \cite{Bloch2015} decays exponentially whereas, in our work it can decay super-exponentially. The super-exponential decay of error exponent turns out be very useful in proving strong secrecy for the reliable deniable communication problem.
	
	\item {\bf Low Probability of Detection \cite{Wang2015}:} Wang {\it et. al.} characterize the exact capacity of reliable deniable communication of discrete memoryless channels and AWGN channels. Wang {\it et. al.}'s deniability measurement are the same as in \cite{hou2013effective, Bloch2015}.

\end{itemize}

A comparison of related works can be found in the following table.

\begin{center}
\begin{tabular}{|l|l|l|l|l|}
\hline
 & shared & channel & security & rate/ \\
 & secret & model   & metric   & throughput \\
\hline
\multicolumn{5}{c}{Deniability/LPD/Covertness/Stealth} \\
\hline
Bash {\it et. al.} \cite{BasGT:12J} & $\Oh{\sqrt{n} \log n}$ & AWGN & $\variation{\cdot}{\cdot}$ & $\Oh{\sqrt{n}}$ \\
\hline
Che {\it et. al.} \cite{Che2013} & No & BSC & $\variation{\cdot}{\cdot}$ & $\Oh{\sqrt{n}}$ \\
\hline
Hou and Kramer \cite{hou2013effective} & No & DMC & $\KL{\cdot}{\cdot}$ and $\mutual{\cdot}{\cdot}$ & $\Oh{1}$ \\
(non-zero  & & & & \\
innocent distribution) & & & & \\
\hline
Che {\it et. al.} \cite{Che2014} & No & Slow-fading BSC & $\variation{\cdot}{\cdot}$ & $\Oh{1}$ \\
\hline
Bloch \cite{Bloch2015} & No ($\pb < \pw$) & DMC & $\KL{\cdot}{\cdot}$ & $\Oh{1}$ \\
& $\Oh{\sqrt{n}}$ ($\pb > \pw$) & & & \\
\hline
Wang \cite{Wang2015} & $\Oh{\sqrt{n} \log n}$ & DMC & $\KL{\cdot}{\cdot}$ & $\Oh{\sqrt{n}}$ \\
& & & & (EXACT) \\
\hline
\multicolumn{5}{c}{Steganography Related Literature} \\
\hline
Ker \cite{ker10} & $\Oh{\sqrt{n} \log n}$ & BSC & $\KL{\cdot}{\cdot}$ & Achievability: \\
non-zero stegotext & & & & $\Oh{1}$ \\
\hline
Wang and Moulin \cite{wang08} & $\Oh{\sqrt{n} \log n}$ & DMC & Distortion & $\Oh{1}$ \\
\hline
\end{tabular}
\end{center}

\section{Model -- Reliable Deniable Communication}
\label{chap:model}

\subsection{Notational Conventions}
Calligraphic symbols such as $\code$ denote sets. Boldface upper-case symbols such as $\vM$ denote random variables, lower-case symbols such as $\msg$ denote particular instantiations of those random variables. Vectors are denoted by an arrow above a symbol, such as in $\vx$. In particular, an arrow above a random variable such as $\vX$ denotes a vector random variable.

For notational convenience, in this work, unless otherwise specified, all vectors are of length $n$, where $n$ corresponds to the {\it block-length} (number of channel uses). Let $\mathbf{A}$ be a random variable and taking values in an alphabet $\mathcal{A}$, and the probability distribution for $\mathbf{A}$ is denoted as $\{p_{\mathbf{A}}(a), a \in \mathcal{A} \}$, where $p_{\mathcal{A}}(a) = \Pr(\mathbf{A} = a)$. The probability $p_{\mathbf{A}}(a)$ is abbreviated as $p(a)$ and the probability distribution $\{p_{\mathbf{A}}(a)\}$ is abbreviated as $p(a)$ if there is no ambiguity. Probabilities of events are denoted with a subscript denoting the random variable(s) over which the probabilities are calculated. For instance,
$$ \Pr_{\mathbf{A}, \mathbf{B}}(\mathbf{C} = c) \triangleq \sum_{a, b} p(a, b) \mathbbm{1}(\mathbf{C}(a, b) = c). $$
All logarithms in this work are binary, unless otherwise stated. { {The {\it Hamming weight} (number of non-zero entries) of a vector $\vx$ is denoted by $\wt{\vx}$, and the {\it Hamming distance} between two vectors $\vx$ and $\vy$ of equal length (the number of corresponding entries in which $\vx$ and $\vy$ differ) is denoted by $\dist(\vx, \vy)$}. The {\it support} of a vector is defined as the set of locations where it is non-zero. For any two numbers $a$ and $b$ in the interval $[0,1]$, we use $a \ast b$ to denote {\it binary convolution} of these two numbers, defined as $a(1-b) + b(1-a)$ -- this corresponds to the noise parameter of the BSC ({\it Binary Symmetric Channel}) comprising of a BSC($a$) followed by a BSC($b$). As is standard in an information-theoretic context, the notation $\entropy{\cdot}$ corresponds to the {\it (binary) entropy function},  $\condentropy{\cdot}{\cdot}$ to {\it conditional entropy}, $\mutual{\cdot}{\cdot}$ to {\it mutual information}, and $\KL{\cdot}{\cdot}$ to the {\it Kullback-Leibler divergence} between two probability distributions. Also, we use $\variation{p}{q}$ to denote the {\it variational distance} between any two probability distributions $p(a)$ and $q(a)$ defined over the same alphabet $\mathbf{A}$, {\it i.e.}, $\variation{p}{q}$ is defined as
$$\variation{p}{q} \triangleq \frac{1}{2}\left (\sum_{a \in \mathbf{A}} \left | p(a) - q(a)\right |\right ).$$
In this work, the alphabet size that we will typically be interested in is $2^n$.

\section{Communication Model}
The transmitter Alice is connected via a binary-input binary-output broadcast medium to the receiver Bob and the warden Willie. The channels from Alice to Bob, and from Alice to Willie, are independent binary symmetric channels with cross-over probabilities $\Pb$ and $\Pw$ respectively.\footnote{In principle the techniques in this work generalize to arbitrary pairs of independent channels from Alice and Bob, and Alice to Willie. However, for ease of presentation of technically intricate results we focus on the binary-input binary-output symmetric noise scenario in this work. Indeed, the case of general DMCs has been treated in recent work by~\cite{Bloch2015, Wang2015}.} Here, the ``noise parameters'' $\Pb$ and $\Pw$ are themselves random variables. In advance of communication, the only knowledge about $\Pb$ and $\Pw$ by all parties is that they are uniformly distributed in their corresponding ranges respectively. We specify two different choices of the ranges of $\Pb$ and $\Pw$ for the two models we consider in this work. That is,
\begin{enumerate}
	\item {\bf Fixed Channel:} In this model, $\Pb = \pb$ and $\Pw = \pw$, where $\pb < \pw$; (that is, the channel noise parameters are {\it precisely} known in advance of communication to all parties [Alice, Bob and Willie].) \label{model:known}
	\item {\bf Slow Fading Channel:} In this model, $\Pb \in (\lb, \ub)$ and $\Pw \in (\lw, \uw)$. (The noise parameters are uniformly and independent distributed over intervals $(\lb, \ub)$ and $(\lw, \uw)$. We specify the relationship between Bob's channel parameters and Willie's channel parameters later in Theorem~\ref{thm:model:unknown:achi}.)\footnote{In principle our techniques can also handle many scenarios wherein the noise parameters are  not necessarily uniformly and independently distributed in intervals. In fact we believe our techniques work for all ``sufficiently slowly fading'' channels. However, for ease of exposition in deriving our technically complex results we focus on this ``base case'' first.} \label{model:unknown}
\end{enumerate} 

In either model, the channel noise parameters independently instantiate as $\Pb$ and $\Pw$ respectively for Bob and Willie, and are fixed for all $n$ channel uses. 

Alice (potentially) wishes to communicate a {\it message} $\msg$ which is uniformly distributed from $\{1, 2, \ldots, N\}$ to Bob -- $\vM$ denotes the random variable corresponding to $\msg$, and $N$ is the number of possible messages Alice wishes to communicate. (If Alice is not transmitting, her message $\vM$ is $0$.) We associate Alice's communication status with a binary variable $\tranrv$ -- if Alice {\it does} wishes to communicate with Bob, $\tranrv$ is set to $1$, else $\tranrv$ equals $0$. Only Alice knows the value of $\tranrv$ {\it a priori}.


\subsection{Reliability (Alice $\longrightarrow$ Bob)}

\noindent {\bf \underline{Alice's Encoder}}:
Alice encodes {{each message $\msg$ into a length-$n$ binary {\it codeword} $\vx(\msg)$ using an {\it encoder} $Enc(\cdot):\{0\} \cup \{1, \ldots, N\} \rightarrow \{0,1\}^n$. This encoder may be a deterministic encoder (each $\msg$ maps to a unique codeword, a length-n vector $\vX$) or may be a {\it stochastic encoder} (using private randomness available {\it only} to her, Alice probabilistically maps her message $m$ to a length-$n$ vector $\vX$). In either case, the encoding function $Enc(\cdot)$ (but not Alice's message, nor her private randomness if any) is fully known in advance to both Bob and Willie -- this corresponds to Alice committing to using a ``public codebook'' $\code$.\footnote{We wish to stress here that unlike what is common in the AVC literature (for instance, see~\cite{Lapidoth-Narayan}), we do not allow {\it randomized encoders}, {\it i.e.} allow Alice and Bob to share common randomness that is private from Willie in advance of communication. Such common randomness, if it were available, would make the problem of deniable communication much easier — for instance, it would allow for Alice and Bob to use a private codebook, and hence make it easier for them to make the fact of their communication from Willie. Indeed, this is leveraged in the work by Bash {\it et al.}~\cite{BasGT:12J}.} To unify notation, we describe only models where Alice uses a stochastic encoder -- models in which Alice uses a deterministic encoder are special cases of this.


If Alice's {\it transmission status} $\tranrv = 0$ ({\it i.e.}, Alice does not wish to transmit), the encoder always maps the $0$ message to the zero-vector $\vec{0}$. If Alice's transmission status $\tranrv = 1$ ({\it i.e.}, Alice wishes to transmit), for each message $\msg$, she chooses a codeword $\vx$ from the codebook $\code$ according to the probability distribution $\Pr_{\vX | \vM}(\cdot | \vM = \msg)$. A codeword $\vx \in \code$ such that $\Pr_{\vX | \vM}(\vX = \vx | \vM = \msg) = 1$ if and only if $\vx$ is the codeword corresponding to the message $\msg$. Alice's encoder is computationally unbounded.


\noindent {\bf \underline{Bob's Decoder}}:
Bob receives the length-$n$ binary vector $\vY_b = \vX \oplus \vZ_b$, where $\vZ_b$ denotes the channel noise (modelled as a Bernoulli random variable $Bern(\pb)$), resulting in the channel from Alice to Bob being a $BSC(\pw)$. Bob uses his {\it decoder} $Dec(\cdot): \{0,1\}^n \rightarrow \{0\} \cup \{1,\ldots,N\}$ to generate his {\it estimate of Alice's message} $\hat{\vM} = Dec(\vY_b)$. We denote the pair $Enc(\cdot)$, $Dec(\cdot)$, Alice's encoder and Bob's decoder as the code $\code$. When Alice's transmission status $\tranrv = 0$, Bob's {\it error probability on no-transmission} is defined as $\Pr_{\vZ_b, \Pb}(\hat{\vM} \neq 0 | \tranrv = 0)$, {\it i.e.} the probability (over randomness in Alice's transmitted message $\vM$, the $Bernoulli(\pb)$ channel noise $\vZ_b$ and any randomness in the channel noise parameter $\pb$ itself) that Bob decodes to an incorrect message $\hat{\vM}$. When Alice's transmission status $\tranrv = 1$, since the message $\vM$ is uniformly distributed from $\{1, \cdots, N\}$, Bob's error probability is defined as $\Pr_{\vM, \vM_s, \vZ_b, \Pb}(\hat{\vM} \neq \vM | \tranrv = 1)$.
Alice's code $\code$ is said to be {\it $(1-\epsreli)$-reliable} if the sum of Bob's two error probabilities $\Pr_{\vZ_b, \Pb}(\hat{\vM} \neq 0 | \tranrv = 0) + \Pr_{\vM, \vM_s, \vZ_b, \Pb}(\hat{\vM} \neq \vM | \tranrv = 1)$ is less than $\epsreli$. Bob's decoder $Dec(\cdot)$ is computationally unbounded.


\subsection{Deniability (Alice $\longrightarrow$ Willie)} \label{sec:a_to_w}

\noindent {\bf \underline{Willie's Estimator}}:
Willie knows both $Enc(\cdot)$ and $Dec(\cdot)$ (and hence Alice's code $\code$) {\it a priori}, and is computationally unbounded. Willie receives the length-$n$ binary vector $\vY_w = \vX \oplus \vZ_w$, where $\vZ_w$ denotes the the channel noise (modelled as $BSC(\pw)$). Willie uses an {\it estimator} $Est(\cdot): \{0,1\}^n \rightarrow \{0,1\}$ to generate his {\it estimate of Alice's transmission status} as $\hat{\tranrv} = Est(\vY_w)$. That is, Willie just wishes to learn a {\it single bit} of information, namely Alice's transmission status $\tranrv$. We explicitly allow the estimator $Est(\cdot)$ that Willie uses to depend on Alice and Bob's code $\code$. Willie's estimator is computationally unbounded.

We use a hypothesis-testing metric to quantify the {\it deniability of Alice's code}.
Let {\it the probability of false alarm} $\Pr_{\vZ_w, \Pw}(\hat{\tranrv} = 1 | \tranrv = 0)$ be denoted by $\alpha(Est(\cdot))$.  Analogously, let {\it the probability of missed detection} $\Pr_{\vM, \vM_s, \vZ_w, \Pw}(\hat{\tranrv} = 0 | \tranrv = 1)$ be denoted by $\beta(Est(\cdot))$.
These quantities 
denote respectively the probabilities  that Willie guesses Alice is transmitting even if she is not, and that Willie guesses Alice is not transmitting even though she is. 
We say Alice's codebook $\code$ is {\it $(1-\epsdeni)$-deniable} if there is no estimator $Est(\cdot)$ such that $\alpha(Est(\cdot)) + \beta(Est(\cdot)) > 1-\epsdeni$. Note that this deniability metric is independent of any prior distribution on Alice's transmission status $\tranrv$. We henceforth denote $\alpha(Est(\cdot)) $ and $ \beta(Est(\cdot))$ simply by $\alpha$ and $\beta$.

\subsection{Capacity} The {\it rate} $\rate$ of Alice's codebook is defined as $(\log{N})/n$. In the Fixed Channel Model, the {\it relative throughput} $\thr$ of Alice's codebook is defined as $(\log{N})/\sqrt{n}$ (see Remark~\ref{remark:thr_replace_rate} for a discussion). For any block-length $n$, we say a corresponding codebook $\code$ is {\it simultaneously $(1-\epsreli)$-reliable and $(1-\epsdeni)$-deniable} if it simultaneously ensures that the sum of Bob's two probabilities of decoding error is at most $\epsreli$, and has deniability $1-\epsdeni$. For a fixed blocklength $n\in\mathbb{Z}^+$, we define $C_{\epsreli,\epsdeni}^{(n)}$ as the supremum over all rates for which a simultaneously $(1-\epsreli)$-reliable and $(1-\epsdeni)$-deniable code exists. For the Fixed Channel Model, we define $c_{\epsreli,\epsdeni}^{(n)}$ as the supremum over all relative throughputs for which a simultaneously $(1-\epsreli)$-reliable and $(1-\epsdeni)$-deniable code exists. Finally, we define the {\em $(1-\epsreli)$-reliable $(1 - \epsdeni)$-deniable capacity} $C_{\epsreli,\epsdeni}$ as $\lim_{n\to\infty}C^{(n)}_{\epsreli,\epsdeni}$ and the {\em $(1-\epsreli)$-reliable $(1 - \epsdeni)$-deniable relative capacity} $c_{\epsreli,\epsdeni}$ as $\lim_{n\to\infty}c^{(n)}_{\epsreli,\epsdeni}$.

\begin{remark} \label{remark:thr_replace_rate}
In the Fixed Channel Model, the rate $\rate$ scales as $1/\sqrt{n}$. So, the rate $\rate \rightarrow 0$ as $n \rightarrow \infty$. Therefore, we consider the throughput $\thr = (\log N) / \sqrt{n}$ which we demonstrate $\thr$ is scales as a constant in our communication schemes (and indeed this is the optimal scaling).
\end{remark}

\section{Main Results}

In the problem of reliable deniable communication, we aim to find the capacity in two different scenarios. In particular, Model~\ref{model:known} (the ``Fixed Channel'' model) corresponds to the channel parameters are known exactly to all parties, whereas Model~\ref{model:unknown} (the ``Slow Fading'' model) corresponds to the scenario wherein all parties know only the {\it distribution} of the Bernoulli channel noise parameters $\pb$ and $\pw$ (which in this work are assumed to be distributed independently and uniformly over certain pre-defined intervals).


{\underline\bf{{Model~\ref{model:known} (Fixed Channel Model)}}:} In Theorems~\ref{thm:model:known:conv} and \ref{thm:model:known:achi} below we provide outer and inner bounds on the reliable deniable capacity for the class of channels defined in Section~\ref{chap:model} that are tight up to constant factors.

\begin{theorem}[Converse for Fixed Channel Model]
\label{thm:model:known:conv}
For every code $\code$ that has $\gamma(\zeta) \in (\epsdeni, 1 - \epsreli)$ fractions of codewords with fractional weight greater than $\zeta$ and simultaneously $(1 - \epsreli)$-reliable and $(1 - \epsdeni)$-deniable, then the optimal throughput $\thr$ satisfies
	\begin{equation}
				\thr_{\epsreli, \epsdeni} \leq \frac{1}{1 - \frac{\epsreli}{1 - \gamma(\zeta)}} \frac{1 - 2\pb}{1 - 2\pw}\sqrt{\frac{8 \pw (1 - \pw)}{1 - \frac{\epsdeni}{\gamma(\zeta)}}} \log \frac{1 - \pb}{\pb} + \Oh{\frac{1}{\sqrt{n}}}.
	\end{equation}
\end{theorem}

\begin{proposition}[Reliability of Random Stochastic Codes $\code(N,S)$]
\label{prop:ran_sto_code}
Suppose that Alice uses a random stochastic code $\code(\thr\sqrt{n}, \thr_s\sqrt{n})$, then the code $\code$ is at least $(1 - \epsreli)$-reliable with probability greater than $1 - \exp\left(-\Omega(\sqrt{n})\right)$ if
	$$ \thr + \thr_s < \thr_U, $$
where
\begin{equation}
	\thr_U = \epsdeni \sqrt{\pw (1 - \pw)} \frac{1 - 2\pb}{1 - 2\pw} \log \frac{1 - \pb}{\pb}.
\end{equation}
\end{proposition}

\begin{theorem}[Achievability for Fixed Channel Model]
\label{thm:model:known:achi}
Let
\begin{equation}
	\thr_L = \epsdeni \sqrt{\pw (1 - \pw)} \log \frac{1 - \pw}{\pw}.
\end{equation}
Then, with probability greater than $1 - \epsreli$ over the codebook generation ensemble $\code$ and the throughput $\thr_L < \thr < \thr_U$ satisfying the following:
	\begin{enumerate}
		\item The codebook $\code$ is at least $(1 - \epsdeni)$-deniable;
		\item The codebook $\code$ is at least $(1 - \epsreli)$-reliable.
	\end{enumerate}
\end{theorem}

\begin{theorem}[Hidability of Fixed Channel Model]
\label{thm:model:known:hide}
Under the same codebook generation as in Theorem~\ref{thm:model:known:achi}, weak secrecy can also be achieved by rearranging the codebook.
\end{theorem}

\begin{theorem}[Converse for Slow Fading Channel Model]
\label{thm:model:unknown:conv}
For every codebook $\code$ satisfies $(1 - \epsreli)$-reliability and $(1 - \epsdeni)$-deniability, the rate
\begin{equation}
	\rate \leq \entropy{\left( \frac{\uw - \lw}{(1 - 2\lw)} \epsdeni \right) \ast \ub} - \entropy{\ub}.
\end{equation}

\end{theorem}

\begin{theorem}[Achievability of Slow Fading Channel Model]
\label{thm:model:unknown:achi}
Given the channel parameters $\lw, \uw, \lb$, and $\ub$ satisfying
			\begin{equation}
				\frac{\uw - \lw}{\ub - \lb} \cdot \frac{1 - 2\lb}{1 - 2\lw} > \frac{1 - \epsreli}{\epsdeni}.
			\end{equation}
Then, with probability greater than $1 - \exp\left( -\Omega(\sqrt{n}) \right)$ over the codebook generation ensemble $\code$, and the rate
	\begin{equation}
		\rate \geq \entropy{\left( \frac{\uw - \lw}{(1 - 2\lw)} \epsdeni \right) \ast \ub} - \entropy{\ub}
	\end{equation}
satisfying the following:
	\begin{enumerate}
		\item The codebook $\code$ is at least $(1 - \epsdeni)$-deniable;
		\item The codebook $\code$ is at least $(1 - \epsreli)$-reliable.
	\end{enumerate}
\end{theorem}




\subsection{Notations and Definitions} \label{sec:notation_def}

\subsubsection{Probability Notation}
As defined in Section~\ref{chap:model}, we use $\code$ to denote the code, an encoder decoder pair. In this section, we abuse notation by overloading $\code$ to also denote Alice's codebook, which contains at most $NS$ codewords, {\it i.e.}, $2^{NS}$ possible length-$n$ binary vectors.

\noindent{\bf\underline{Random Stochastic Codebook $\code(N,S)$}:}\\
We first define Alice's codebook generation ensemble, denoted $p_{\bm{\code}}(\code) \triangleq \Pr(\bm{\code} = \code)$ (henceforth abbreviated as $p_{\bm{\code}}$). As is common in information theory, Alice generates her codebook $\bm{\code}$ by choosing each codeword $\vx$ according to an independent and identically distribution. Specifically, there are $S$ codewords correspond to each message $\msg$, and the probability $\Pr_{\vX | \vM}(\vX = \vx | \vM = \msg) = \frac{|\{\msg_s : \vx(\msg, \msg_s) = \vx\}|}{S}$. For a random deterministic codebook, we set the value of $S$ to $1$, that is, $\code(N,1)$.


\noindent{\bf\underline{Codeword distribution}:}\\
As already mentioned in the Section~\ref{chap:model}, Alice's ``silent'' codeword distribution (corresponding to the situation when she does not transmit, {\it i.e.}, when $\tranrv = 0$), $p_{\vX | \tranrv = 0}(Enc(\msg = 0)) \triangleq \Pr(\vX = Enc(0) | \tranrv = 0)$, is a singleton, with the corresponding codeword $\vec{0}$ having probability mass $1$.

The probability distribution $p_{\vX | \tranrv = 1}(\vx) \triangleq \Pr(\vX = \vx | \tranrv = 1)$, corresponding to the probability distribution on Alice's transmission, is more complicated. It depends on Alice's code $\bm{\code}$. In particular, a probability mass of $\rho^{\wt{\vx}}(1 - \rho)^{n - \wt{\vx}}$ is assigned to each length-$n$ binary vector $\vx$ corresponding to $Enc(\msg, \priran)$ (an encoding of message $\msg$ with key $\priran$). Since Alice's code is randomly generated, note that in case there are ``collisions'' in the codebook (the same length-$n$ vector $\vx$ corresponds to multiple pairs $(\msg, \priran)$), then the probability mass assigned to that particular $\vx$ is in general an integer multiple of $\rho^{\wt{\vx}}(1 - \rho)^{n - \wt{\vx}}$.


\noindent{\bf\underline{Willie's received vector distribution}:}\\
We first define three ``$n$-letter''-probability distributions on the $2^n$-dimensional space of Willie's received vector $\vY_w$. 

{\it The silent distribution} on Willie's received vectors $\vY_w$, $p_{\vY_w | \tranrv = 0}(\vy_w) \triangleq \Pr(\vY_w = \vy_w | \tranrv = 0)$, corresponds to the probability distribution on $\vY_w$ when Alice's transmission status $\tranrv = 0$ (she stays silent). So, for any $\vy_w$, $p_{\vY_w | \tranrv = 0}(\vy_w)$ equals $\pw^{\wt{\vy_w}}(1 - \pw)^{n - \wt{\vy_w}}$.

{\it The active distribution} $p_{\vY_w | \tranrv = 1}(\vy_w) \triangleq \Pr(\vY_w = \vy_w | \tranrv = 1)$ on Willie's received vectors $\vY_w$, corresponds to the probability distribution on $\vY_w$ when Alice's transmission status $\tranrv = 1$ (she transmits using code $\code$). So, for any $\vy_w$, $p_{\vY_w | \tranrv = 1}(\vy_w)$ equals 
\begin{multline*}
\sum_{\vx \in \code} p_{\vY_w|\vX}(\vy_w | \vx) p_{\vX | \tranrv = 1}(\vx) \\
	= \sum_{\vx \in \code} \pw^{\dist(\vx, \vy_w)} (1 - \pw)^{n - \dist(\vx, \vy_w)} \rho^{\wt{\vx}} (1 - \rho)^{n - \wt{\vx}}.
\end{multline*} 

\noindent{\bf\underline{Willie's ensemble average received vector distribution}:}\\
Finally, we define the {\it ensemble average active} distribution on Willie's received vectors $\vY_w$, $\Exp_\code \left( p_{\vY_w | \tranrv = 1}(\vy_w) \right) \triangleq \Exp_\code ( \Pr(\vY_w = \vy_w | \tranrv = 1) )$, is an ensemble average (over $p_\code$) over all possible codes that Alice could use. Specifically, $\Exp_\code \left( p_{\vY_w | \tranrv = 1}(\vy_w) \right)$ equals $(\rho \ast \pw)^{\wt{\vy_w}} (1 - \rho \ast \pw)^{n - \wt{\vy_w}}$. Note that this equality follows from the fact that the ensemble average distribution on $\vY_w$ is the same as the distribution obtained by passing the all-zero vector through two successive Binary Symmetric Channels -- one a $BSC(\rho)$ (corresponding to the codebook generation parameter), and one a $BSC(\pw)$ (corresponding to the channel noise from Alice to Willie).

We have similar notation for Bob's probability distribution.

\noindent{\bf\underline{Bob's received vector distribution}:}\\
The {\it silent} distribution on Bob's received vectors $\vY_b$, $p_{\vY_b | \tranrv = 0}(\vy_b) \triangleq \Pr( \vY_b = \vy_b | \tranrv = 0)$, equals $\pb^{\wt{\vy_b}}(1 - \pb)^{n - \wt{\vy_b}}$.

The {\it active} distribution $p_{\vY_b | \tranrv = 1}(\vy_b) \triangleq \Pr(\vY_b = \vy_b | \tranrv = 1)$ on Bob's received vectors $\vY_b$, equals
\begin{multline} \nonumber
\sum_{\vx \in \code} p_{\vY_b|\vX}(\vy_b | \vx) p_{\vX | \tranrv = 1}(\vx) \\ = \sum_{\vx \in \code} \pb^{\dist(\vx, \vy_b)} (1 - \pb)^{n - \dist(\vx, \vy_b)} \rho^{\wt{\vx}} (1 - \rho)^{n - \wt{\vx}}.
\end{multline}

Finally, we define the {\it ensemble average active} distribution on Bob's received vectors $\vY_w$, $\Exp_\code \left( p_{\vY_b | \tranrv = 1}(\vy_b) \right) \triangleq \Exp_\code (\Pr(\vY_b = \vy_b | \tranrv = 1))$, equals $(\rho \ast \pb)^{\wt{\vy_b}} (1 - \rho \ast \pb)^{n - \wt{\vy_b}}$. 
\begin{center}\begin{table}\begin{center}
    \begin{tabular}{| p{3.5cm} | p{2cm} | p{6.3cm} |}
    \hline
    Probability notation & Simplified notation & Mathematical expression \\ \hline
    $p_{\bm{\code}}(\code)$ & / & $\prod_{\vx \in \code} \rho^{\wt{\vx}} (1 - \rho)^{n - \wt{\vx}}$ \\ \hline
    $p_{\vX | \tranrv = 0}(\vec{0})$ & $p(\vec{0})$ & $p(\vec{0}) = 1$ \\ \hline
    $p_{\vX | \tranrv = 1}(\vx)$ & $p(\vx)$ & $\rho^{\wt{\vx}}(1 - \rho)^{n - \wt{\vx}}$ \\  \hline
    $p_{\vY_w | \vX}(\vy_w | \vx)$ & $p(\vy_w | \vx)$ & $\pw^{\dist(\vx, \vy_w)} (1 - \pw)^{n - \dist(\vx, \vy_w)}$ \\ \hline
    $p_{\vY_w | \tranrv = 0}(\vy_w)$ & $p_0(\vy_w)$ & $\pw^{\wt{\vy_w}}(1 - \pw)^{n - \wt{\vy_w}}$  \\ \hline
    $p_{\vY_w | \tranrv = 1}(\vy_w)$ & $p_1(\vy_w)$ & $\sum_{\vx \in \code} \pw^{\dist(\vx, \vy_w)} (1 - \pw)^{n - \dist(\vx, \vy_w)} \times \rho^{\wt{\vx}}(1 - \rho)^{n - \wt{\vx}}$ \\ \hline
    $\Exp_\code(p_{\vY_w | \tranrv = 1}(\vy_w))$ & $\Exp_\code(p_1(\vy_w))$ & $(\rho \ast \pw)^{\wt{\vy_w}}(1 - \rho \ast \pw)^{n - \wt{\vy_w}}$ \\ \hline
    $p_{\vY_b|\vX}(\vy_b | \vx)$ & $p(\vy_b | \vx)$ & $\pb^{\dist(\vx, \vy_b)} (1 - \pb)^{n - \dist(\vx, \vy_b)}$ \\ \hline
    $p_{\vY_b | \tranrv = 0}(\vy_b)$ & $p_0(\vy_b)$ & $\pb^{\wt{\vy_b}}(1 - \pb)^{n - \wt{\vy_b}}$ \\ \hline
    $p_{\vY_b | \tranrv = 1}(\vy_b)$ & $p_1(\vy_b)$ & $\sum_{\vx \in \code} \pb^{\dist(\vx, \vy_b)} (1 - \pb)^{n - \dist(\vx, \vy_b)} \times \rho^{\wt{\vx}}(1 - \rho)^{n - \wt{\vx}}$ \\ \hline
    \hline
	Probability distribution & Simplified notation &  \\ \hline
	$\{ p_{\vY_w | \tranrv = 0}(\vy_w) \}$ & $\Pz$ &  \\ \hline
	$\{ p_{\vY_w | \tranrv = 1}(\vy_w) \}$ & $\Po$ &  \\ \hline
	$\{ \Exp_\code(p_{\vY_w | \tranrv = 1}(\vy_w)) \}$ & $\Exp_\code(\Po)$ &  \\ \hline
    \end{tabular}
	\caption[Probability and Distribution Notations]{Probability and Distribution Notations}\end{center}
\end{table}
\end{center}

\subsubsection{Definitions}
In the direct part of the proof, we need the following definitions.

\noindent{\underline{\bf Definitions used for deniability}}

For notational convenience we denote $\wt{\vy_w}/n$, the fractional Hamming weight of $\vy_w$, by $f_{\vy_w}(1)$, but in most usage, to further simplify notation, we shall simply write it as $\fyw$.
Following the definition of ``robust typicality''\footnote{Note that this definition is slightly different than that of ``strong typicality'', for instance in~\cite{cover2012book}, since the ``width'' of the set is proportional to corresponding values of the probability distribution for robustly typical sets. Such a definition is useful when (as in our case) some values in the $n$-letter probability distribution itself might be ``very small''.} (for instance, see~\cite{OrlR:01}) we define $\typyw$ as the {\it narrow typical set of $\vy_w$} as the set of sequences whose Hamming weight is in a range around $\rho \ast \pw$. Here the subscript $1$ in $\typyw$ denotes the fact that Alice is transmitting ($\tranrv = 1$). That is, 

\begin{definition}[Narrow typical set of $\vY_w$ when $\tranrv = 1$]
\begin{equation}
\typyw \triangleq \left \{ \vy_w : \fyw \in ( \rho \ast \pw (1 - \delyw), \rho \ast \pw (1 + \delyw) \right \}. 
\label{def_typyw}
\end{equation}
\end{definition}

\begin{remark}
One could also define the set $\mathcal{A}_0(\vY_w)$, {\it i.e.}, the set of all typical sequences conditioned on the fact that Alice is silent ($\tranrv = 0$). This would comprise of the set of all length-$n$ binary vectors of weight approximately $n \pw$. However, since this set will not be used in our proofs, we do not define it explicitly.
\end{remark}

\begin{remark}
Later in Lemma~\ref{lemma:achi_deni_EcP1_P1} equation~\eqref{eq:mo1_delyw}, we choose $\delyw$ to scale as $\Oh{1/\sqrt{n}}$ -- this is about as narrow as a typical set can get, and still be a high probability set. Indeed, the choice of $\delyw$ determines ``how typical'' this set is.
\end{remark}

Also, for each $j,j' \in \{0,1\}$ we use $f_{\vx, \vy_w}(j,j')$ to denote the fraction of indices $i \in \{1, \ldots, n\}$ such that the $i$-th components of $\vx$ and $\vy_w$ are respectively $j$ and $j'$. Hence $f_{\vx, \vy_w}(0,0), f_{\vx, \vy_w}(0,1), f_{\vx, \vy_w}(1,0)$ and $f_{\vx, \vy_w}(1,1)$ respectively denote the fractions of $(0,0), (0,1), (1,0)$ and $(1,1)$ pairs in $(\vx,\vy_w)$, but in most usage, to further simplify notation, we shall simply write them as $\fzzw, \fzow,\fozw$ and $\foow$. Next, we define $\typexgyw$ to be the {\it conditionally type-class} of $\vX$ given a particular $\vy_w$ and $\typxgyw$ to be the {\it narrow conditionally typical set} of $\vX$ given a particular $\vy_w$  as follows

\begin{definition}[Conditional type of $\vX$ given $\vy_w$]
\begin{equation}
	\typexgyw \triangleq \left \{ (\vx,\vy_w): 
	\begin{array}{ll}
		\frac{|\{i: (x_i, y_{w,i})=(1,0)|}{n} = \fozw ,\\ 
		\frac{|\{i: (x_i, y_{w,i})=(1,1)|}{n} = \foow
	\end{array}	   
    \right \}.
\label{def:typexgyw}
\end{equation}
\end{definition}

Next, we define a set $\typFw$ to be the set of pairs $(\fozw, \foow)$ such that the type-classes are typical with respect to $(\fozw, \foow)$ as follows,
\begin{small}
\begin{equation} \nonumber
	\typFw \triangleq \left\{ (\fozw, \foow):
		\begin{array}{ll} 
			\fozw &  \in ( \rho \pw (1 - \delozw),  \rho \pw (1 + \delozw))  ,\\ 
			\foow &  \in ( \rho (1 - \pw) (1 - \deloow),  \rho (1 - \pw) (1 + \deloow)), \\
			n \fozw & \in \mathcal{Z}, \\
			n \foow & \in \mathcal{Z}.
		\end{array}
	\right\}.
\end{equation}
\end{small}

Therefore, we define the narrow conditionally typical set of $\vX$ given $\vy_w$ when $\tranrv = 1$ in the following definition.

\begin{definition}[Narrow conditionally typical set of $\vX$ given $\vy_w$ when $\tranrv = 1$] 
\begin{equation}
	\typxgyw \triangleq \left \{ \vx: (\fozw, \foow) \in \typFw \right \}.
\label{eq:defn_typxgy}
\end{equation}
\end{definition}

Here the subscript $1$ in $\typxgyw$ denote the fact that Alice is transmitting ($\tranrv = 1$).

\begin{remark}
This definition makes sense since the ``expected'' value of $\fozw$ and $\foow$ are respectively $n \rho \pw$ and $n \rho (1 - \pw)$. Also, note that given a tuple $(\fyw, \fozw, \foow)$, the values of $\fzow$ and $\fzzw$ can be computed (as $\fyw - \foow$ and $1-\fyw-\fozw$ respectively) -- hence the joint type is completely determined by $(\fyw, \fozw, \foow)$. For this set $\typxgyw$, $\delozw$ and $\deloow$ shall be chosen to scale as $\Oh{n^{-3/4}}$. As we shall see in Lemma~\ref{lemma:achi_deni_EcP1_P1} equation~\eqref{eq:mo1_delozw}, this is again about as narrow a choice as can be made that still guarantees concentration of measure.
\end{remark}

Note that $\typxgyw$ can be written as the union of the conditional types of $\vx$ given $\vy_w$ such that each component of the pair $(\fozw, \foow)$ is within a certain interval. Mathematically,
\begin{equation}
	\typxgyw = \bigcup_{(\fozw, \foow) \in \typFw} \typexgyw.
\end{equation}

\noindent {\underline{\bf Definitions used for reliability}}

Similar to the definitions in the previous section on deniability, we have the following definitions.

\begin{definition}[Narrow typical sets of $\vY_b$]
Here we define two narrow typical sets for $\vY_b$, depending on whether Alice was transmitting or not.
\begin{itemize}
	\item Narrow typical set of $\vY_b$ when $\tranrv = 0$,
	\begin{equation}
		\bobtypyz = \{ \vy_b : \fyb \in (\pb (1  - \delybz), \pb (1 + \delybz)) \}.
		\label{def:bobtypyz}
	\end{equation}

	\item Narrow typical set of $\vY_b$ when $\tranrv = 1$,
	\begin{equation}
		\bobtypyo = \{ \vy_b : \fyb \in (\rho \ast \pb (1  - \delybo), \rho \ast \pb (1 + \delybo)) \}.
		\label{def:bobtypyo}
	\end{equation}
\end{itemize}
\end{definition}

\begin{remark}
Later in Claim~\ref{claim:delybz_val} and Claim~\ref{claim:delybo_val}, we choose $\delybz$ and $\delybo$ to scale as $\Oh{1/\sqrt{n}}$, where the superscript corresponds to different transmission status to be $0$ and $1$ respectively. The reason for choosing such narrow typical sets is because we are trying to distinguish a very low-weight (here and everywhere else in the thesis) codeword from relatively high-weight background noise.
\end{remark}

Let $\fzzb, \fzob,\fozb$ and $\foob$ respectively denote the fractions of $(0,0), (0,1), (1,0)$ and $(1,1)$ pairs in $(\vx,\vy_b)$. Then the {\it conditionally type-class} of $\vX$ given a particular $\vy_b$ when $\tranrv = 1$ $\bobtypexgyb$ and the {\it narrow conditionally typical set} of $\vX$ given a particular $\vy_b$ when $\tranrv = 1$, $\bobtypxgyb$, are defined as follows

\begin{definition}[Conditional type of $\vX$ given $\vy_b$]
\begin{equation}
	\bobtypexgyb \triangleq \left \{ (\vx,\vy_b): 
	\begin{array}{ll} 
		\frac{|\{i:(x_i, y_{b,i}) = (1,0)\}|}{n} = \fozb ,\\ 
		\frac{|\{i:(x_i, y_{b,i}) = (1,1)\}|}{n} = \foob
	\end{array}	   
    \right \}.
\end{equation}
\end{definition}

We define a set $\typFb$ to be the set of pairs $(\fozb, \foob)$ such that the type-classes are typical with respect to $(\fozb, \foob)$ as follows,
\begin{small}
\begin{equation} \nonumber
	\typFb \triangleq \left\{ (\fozb, \foob):
		\begin{array}{ll} 
			\fozb &  \in ( \rho \pb (1 - \delozb),  \rho \pb (1 + \delozb))  ,\\ 
			\foob &  \in ( \rho (1 - \pb) (1 - \deloob),  \rho (1 - \pb) (1 + \deloob)), \\
			n \fozb & \in \mathcal{Z}, \\
			n \foob & \in \mathcal{Z}.
		\end{array}
	\right\}.
\end{equation}
\end{small}

Therefore, we define the narrow conditional typical set of $\vX$ given $\vy_b$ given $\tranrv = 1$ as follows,
\begin{definition}[Narrow conditional typical set of $\vX$ given $\vy_b$]
\begin{equation}
 \bobtypxgyb \triangleq \{ \vx: (\fozb, \foob) \in \typFb \}.
\label{def:bobtypxgyb}
\end{equation}
\end{definition}

\begin{remark}
For this set $\bobtypxgyb$, $\delozb$ and $\deloob$ shall be chosen to scale as $\Oh{n^{-3/4}}$. As we shall see in Lemma~\ref{lemma:achi_deni_EcP1_P1}, this is again about as narrow a choice as can be made that still guarantees concentration of measure.
\end{remark}

Using the above definitions, the conditionally typical set of $\vX$ given $\vy_b$, $\bobtypxgyb$, which equals the union of the conditional types of $\vX$ given $\vy_b$ such that each component of the pair $(\fozb, \foob)$ is within a certain interval. Mathematically,
\begin{equation}
	\bobtypxgyb = \bigcup_{(\fozb, \foob) \in \typFb} \bobtypexgyb
\end{equation}


\begin{definition}[Empirical entropy, empirical conditional entropy, empirical mutual information and empirical Kullback-Leibler divergence]
Given length-$n$ vectors $\vx$ and $\vy$, denote the fractional weight of $\vx$ and $\vy$ as $f_{1*}$ and $f_{*1}$ respectively. Note that $f_{*1}$ equals $f_{11} + f_{01}$, and similarly $f_{1*} = f_{11} + f_{10}$.

In later sections we shall be interested in properties of combinatorial objects such as sizes of type-classes, and the probability of that a ``randomly chosen vector'' falls within a specific type-class. As well-known classically due to the work of Csiszar and his ``method of types''~\cite{csiszar1998method}, the calculations of these properties of combinatorial objects are related to the specific functions on the empirical distribution of the type of some $\vx$, $\vy_b$ or $\vy_w$ vectors under consideration. Therefore, we define the following empirical information-theoretic quantities to help with the calculations. 

For $j,j' \in \{0, 1\}$, we use $f_{jj'}$ to denote the fraction of indices $i \in \{1, \ldots, n\}$ such that the $i$-th components of $\vx$ and $\vy$ are respectively $j$ and $j'$. Then,
\begin{enumerate}
	\item \underline{Empirical entropy}: The empirical entropy of a vector $\vx$ is the entropy of the empirical distribution of $\vx$. The ``physical meaning'' of the empirical entropy of a vector corresponds to the fact that the normalized log-volume of the type-class containing is the empirical entropy of $\vx$, {\it i.e.}, $|\mathcal{T}(\vx)| \doteq 2^{n \entropy{f_1*}}$, where $\mathcal{T}(\vx) = \{\vx' : \wt{\vx'} = \wt{\vx}\}$.	
	$$ \entropy{\vx} \triangleq \sum_{j \in \{0,1\}}f_{1*} \log \frac{1}{f_{1*}}; $$
	\item \underline{Empirical conditional entropy}: $$ \condentropy{\vx}{\vy} \triangleq \sum_{j' \in \{0,1\}} f_{*j'} \entropy{\frac{f_{1j'}}{f_{*j'}}}; $$
	\item \underline{Empirical Kullback-Leibler divergence}: The empirical Kullback-Leibler divergence of a vector $\vx$ is the Kullback-Leibler divergence between the empirical distribution of $\vx$ and the generation parameter of $\vX$, {\it i.e.},
	$$ \KL{\vx}{\rho} \triangleq f_{0*} \log \frac{f_{0*}}{1-\rho} + f_{1*} \log \frac{f_{1*}}{\rho}; $$
	\item \underline{Empirical mutual information}: The empirical mutual information between $\vx$ and $\vy$ is the Kullback-Leibler divergence between the empirical distribution $f_{jj'}$ and $f_{j*} f_{*j'}$, {\it i.e.},
	$$ \mutual{\vx}{\vy} \triangleq \sum_{(j, j') \in \{0,1\} \times \{0,1\}} f_{jj'} \log \frac{f_{jj'}}{f_{j*}f_{*j'}}. $$
\end{enumerate}
\end{definition}


\subsection{High Level Intuition}

\subsubsection{Converse of Fixed Channel Model (Theorem~\ref{thm:model:known:conv})}

\begin{figure}
		\centering
		\includegraphics[width=1.0\columnwidth]{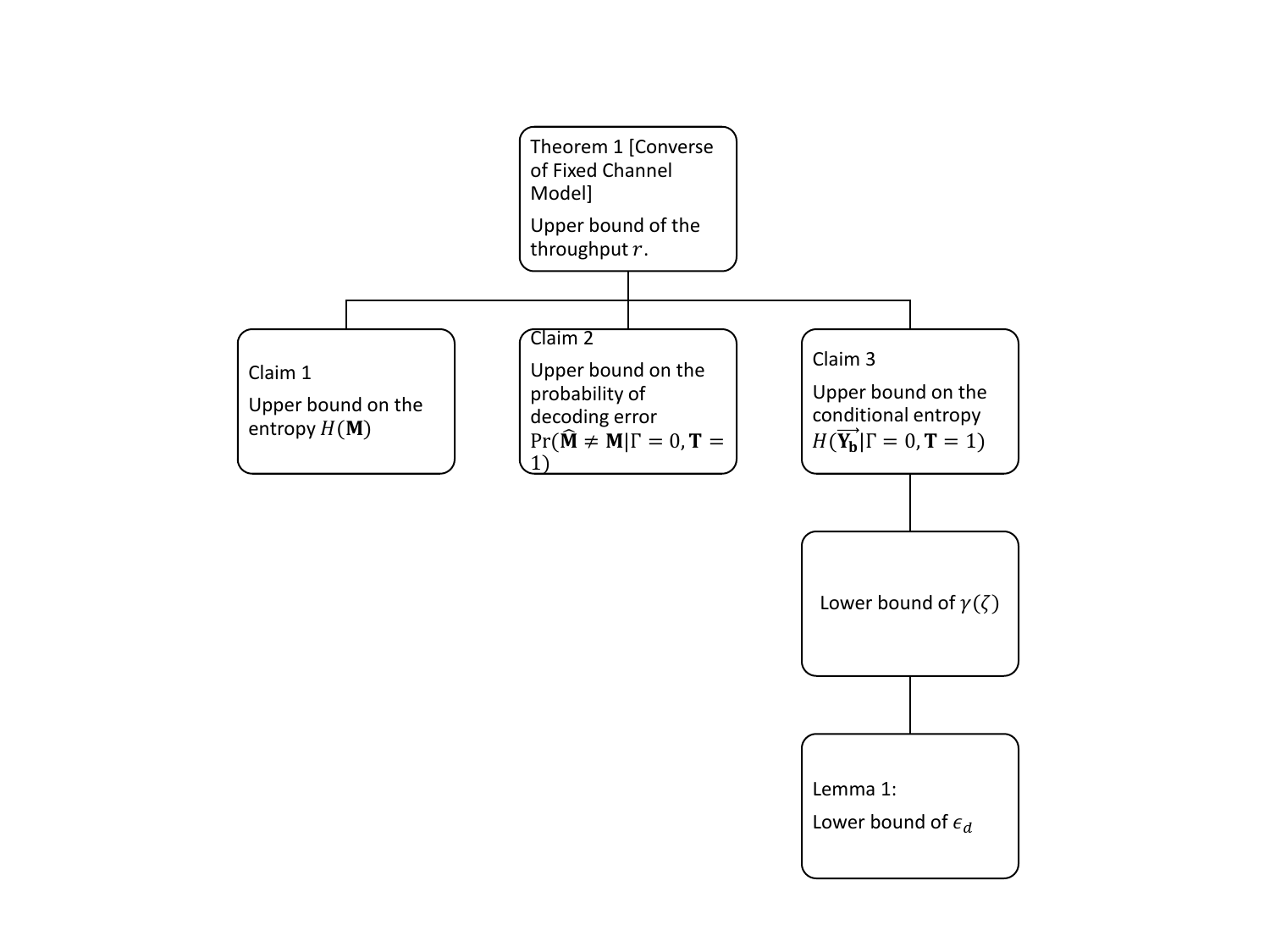}
		\caption[Structure describing the proof of Theorem~\ref{thm:model:known:conv}]{The structure of proving Theorem~\ref{thm:model:known:conv}.}
		\label{fig:model:known:conv:flow}
\end{figure}

The proof of Theorem~\ref{thm:model:known:conv} consists of two major parts, the first part is the deniability of the converse, and the second part is the reliability of the converse:
\begin{enumerate}
	\item (Lemma~\ref{lemma:low_bd_epsdeni}: {\bf Lower bound on the deniability parameter $\epsdeni$}) Roughly speaking, if the codebook $\code$ has too many ``high weight'' codewords, Willie can easily distinguish Alice's transmission status by a simple ``threshold detector''-based estimator. So, most of the probability mass of a codebook that is deniable from Willie must be in ``low-weight codewords''. In particular, if the codebook $\code$ has a probability $\gamma(\zeta)$ of codewords with fractional weight greater than $\zeta$, where
	\begin{equation} \label{eq:def:gamma_rho}
		\gamma(\zeta) = \sum_{\vx \in \code} \Pr(\vX = \vx) \mathbbm{1}(\wt{\vx} \geq \zeta n).
	\end{equation}
\begin{remark}
For a fixed codebook, $\gamma$ is a function of $\zeta$.
\end{remark}
	
	The probability $\Pr(\vX = \vx)$ is over any stochasticity in the encoder. (Since this is a converse argument, it must hold for all codes, including those using private randomness at the encoder to generate possibly many non-uniformly distributed codewords for even a single message. Indeed, stochasticy at the encoder is an important component of many information-theoretically secure schemes — see, for instance, the classical text~\cite{csiszar81}).
	
	We show in Lemma~\ref{lemma:low_bd_epsdeni} below that Willie's deniability parameter $\epsdeni$ must satisfy
	\begin{equation} \label{eq:upbd:epsdeni}
		\epsdeni \geq \gamma(\zeta) \cdot \left( 1 - \frac{8 \pw (1 - \pw)}{n \zeta^2 (1 - 2\pw)^2} \right).
	\end{equation}
	
	\item ({\bf Upper bound on the throughput $\thr(\epsreli, \epsdeni)$}) Using the result of Lemma~\ref{lemma:low_bd_epsdeni} as a constraint on any ``good'' code (one that is both highly deniable and highly reliable), together with three claims, we use standard information-theoretic converse arguments to bound Alice's optimal rate of $(1-\epsdeni)$-deniable $(1-\epsreli)$-reliable communication with Bob.
\end{enumerate}


\subsubsection{Achievability of Fixed Channel Model (Theorem~\ref{thm:model:known:achi})}

\begin{figure}
		\centering
		\includegraphics[width=1.0\columnwidth]{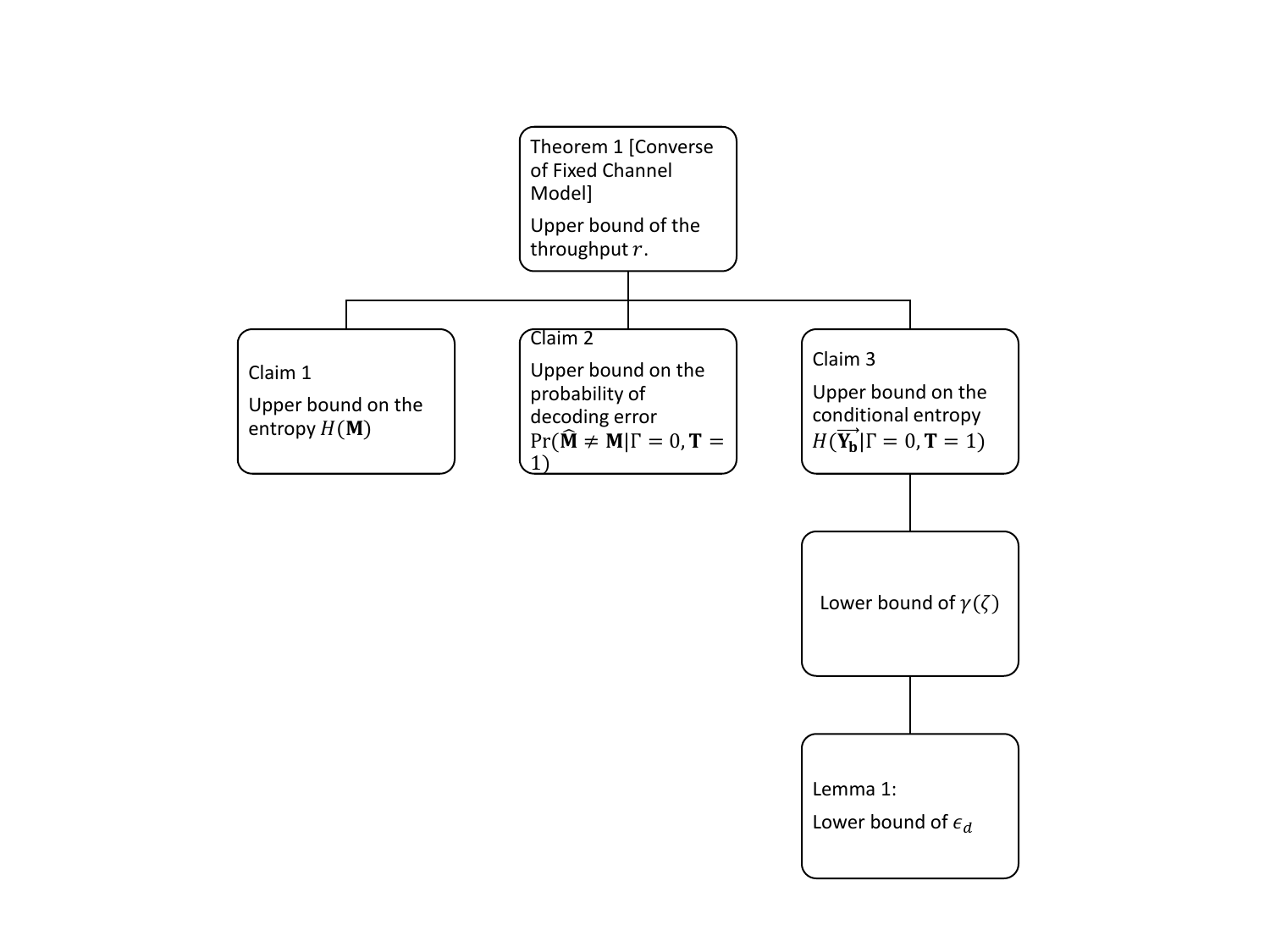}
		\caption[Structure describing the proof of Theorem~\ref{thm:model:known:achi}]{The structure of  proving Theorem~\ref{thm:model:known:achi}.}
		\label{fig:model:known:achi:flow}
\end{figure}

The achievability consists of two parts:
\begin{enumerate}
	\item ({\bf Deniability}) We first prove that an appropriately chosen random code $\code$ {\it also} has overwhelming probability of being highly deniable.

Recall that a code $\code$ is $(1 - \epsdeni)$-deniable if for every estimator $Est_\code(\cdot)$ of Willie, 
\begin{equation}
	\alpha(Est_\code(\vY_w)) + \beta(Est_\code(\vY_w)) \geq 1 - \epsdeni. 
	\label{eq:est}
\end{equation}
But by ``standard statistical arguments''~\cite[Theorem 13.1.1]{lehmann2006testing} (reprised in~\cite{BasGT:12J} as Fact~$1$), (\ref{eq:est}) is implied by the condition that 
\begin{equation}
	\variation{\Pz}{\Po} \leq \epsdeni,
	\label{eq:var}
\end{equation}
where $\Pz$ corresponds to the probability distribution of $\vy_w$ when Alice's transmission status $\tranrv = 0$, and $\Po$ corresponds to the probability distribution of $\vy_w$ when Alice's transmission status $\tranrv = 1$.

The following three figures shows the brief idea about proving $\variation{\Pz}{\Po}$ is small. Figure~\ref{fig:pw} shows the distribution that Alice does not transmit, and Figure~\ref{fig:ps} show the distribution when Alice does transmit. We see that the distribution when Alice does transmit is not ``regular'', it is hard to show $\variation{\Pz}{\Po}$ is small directly. Therefore, we introduce the ``ensemble-averaged'' distribution when Alice does transmit over all codebooks in Figure~\ref{fig:pmfs}. Hence, we see that it is easier to show $\variation{\Pz}{\Exp_\code(\Po)}$ and $\variation{\Exp_\code(\Po)}{\Po}$ is small respectively, where the ``ensemble-averaged'' distribution by $\Exp_\code(\Po)$, {\it i.e.}, $\Exp_\code(\Po)(\vy_w) = \Exp_\code(\Po(\vy_w))$ for all $\vy_w \in \{0,1\}^n$.

\begin{figure}
  \centering
  \includegraphics[width=0.8\columnwidth]{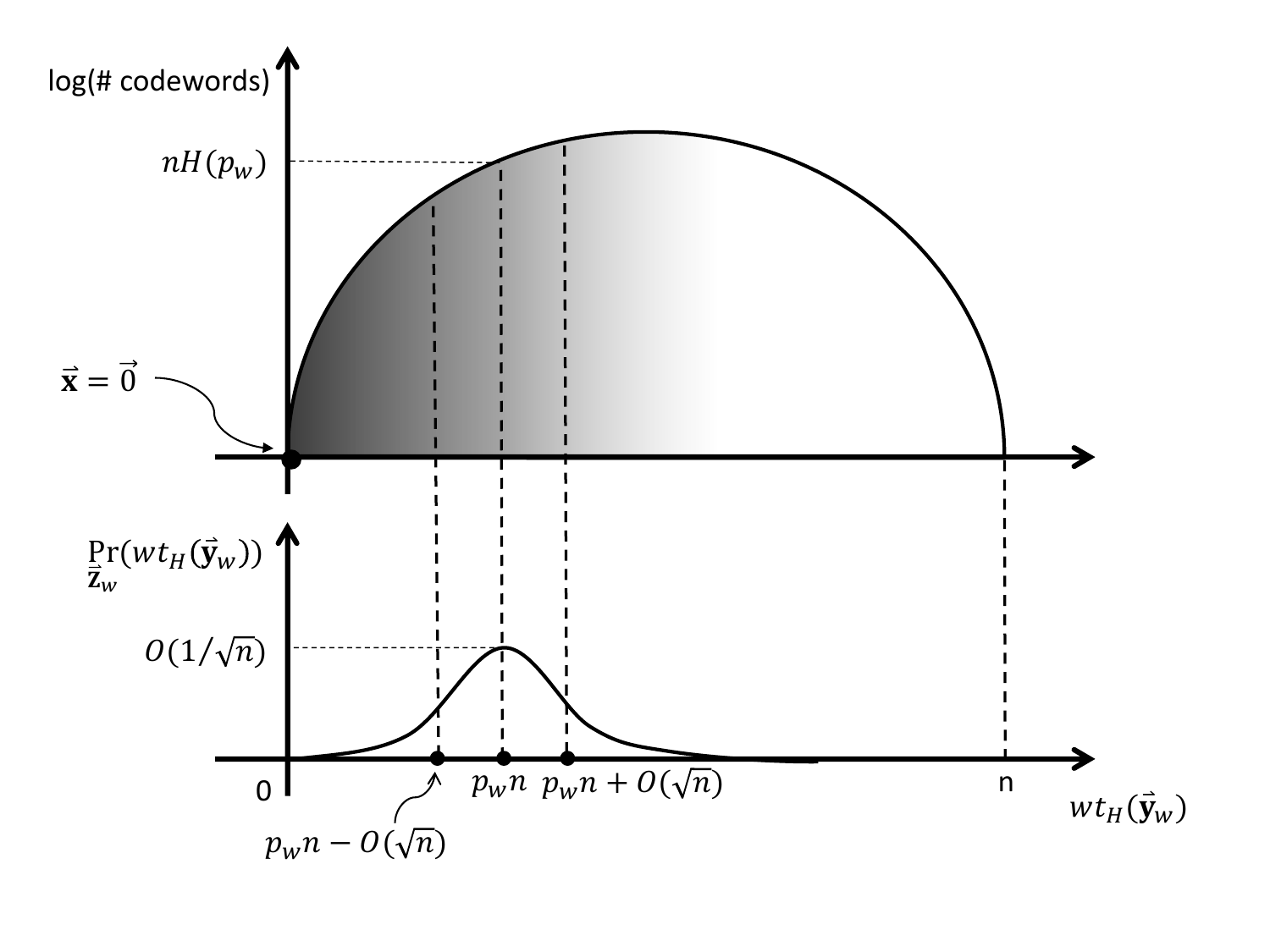}
  \caption[Willie's observation if Alice does not transmit]{{\small{{\bf\underline{Willie's observation if Alice does not transmit:}} The upper curve represents the set of all possible $\vy_w$ that Willie may observe if Alice transmits nothing. The $\vy_w$ are arranged so that vectors with lower Hamming weight are to the left of vectors with higher Hamming weight, and the height of the enclosing curve (the binary entropy function) denotes the (logarithm of the) number of binary vectors of a particular Hamming weight. Hence the shaded region denotes the set of ``likely'' $\vy_w$ that Willie observes, with the density denoting probability of observing corresponding $\vy_w$s. The curve at the bottom plots the probability distribution of observing $\vy_w$ of a particular Hamming weight. 
Since Alice's transmitted codeword is $\mathbf{\vec{0}}$, the ``typical'' $\vy_w$ that Willie observes are of weight approximately $\pw n$ (with a variation of $\Oh{\sqrt{n}}$). This curve is ``smooth'' and follows a binomial distribution.}}}
  \label{fig:pw}
\end{figure}

\begin{figure}
  \centering
  \includegraphics[width=0.8\columnwidth]{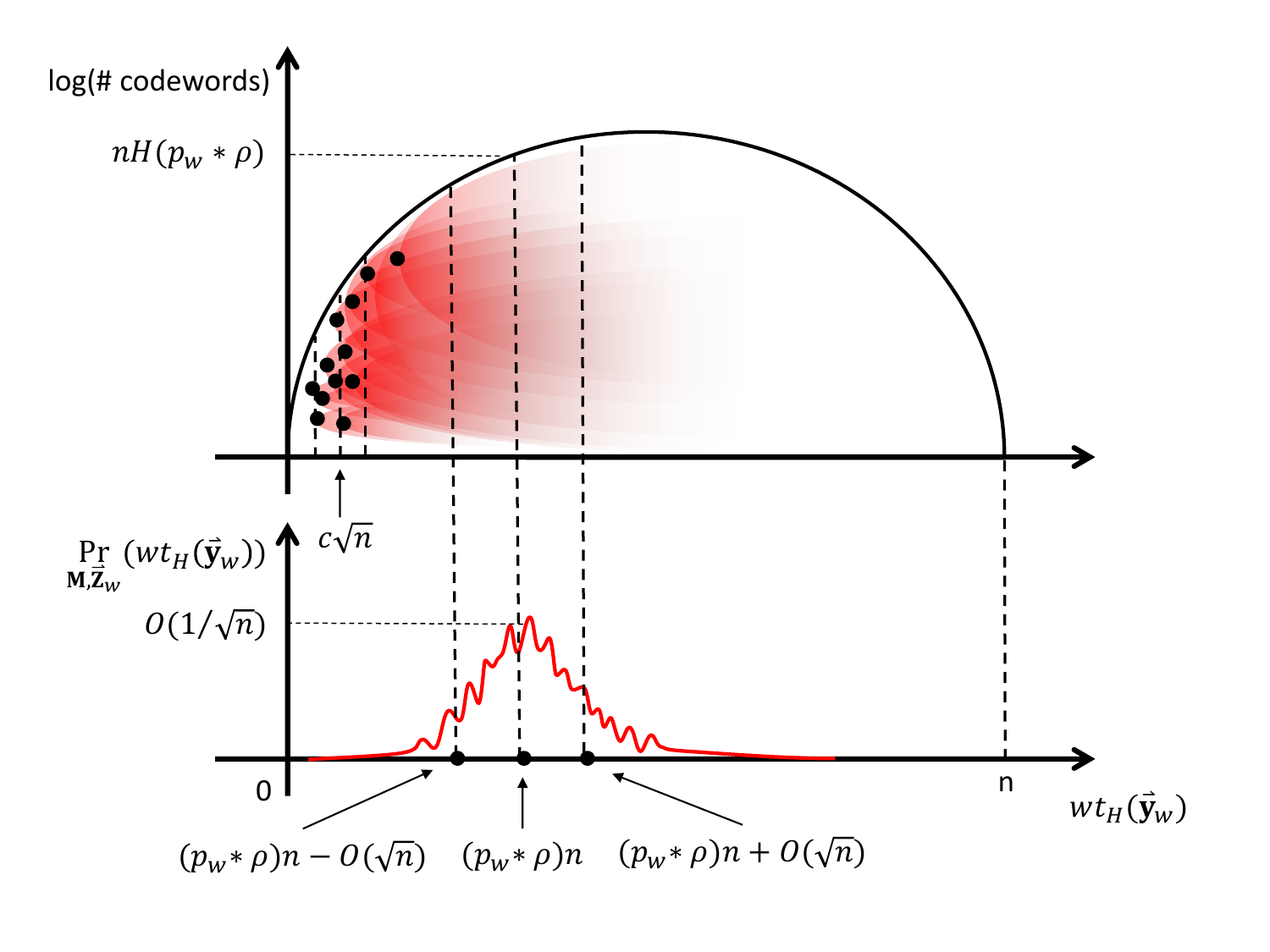}
  \caption[Willie's observation if Alice transmits]{{\small{{\bf\underline{Willie's observation if Alice transmits:}} The red region denotes the set of $\vy_w$ that Willie may observe if Alice transmits a codeword. The black dots on the left denote codewords of $\code$. If Alice transmits a particular $\vx$, the set of $\vy_w$ that Willie is likely to observe is shown by the red paraboloid region extending rightwards from that $\vx$. The overall probability distribution over Willie's observed $\vy_w$ is hence the ``average'' of the paraboloid regions. 
  In this case 
  the probability distribution on $\vy_w$ is somewhat ``lumpy'', since the probability that Willie observes a particular $\vy_w$ depends on the distribution of the Hamming distance between that particular $\vy_w$ and the set of codewords $\vx \in \code$. 
  So the weight distribution of $\vy_w$ is a weighted sum of binomial distributions. }}}
  \label{fig:ps}
\end{figure}

\begin{figure}
  \centering
  \includegraphics[width=0.8\columnwidth]{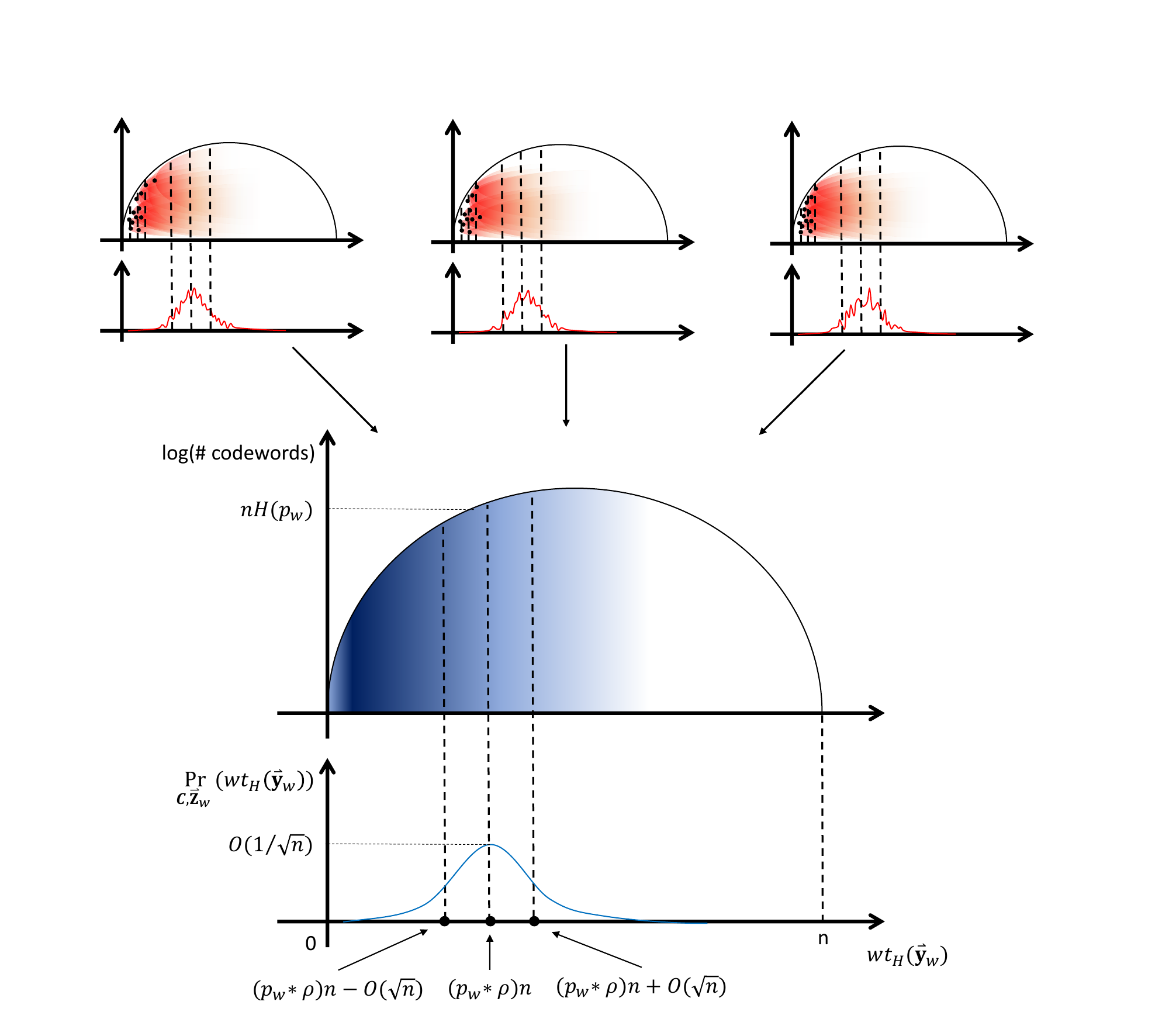}
  \caption[Deniability from Willie]{\small{{\bf\underline{Deniability from Willie:}} Our proof that a random codebook $\code$ chosen with the ``right'' parameters (number of codewords, expected weight of codewords) proceeds as follows. We need to demonstrate that the probability distributions $\prob_{\vZ_w}(\vy_w|\tranrv= 0)$ and $\prob_{\messrv,\vZ_w}(\vy_w|\tranrv= 1)$ are ``close'' (in variational distance). However, since the latter distribution is complex (due to its dependence on the specific codebook $\code$), we do this comparison in two stages. We first compute the {\it ensemble distribution} of $\vy_w$, {\it i.e.}, the ``smooth blue'' region/curve denoting the ``ensemble average'' (over all suitably chosen random codebooks) of the probability distribution on $\vy_w$.
We then  demonstrate that the probability distribution $\prob_{\vZ_w}(\vy_w|\tranrv= 0)$ and the ensemble distribution $\prob_{\code,\messrv,\vZ_w}(\vy_w|\tranrv= 1)$ ({\it i.e.} the weighted average over all possible codebooks $\code$ of the latter distribution) are ``close''. Finally, we prove that with high probability over the choice of codebooks $\code$, the distribution of $\prob_{\messrv,\vZ_w}(\vy_w|\tranrv= 1)$ is tightly concentrated around its expectation $\prob_{\code,\messrv,\vZ_w}(\vy_w|\tranrv= 1)$. This figure visually depicts deniability in Theorem~\ref{thm:model:known:achi}.
}}
  \label{fig:pmfs}
\end{figure}

By the triangle inequality, the left hand side of~\eqref{eq:var} can be bounded from above as follows
\begin{equation}
	\variation{\Pz}{\Po} \leq \variation{\Pz}{\Exp_\code(\Po)} + \variation{\Exp_\code(\Po)}{\Po}.
\label{eq:var_dist_tri_ineq}
\end{equation}
So, instead of showing~\eqref{eq:var} directly, we show $\variation{\Pz}{\Exp(\Po)} < \epsdeni$ and $\Pr_\code\left[\variation{\Exp_\code(\Po)}{\Po} < 2^{-\Omega(n^\delta)}\right]$ with exponentially high probability for some $\delta > 0$.

	\item ({\bf Reliability}) Since the proof of the reliability in the achievability is a special case of Proposition~\ref{prop:ran_sto_code}, we will describe briefly of the proof of Proposition~\ref{prop:ran_sto_code}.
	
	We show that if Bob uses a carefully chosen typicality-based decoder with the decoding rule, then his probability of decoding error (regardless of Alice's transmission status) is exponentially small.
\end{enumerate}

\subsubsection{Converse of Slow Fading Channel Model (Theorem~\ref{thm:model:unknown:conv})}

\begin{figure}
		\centering
		\includegraphics[width=1.0\columnwidth]{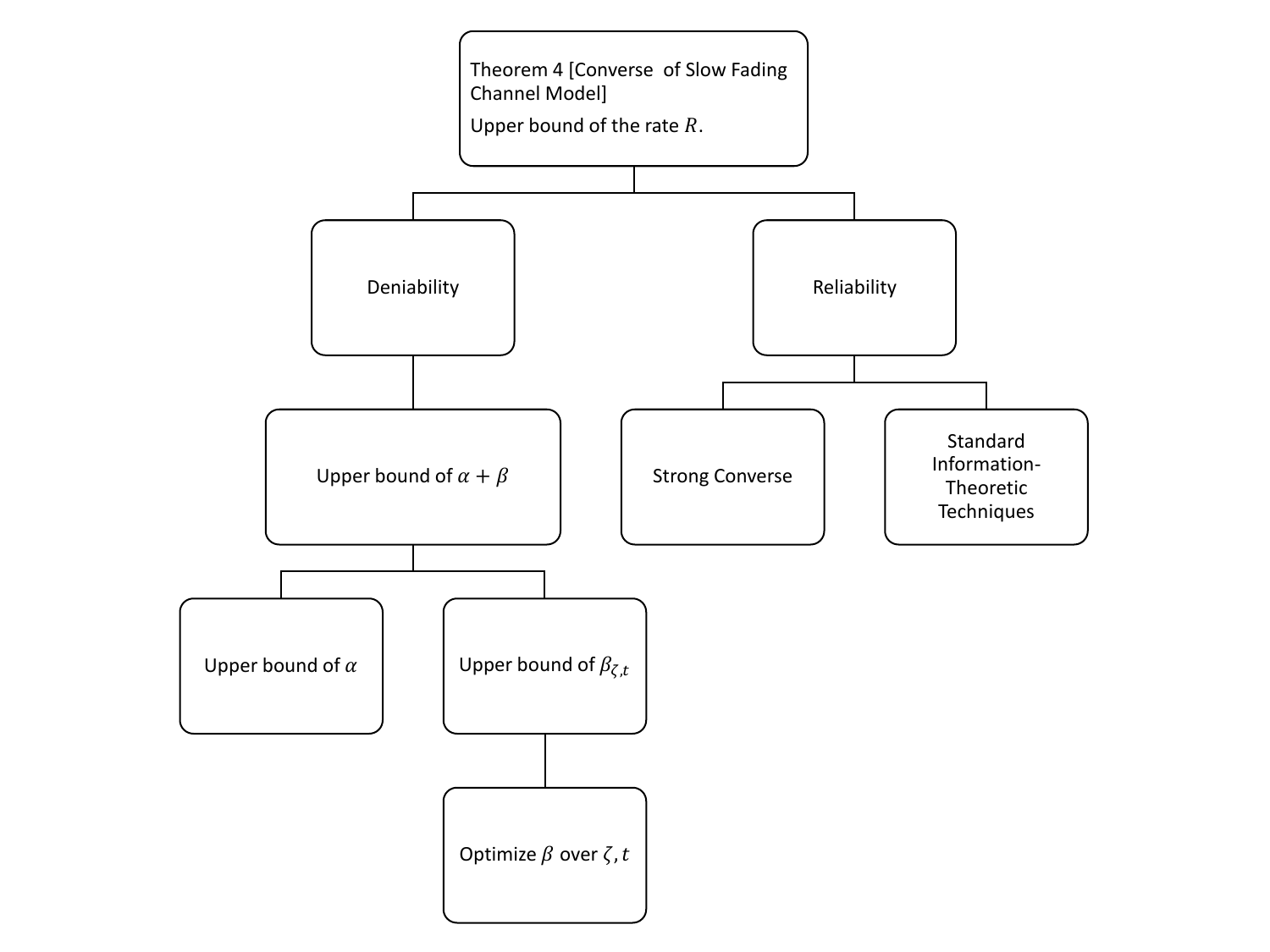}
		\caption[Structure describing the proof of Theorem~\ref{thm:model:unknown:conv}]{The structure of proving Theorem~\ref{thm:model:unknown:conv}.}
		\label{fig:model:unknown:conv:flow}
\end{figure}

Similar to the Fixed Channel Model, we proceed the proof of the outer bound of the reliable-deniable rate in Slow Fading Channel Model in two stages.
\begin{enumerate}
	\item ({\bf Upper bound on $\alpha + \beta$}) As in the Fixed Channel model, given any code with more than a certain probability mass $\gamma(\zeta)$ of codewords with Hamming weight above a certain threshold $\zeta$, we demonstrate that the deniability of the code is bounded from above by an explicitly computable function of $\zeta$ and $\gamma(\zeta)$.
	\item ({\bf Upper bound on the rate $\rate$}) The idea of the calculation of the rate $\rate$ is to find a ``sweet spot'' of $\pb \in (\lb, \ub)$ and $\epsreli$. At this ``sweet spot'', we could apply strong converse and information-theoretic inequalities to find the upper bound of the rate $\rate$.
\end{enumerate}

\subsubsection{Achievability of Slow Fading Channel Model (Theorem~\ref{thm:model:unknown:achi})}

\begin{figure}
		\centering
		\includegraphics[width=1.0\columnwidth]{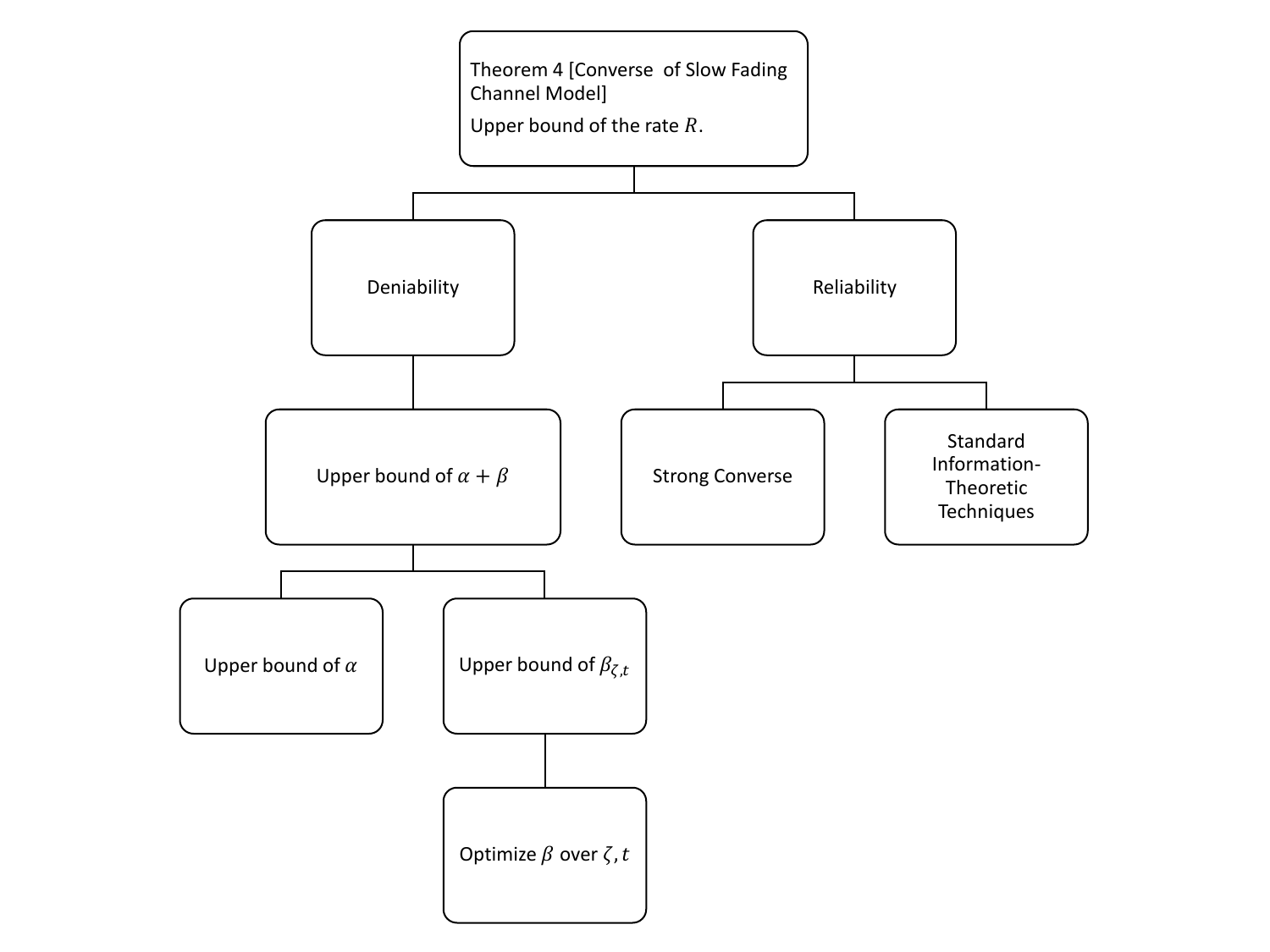}
		\caption[Structure describing the proof of Theorem~\ref{thm:model:unknown:achi}]{The structure of proving Theorem~\ref{thm:model:unknown:achi}.}
		\label{fig:model:unknown:achi:flow}
\end{figure}

We prove Theorem~\ref{thm:model:unknown:achi} in two parts, deniability and reliability.
\begin{enumerate}
	\item ({\bf Deniability}) The proof of this part is similar to the deniability part of Theorem~\ref{thm:model:known:achi}. We break the variational distance $\variation{\Pz}{\Po}$ into two parts $\variation{\Pz}{\Exp_\code(\Po)}$ and $\variation{\Exp_\code(\Po)}{\Po}$. We show that the variational distance $\variation{\Pz}{\Exp_\code(\Po)} < \epsdeni$ if the codebook generation parameter $\rho < \frac{\uw - \lw}{1 - 2\lw} \epsdeni$. Then, we show that $\variation{\Exp_\code(\Po)}{\Po} < 2^{-\Omega(n^\delta)}$ with high probability. Therefore, combining two parts, we have $\variation{\Pz}{\Po} < \epsdeni$ with high probability.
	\item ({\bf Reliability}) In this proof, Bob first uses the same technique as described in the deniability proof in the converse (Theorem~\ref{thm:model:unknown:conv} to determine whether Alice's transmission status $\tranrv$ is $0$ or $1$. If Alice's transmission status is $0$, then Bob decodes $\hat{\vM} = 0$. Otherwise, Bob uses the standard maximum likelihood decoding rule to find the message $\hat{\vM}$.
\end{enumerate}

\subsection{Converse of Fixed Channel Model (Theorem~\ref{thm:model:known:conv})}
\label{sec:fixed_channel:conv}

\begin{lemma}[{\bf Lower Bound on the Deniability Parameter $\epsdeni$}]
\label{lemma:low_bd_epsdeni}
The code $\code$ has a probability $\gamma$ of codewords with fractional weight greater than $\rho(\gamma)$, then
	\begin{equation}
		\epsdeni \geq \gamma(\zeta) \cdot \left(1 - \frac{8 \pw (1 - \pw)}{n \zeta^2(1 - 2\pw)^2}\right).
	\end{equation}
\end{lemma}

\begin{proof}
The goal of this lemma is to give a lower bound on the deniability parameter $\epsdeni$ for any code that has ``too many'' high-weight codewords (more precisely, too much of the probability mass of the codebook is concentrated in high-weight codewords). To prove this, we assume that Willie uses a simple threshold estimator based on the fractional weight of his received vector $\vY_w$. Note that since we are proving an outer bound on throughput of any code in this section, analysis of a non-optimal estimator used by Willie suffices for our purposes, as long as it gives a non-trivial outer bound bound on the deniable throughput Alice manages to get through.
\footnote{Indeed, the optimal estimator seems hard to analyze. The reason is that the optimal estimator is based on hypothesis testing, and depends non-trivially on the idiosyncrasies of the codebook use. Also, the computational complexity of such an estimator may be high for Willie. Instead, we analyze a sub-optimal estimator, in which Willie only checks whether the fractional weight of $\vY_w$ exceed a certain explicitly computed threshold $\thre$ or not; this estimator has the added advantage (for Willie) of having computational complexity that is essentially linear in the block-length $n$. It is also a very ``natural'' estimator, trying to estimate the ``energy'' of the observed signal $\vY_w$. Indeed, this is the intuition that leads Bash {\it et. al.} to consider a similar estimator in \cite{BasGT:12J}. In this theorem, we broadly follow their lead.}

Let $\vS$ denote the fraction of $1$'s in Willie's received vector $\vY_w$. Let $\thre$ denote the {\it threshold} that estimates Alice's transmission status $\tranrv$. More precisely, Willie estimates $\hat{\tranrv}$ as $0$ if $\vS < \thre$, and $\hat{\tranrv}$ as $1$ otherwise. That is, the estimator
$$ Est(\vY_w) = \left\{ \begin{array}{ll}
1, & \mbox{if } \vS \geq \thre, \\
0, & \mbox{otherwise.}
\end{array} \right. $$

For an arbitrary threshold $\thre$, we use $\alpha(\thre)$ to denote the false alarm probability, that is, $\alpha(\thre) = \Pr_{\vZ_w}(\vS \geq \thre | \tranrv = 0)$. We then let $\beta(\thre,\vx)$ denote the missed detection probability for a particular transmitted codeword $\vx$, that is, $\Pr_{\vZ_w}(\vS < \thre | \tranrv = 1, \vX = \vx)$. And we use $\beta(\thre)$ to denote the overall missed detection probability, that is, $\beta(\thre) = \sum_{\vx \in \code} \Pr(\vX = \vx) \beta(\thre,\vx)$. We first compute $\alpha(\thre)$ and $\beta(\thre,\vx)$, and obtain $\beta(\thre)$ by averaging $\beta(\thre,\vx)$ over $\vx$. We then bound $\alpha(t) + \beta(t)$ from above using Chevbyshev's inequality, and finally optimize the upper bound on $\alpha(t) + \beta(t)$ via standard calculus techniques.

Note that when Alice does not transmit, Willie's received vector $\vY_w$ equals Willie's noise $\vZ_w$. Hence, the expected fractional weight of $\vY_w$, denoted by $\Exp_{\vZ_w}(\vS | \tranrv = 0)$, equals $\pw$, and the variance $\Var_{\vZ_w}(\vS | \tranrv = 0)$ equals $\frac{\pw (1 - \pw)}{n}$ -- these correspond respectively to the expected mean and variance of a Bernoulli $Bern(\pw)$ random variable.

On the other hand, if Alice transmits a codeword $\vx$ with fractional weight $\rho(\vx)$, then $\vY_w = \vX \oplus \vZ_w$. Hence, we have $\Exp_{\vZ_w}(\vS | \tranrv = 1, \vX = \vx) = \rho(\vx) \ast \pw $ (recall that $a \ast b$ denotes the binary convolution of $a$ and $b$), and $\Var_{\vZ_w}(\vS | \tranrv = 1, \vX = \vx) = \frac{\pw (1 - \pw)}{n}$ (which corresponds to the variance of $\vZ_w$.)

By Chebyshev's inequality, we have
\begin{equation}
	\alpha(\thre) 	= 		\prob(\bS \geq \thre) 
			\leq	\prob(|\bS - \pw| \geq \thre - \pw) 
			\leq 	\frac{\pw(1-\pw)}{n (\thre - \pw)^2}.
\label{eq:alpha}
\end{equation}

Similarly, when Alice transmits a codeword $\vx$ with fractional weight $\zeta(\vx) = \frac{\wt{\vx}}{n}$ we have via Chebyshev's inequality
\begin{align}
	\beta(\thre, \vx) &= \prob(\bS < \thre) \nonumber \\
		& \leq 			\prob(|\bS - \zeta(\vx) \ast \pw|) \geq \zeta(\vx) \ast \pw - \thre) \nonumber \\
		& \leq 	\frac{\pw(1 - \pw)}{n(\zeta(\vx) \ast \pw - \thre)^2}. \label{eq:beta}
\end{align}
Notice that for a fixed $\thre$, the RHS of equation~\eqref{eq:alpha} is fixed and increases as $\zeta(\vx)$ decreases.

By optimizing the sum of~\eqref{eq:alpha} and~\eqref{eq:beta} with respect to $\thre$, we have that an outer bound is attained when $\thre$ equals $\frac{\pw + \zeta(\vx) \ast \pw}{2}$. This implies that
\[ \alpha + \beta(\vx) \leq \frac{8 \pw (1 - \pw)}{n \zeta(\vx)^2 (1 - 2\pw)^2}. \]
Therefore, the codebook $\code$ has a probability $\gamma(\zeta)$ of codewords with fractional weight greater than $\zeta$, we have
\begin{align}
	\alpha + \beta &= (1 - \gamma(\zeta)) (\alpha + \beta)|_{\wt{\vx} < \zeta n} + \gamma(\zeta) (\alpha + \beta)|_{\wt{\vx} \geq \zeta n} \nonumber \\
		& \leq  (1 - \gamma(\zeta)) \cdot 1 + \gamma(\zeta) \frac{8 \pw (1 - \pw)}{n \zeta^2 (1 - 2\pw)^2} \nonumber \\
		&= 1 - \gamma(\zeta) + \gamma(\zeta) \frac{8 \pw (1 - \pw)}{n \zeta^2 (1 - 2\pw)^2} \nonumber \label{eq:eta_eq_1}
\end{align}
Therefore, by our definition of the deniability parameter $\epsdeni$ (see Section~\ref{sec:a_to_w}), we obtain $\epsdeni \geq \gamma(\zeta) \left(1 - \frac{8 \pw (1 - \pw)}{n \zeta^2 (1 - 2\pw)^2}\right)$.
\end{proof}

From Lemma~\ref{lemma:low_bd_epsdeni}, we note that
\[ 1 - \epsdeni < \alpha + \beta \leq 1 - \gamma(\zeta) + \gamma(\zeta) \left(1 - \frac{8 \pw (1 - \pw)}{n \zeta^2 (1 - 2\pw)^2}\right). \]
So, a sufficient condition for $(1 - \epsdeni)$-deniability is
\begin{equation}
1 - \epsdeni < 1 - \gamma(\zeta) + \gamma(\zeta) \left(1 - \frac{8 \pw (1 - \pw)}{n \zeta^2 (1 - 2\pw)^2}\right),
\label{eq:mo1:deni:cond}
\end{equation}
which is equivalent to
\begin{equation}
	\gamma(\zeta) > \frac{\epsdeni}{1 - \frac{\zeta^2(1 - 2\pw)^2 n}{8 \pw (1 - \pw)}}.
\label{eq:mo1:bd:rho}
\end{equation}

Now, we define the indicator function $\Gamma$ as,
\[ \Gamma = \left\{ \begin{array}{ll} 1, & \mbox{if $\wt{\vx} > \zeta n$;} \\
0, & \mbox{otherwise.}
\end{array}\right. \]
So from the definition of $\zeta$, we have that $\Pr(\Gamma = 1) = \gamma(\zeta)$. Here the probability is taken over any stochasticity in the encoder.

Next, we show that an outer bound on the throughput of any highly reliable and deniable scheme is $\Oh{1/\sqrt{n}}$, as an explicitly calculated constant factor of $1/\sqrt{n}$. Before getting into the proof, we need the following three lemmas. Roughly speaking, the lemmas demonstrate connections between properties (rate, probability of error, entropy of received vectors $\vY_w$) of Alice's original code $\code$, and a modified code containing only the low-weight codewords from the original code $\code$.

Claim~\ref{claim:cond_ent_gamma} below shows a relationship between the entropy (and hence rate) of Alice's message variable $\vM$, and the entropy of messages corresponding to low-weight codewords.
\begin{claim}
\label{claim:cond_ent_gamma}
$\condentropy{\vM}{\tranrv = 1} \leq \condentropy{\vM}{\Gamma = 0, \tranrv = 1} + \log \frac{1}{1 - \gamma(\zeta)}$.
\end{claim}

\begin{proof}
	\begin{align}
		\lefteqn{\condentropy{\vM}{\Gamma = 0, \tranrv = 1}} \nonumber \\
			&= \sum_\msg p(\msg | \Gamma = 0, \tranrv = 1) \log \frac{1}{p(\msg | \Gamma = 0, \tranrv = 1)} \nonumber \\
			&= \sum_\msg p(\msg | \Gamma = 0, \tranrv = 1) \log \frac{p(\Gamma = 0 | \tranrv = 1)}{p(\Gamma = 0 | \msg, \tranrv = 1) p(\msg | \tranrv = 1)} \nonumber \\
			&= \sum_\msg p(\msg | \Gamma = 0, \tranrv = 1) \log \frac{N (1 - \gamma(\zeta))}{p(\Gamma = 0 | \msg, \tranrv = 1)} \nonumber \\
			&= \sum_\msg p(\msg | \Gamma = 0, \tranrv = 1) \log N + \sum_\msg p(\msg | \Gamma = 0) \log (1 - \gamma(\zeta)) \nonumber \\
			& \eqspacing  - \sum_\msg p(\msg | \Gamma = 0, \tranrv = 1) \log p(\Gamma = 0 | \msg, \tranrv = 1) \nonumber \\
			&= \condentropy{\vM}{\tranrv = 1} + \log (1 - \gamma(\zeta)) \nonumber \\
			& \eqspacing - \sum_\msg p(\msg | \Gamma = 0, \tranrv = 1) \log p(\Gamma = 0 | \msg, \tranrv = 1) \label{eq:logN_eq_entM} \\
			& \geq  \condentropy{\vM}{\tranrv = 1} + \log (1 - \gamma(\zeta)), \nonumber
	\end{align}
where equation~\eqref{eq:logN_eq_entM} holds since $\log N = \entropy{\vM}$. Therefore, we obtain $\entropy{\vM} \leq \condentropy{\vM}{\Gamma = 0} + \log \frac{1}{1 - \gamma(\zeta)}$.
\end{proof}

\begin{remark}
In general, we have $\entropy{\vM} \geq \condentropy{\vM}{\Gamma = 0}$. From Claim~\ref{claim:cond_ent_gamma}, we see that when $\gamma(\zeta)$ is small, the difference between $\condentropy{\vM}{\Gamma = 0}$ and $\entropy{\vM}$ is small. When $\gamma(\zeta) \rightarrow 1$, the bound in Claim~\ref{claim:cond_ent_gamma} is trivial.
\end{remark}

Claim~\ref{claim:cond_reli_err} below shows a relationship between Bob's probability of decoding error if Alice uses her original code, and his probability of error if Alice uses the ``sub-code'' of the original code with only low-weight codewords.
\begin{claim}
\label{claim:cond_reli_err}
$\Pr_{\vM, \vZ_b}(\hat{\vM} \neq \vM | \Gamma = 0, \tranrv = 1) \leq \frac{\epsreli}{1 - \gamma(\zeta)}$.
\end{claim}
\begin{proof}
	\begin{align}
		\epsreli &= \Pr_{\vZ_b}(\hat{\vM} \neq 0 | \tranrv = 0) + \Pr_{\vM, \vZ_b}(\hat{\vM} \neq \vM | \tranrv = 1) \nonumber \\
			&\geq \Pr_{\vM, \vZ_b}(\hat{\vM} \neq \vM | \tranrv = 1) \nonumber \\
			&= \Pr_{\vM}(\Gamma = 0) \prob_{\vM, \vZ_b}(\hat{\vM} \neq \vM | \Gamma = 0, \tranrv = 1) \nonumber \\
			& \eqspacing + \prob_{\vM}(\Gamma = 1) \Pr_{\vM, \vZ_b}(\hat{\vM} \neq \vM | \Gamma = 1, \tranrv = 1) \label{eq:fix_code_fix_dec} \\
			& \geq  (1 - \gamma(\zeta)) \Pr_{\vM, \vZ_b}(\hat{\vM} \neq \vM | \Gamma = 0, \tranrv = 1)	\nonumber \\
			& \geq (1 - \gamma(\zeta)) \Pr_{\vM, \vZ_b}(\hat{\vM} \neq \vM | \Gamma = 0, \tranrv = 1) \label{eq:pr_err_opt_dec} \end{align}
where equation~\eqref{eq:fix_code_fix_dec} holds since for a fixed code $\code$, Alice's encoder and Bob's decoder are fixed, $\Pr_{\vM, \vZ_b}(\hat{\vM} \neq \vM | \Gamma = 0, \tranrv = 1)$ corresponds to the probability of decoding error for the original decoder. In equation~\eqref{eq:pr_err_opt_dec}, the probability of decoding error $\Pr_{\vM, \vZ_b}(\hat{\vM} \neq \vM | \Gamma = 0, \tranrv = 1)$ corresponds to the probability of decoding error for the optimal decoder of the sub-code with the condition $\Gamma = 0$. Therefore, we obtain $\Pr_{\vM, \vZ_b}(\hat{\vM} \neq \vM | \Gamma = 0, \tranrv = 1) \leq \frac{\epsreli}{1 - \gamma(\zeta)}$.
\end{proof}

Claim~\ref{claim:cond_ent_up_bd} below shows a relationship between the entropy of Willie's received vector $\vY_w$ if Alice uses her original code, and the entropy of his $\vY_w$ if Alice uses the ``sub-code'' of the original code with only low-weight codewords.

\begin{claim}
\label{claim:cond_ent_up_bd}
$\condentropy{\vY_w}{\Gamma = 0, \tranrv = 1} \leq n \entropy{\zeta \ast \pw}$
\end{claim}

\begin{proof}
Note that
	\begin{align}
		\lefteqn{\frac{1}{n} \Exp_{\vY_b} \left\{ \wt{\vY_b} | \Gamma = 0, \tranrv = 1 \right\} } \nonumber \\
			&= \Exp_{\mathbf{U}} \left\{ \Exp_{\bar{\mathbf{Y}}_{b, \mathbf{U}}} \left[ \wt{\bar{\mathbf{Y}}_{b, \mathbf{U}}} | \Gamma = 0, \tranrv = 1  \right] \right\} \label{eq:cond_exp_unif}\\
			&= \Exp_{\mathbf{U}} \left\{ \Exp_{\bar{\mathbf{X}}_\mathbf{U}, \bar{\mathbf{Z}}_{b, \mathbf{U}}} \left[ \wt{\bar{\mathbf{X}}_\mathbf{U} \oplus \bar{\mathbf{Z}}_{b, \mathbf{U}}} | \Gamma = 0, \tranrv = 1 \right] \right\} \label{eq:cond_exp_vx_vz} \\
			&= \Exp_{\mathbf{U}} \Bigg\{ \Exp_{\bar{\mathbf{X}}_\mathbf{U}, \bar{\mathbf{Z}}_{b, \mathbf{U}}} \Bigg[ \sum_{\bar{x}, \bar{z}_b} p_{\bar{\mathbf{X}}_\mathbf{U}, \bar{\mathbf{Z}}_{b, \mathbf{U}} }(\bar{x}, \bar{z}_b | \Gamma = 0, \tranrv = 1 ) \nonumber \\
			& \eqspacing \times \wt{\bar{x} \oplus \bar{z}_b} \Bigg] \Bigg\} \nonumber  \\
			&= \Exp_{\mathbf{U}} \left\{ \Pr_{\bar{\mathbf{X}}_\mathbf{U}, \bar{\mathbf{Z}}_{b, \mathbf{U}}}( \bar{\mathbf{X}}_\mathbf{U} \oplus \bar{\mathbf{Z}}_{b, \mathbf{U}} = 1 | \Gamma = 0, \tranrv = 1 ) \right\} \nonumber \\
			&= \Exp_{\mathbf{U}}\left\{ \Pr_{\bar{\mathbf{X}}_\mathbf{U}}(\bar{\mathbf{X}}_\mathbf{U} = 1 | \Gamma = 0, \tranrv = 1 ) (1 - \pb) \right. \nonumber \\
			& \eqspacing \left. + \Pr_{\bar{\mathbf{X}}_\mathbf{U}}(\bar{\mathbf{X}}_\mathbf{U} = 0 | \Gamma = 0, \tranrv = 1 ) \pb \right\} \nonumber \\
			&= \frac{\wt{\vx}}{n} \ast \pb \nonumber \\
			&\leq \zeta \ast \pb \label{eq:cond_exp_up_bd},
	\end{align}
where equation \eqref{eq:cond_exp_unif} holds by taking $\mathbf{U}$ as a uniform random variable in the set $\{1, 2, \ldots, n\}$. Using the fact that $\vY_b = \vX \oplus \vZ_b$ where ``$\oplus$'' is exclusive OR operation, we can obtain \eqref{eq:cond_exp_vx_vz}.

Therefore, by the concavity of the entropy function, and the inequality \eqref{eq:cond_exp_up_bd}, we obtain
	\begin{align}
		\lefteqn{\condentropy{\vY_b}{\Gamma = 0, \tranrv = 1}} \nonumber \\
		&\leq \sum_{i=1}^n \condentropy{\mathbf{Y}_{b,i}}{\Gamma = 0, \tranrv = 1} \nonumber \\
		&= \sum_{i=1}^n \entropy{\zeta_{\mathbf{x}_i} \ast \pb} \nonumber \\
		&\leq n \entropy{\frac{1}{n} \sum_{i=1}^n \zeta_{\mathbf{x}_i} \ast \pb} \nonumber \\
		&= n \entropy{\frac{1}{n} \sum_{i=1}^n \wt{\mathbf{x}_i} \ast \pb} \nonumber \\
		&= n \entropy{\left( \frac{1}{n} \sum_{i=1}^n \wt{\mathbf{x}_i} \right) \ast \pb} \nonumber \\
		&= n \entropy{\zeta \ast \pb}.
	\end{align}
\end{proof}

We now find an upper bound on the throughput $\thr$ through the following series of inequalities,
\begin{align}
	\lefteqn{\thr \sqrt{n}} \nonumber \\
		&= \condentropy{\vM}{\tranrv = 1} \nonumber \\
		& \leq \condentropy{\vM}{\Gamma = 0, \tranrv = 1} + \log \frac{1}{1 - \gamma(\zeta)} \label{eq:up_bd_ent_m} \\
		&= \condentropy{\vM}{\hat{\vM}, \Gamma = 0, \tranrv = 1}  \nonumber \\
		& \eqspacing + \condmutual{\vM}{\hat{\vM}}{\Gamma = 0, \tranrv = 1} + \log \frac{1}{1 - \gamma(\zeta)} \nonumber \\
		& \leq 1 + \thr \sqrt{n} \Pr_{\vM, \vZ_b}(\hat{\vM} \neq \vM | \Gamma = 0, \tranrv = 1) \nonumber \\
		& \eqspacing + \condmutual{\vX}{\vY_b}{\Gamma = 0, \tranrv = 1} + \log \frac{1}{1 - \gamma(\zeta)} \label{eq:fano_and_data} \\
		& \leq 1 + \thr \sqrt{n} \frac{\epsreli}{1 - \gamma(\zeta)} \nonumber \\
		& \eqspacing + \left[ \condentropy{\vY_b}{\Gamma = 0, \tranrv = 1} - \condentropy{\vY_b}{\vX, \Gamma = 0, \tranrv = 1} \right] \nonumber \\
		& \eqspacing \eqspacing + \log \frac{1}{1 - \gamma(\zeta)} \label{eq:fano_simp} \\
		& \leq 1 + \thr \sqrt{n} \frac{\epsreli}{1 - \gamma(\zeta)} + n\left( \entropy{\zeta \ast \pb} - \entropy{\pb} \right) + \log \frac{1}{1 - \gamma(\zeta)} \label{eq:concavity} \\
		&= 1 + \thr \sqrt{n} \frac{\epsreli}{1 - \gamma(\zeta)} \nonumber \\
		& \eqspacing + n \left( \KL{\pb}{\zeta \ast \pb} + \zeta (1 - 2\pb) \log \frac{1 - \zeta \ast \pb}{\zeta \ast \pb} \right) \nonumber \\
		& \eqspacing \eqspacing + \log \frac{1}{1 - \gamma(\zeta)} \label{eq:ent_diff} \\
		& \leq 1 + \thr \sqrt{n} \frac{\epsreli}{1 - \gamma(\zeta)} \nonumber \\
		& \eqspacing + n \left( \frac{\zeta^2(1 - 2\pb)^2}{2 \pb (1 - \pb) \ln 2} + \zeta (1 - 2\pb) \log \frac{1 - \pb}{\pb} \right) \nonumber \\
		& \eqspacing \eqspacing + \log \frac{1}{1 - \gamma(\zeta)}, \label{eq:appendix_KL}
\end{align}
where inequality~\eqref{eq:up_bd_ent_m} follows from Claim~\ref{claim:cond_ent_gamma}, which shows that conditioning on the event that Alice uses a low-weight sub-code ($\Gamma = 0$) does not change the uncertainty in $\vM$ ``much''. Inequality \eqref{eq:fano_and_data} holds by using Fano's inequality (recall that $\Pr_{\vM, \vZ_b}(\hat{\vM} \neq \vM | \Gamma = 0, \tranrv = 1)$ is the probability of decoding error corresponds to the optimal decoder for the sub-code) and data-processing inequality over the probability distribution $\tilde{p}(\cdot) \triangleq p(\cdot|\Gamma = 0, \tranrv = 1)$. Using Claim~\ref{claim:cond_reli_err}, we obtain equation~\eqref{eq:fano_simp}. We obtain equation \eqref{eq:concavity} using Claim~\ref{claim:cond_ent_up_bd} and $\condentropy{\vY_b}{\vX, \Gamma = 0, \tranrv = 1} = \entropy{\pb}$ by noting that the channel from Alice to Bob is an {\it i.i.d.} DMC (Discrete Memeoryless Channel). Equation~\eqref{eq:ent_diff} is obtained from Claim~\ref{clm:ent_diff} in the Appendix and equation~\eqref{eq:appendix_KL} is obtained from Claim~\ref{clm:pinsker} in the Appendix.

Therefore, from \eqref{eq:appendix_KL} we have the upper bound on the throughput $\thr$,
\begin{align}
	\thr & \leq \frac{1}{(1 - \epsreli/(1 - \gamma(\zeta)))\sqrt{n}}\Bigg[1 + n \left( \frac{\zeta^2(1 - 2\pb)^2}{2 \pb (1 - \pb) \ln 2} \right. \nonumber \\
	& \eqspacing \left. + \zeta (1 - 2\pb) \log \frac{1 - \pb}{\pb} \right) + \log \frac{1}{1 - \gamma(\zeta)} \Bigg] \nonumber \\
		& \leq \frac{1}{1 - \epsreli/(1 - \gamma(\zeta))} \frac{1 - 2\pb}{1 - 2\pw}\sqrt{\frac{8 \pw (1 - \pw)}{1 - \epsdeni/\gamma(\zeta)}} \log \frac{1 - \pb}{\pb} + \Oh{\frac{1}{\sqrt{n}}} \label{eq:up_bd_thr}
\end{align}


\subsection{Proof of Proposition~\ref{prop:ran_sto_code}}

Recall that the codebook $\code$ was generated by choosing $2^{(\thr + \thr_s) \sqrt{n}}$ codewords, with each bit of each codeword generated {\it i.i.d.} according to a Bernoulli($\rho$) distribution, where $\rho$ is a code-design parameter specified in Subsection~\ref{subsec:achi:deni}. 

Note that a code $\code$ is $(1 - \epsreli)$-reliable regardless of whether Alice transmitted a non-zero codeword or not if the probability (only over channel noise $\vZ_b$) that Bob's decoded message $\hat{\vM}$ differs from the true message $\vM$ (which may equal either $0$ if Alice is silence, or a value from $\{ 1, \ldots, 2^{(\thr + \thr_s) \sqrt{n}} \}$ if Alice is indeed transmitting).


\noindent \underline{\bf Bob's decoding rule} \\
Suppose that Bob receives a vector $\vy_b$, then Bob's decoding rule is shown as follows,
\begin{enumerate}
	\item[1.] If the received vector $\vy_b$ is in the typical set of $\vY_b$ corresponding to Alice being silent. That is, if $\vy_b \in \bobtypyz \setminus \bobtypyo$, then Bob decodes $\msg = 0$;
	\item[2.] If the received vector $\vy_b$ is in the typical set of $\vY_b$ corresponding to Alice being transmitting, {\it i.e.}, if $\vy_b \in \bobtypyo$, then
		\begin{enumerate}
			\item[2.1.] If there is only one message $\msg$ in $\{1, \ldots, N\}$ such that the codewords $\vx(\msg, \msg_s)$ in the decoding ball $\bobtypxgyb$ for some $\msg_s \in \{1, \ldots, S\}$, then Bob decodes $\msg$. That is, if $\exists ! \msg \in \{1, \ldots, N\}$, and $\exists \msg_s \in \{1, \ldots, S\}$ s.t. $\vx(\msg, \msg_s) \in \code \cap \bobtypxgyb$, then Bob decodes $\msg$;
			\item[2.2.] If there are more than one codewords $\vx(\msg, \msg_s), \vx'(\msg', \msg_s')$ in $\code \cap \bobtypxgyb$ where $\msg \neq \msg'$, then Bob outputs an error;
			\item[2.3.] If there is no codeword in $\code \cap \bobtypxgyb$, then Bob decodes the zero message $0$. That is, if $\not{\exists} \vx \in \bobtypxgyb$, then Bob decodes $\msg = 0$, which means Alice did not transmit;
		\end{enumerate}
	\item[3.] If Bob's received vector $\vy_b$ is neither in $\bobtypyz$ nor $\bobtypyo$, then Bob outputs error.
\end{enumerate}

\noindent \underline{\bf Error analysis of the decoding rule} \\
In this part, we show that Bob can decode the original message correctly with high probability.

We first define the probability of error. Note that there are two types of errors according to Alice's transmission status $\tranrv = 0$.

When Alice is silent, {\it i.e.}, $\tranrv = 0$, the probability of decoding error is defined as follows,
\begin{equation}
	\Pr(\hat{\vM} \neq 0 | \tranrv = 0).
\end{equation}

When Alice is indeed transmitting, without loss of generality, we assume that the codeword $\vx(\msg, \msg_s)$ is sent. We define the {\it maximal probability of decoding error} as follows,
\begin{equation}
	\max_{\msg \in \{1, \ldots, N\}} \Pr(\hat{\vM} \neq \msg | \vX = \vx(\msg, \msg_s), \tranrv = 1).
\end{equation}

\begin{remark}
In the case $\tranrv = 1$, we show that the maximal probability of decoding error is small with super-exponential high probability.
\end{remark}


\noindent \underline{\it Analysis of $\Pr(\hat{\vM} \neq 0 | \tranrv = 0)$:}\\

We note that the error given Alice's transmission status $\tranrv = 0$ can be separated in to two parts. That is,
\begin{align}
	\lefteqn{\Pr_{\vZ_b}(\vM \neq 0 | \tranrv = 0)} \nonumber \\
	& \leq \Pr_{\vZ_b}(\vY_b \notin \bobtypyz | \tranrv = 0) \nonumber \\
	& \eqspacing + \Pr_{\vZ_b}\left(\exists \vx \in \code \mbox{ s.t. } \vx \in \bobtypxgybrv, \right. \nonumber \\
	& \eqspacing \eqspacing \left. \vY_b \in \bobtypyz \setminus \bobtypyo | \tranrv = 0 \right).
	\label{eq:mo1_errbz}
\end{align}
We now prove two claims to show that $\Pr(\hat{\vM} \neq 0 | \tranrv = 0)$ is small. The following Claim~\ref{claim:delybz_val} says that when Alice's transmission status $\tranrv$ is $0$, the probability of Bob's received vector being atypical is small, which corresponds to the first term in equation~\eqref{eq:mo1_errbz}.

\begin{claim}
\label{claim:delybz_val}
When Alice does not transmit,
$$ \Pr_{\vZ_b}(\vY_b \notin \bobtypyz | \tranrv = 0) < 2^{-\tilde{c} n^\delta} $$
for some constant $\tilde{c}$ if $\delybz = \tilde{k} n^{-1/2 + \delta/2}$.
\end{claim}

\begin{proof}
Note that
	\begin{align}
		\lefteqn{\Pr_{\vZ_b}(\vY_b \notin \bobtypyz | \tranrv = 0)} \nonumber \\
			&= \Pr_{\vZ_b}(\fyb \notin (\pb(1 - \delybz), \pb(1 + \delybz)) | \tranrv = 0) \nonumber \\
			&< 2\exp\left( -\frac{1}{3} (\delybz)^2 \pb n \right). \nonumber
	\end{align}
Therefore, choosing $\delybz = \tilde{k} n^{-1/2 + \delta/2}$, we obtain the desired result.
\end{proof}

In Claim~\ref{claim:exist_code_err_no_tran}, we show that when Alice does not transmit, given that Bob's received vector $\vy_b \in \bobtypyz$, the probability of Bob's decoding ball contains more than $1$ codeword is small. This lemma corresponds to the second term of equation~\eqref{eq:mo1_errbz}.

\begin{claim}
\label{claim:exist_code_err_no_tran}
When Alice does not transmit,
\begin{multline*}
	\Pr_{\vZ_b}\left(\exists \vx \in \code \mbox{ s.t. } \vx \in \bobtypxgybrv, \vY_b \in \bobtypyz \setminus \bobtypyo | \tranrv = 0 \right) \\
		< 2^{-\Omega(\sqrt{n})}.
\end{multline*}
with probability over $\code$ at least $1 - 2^{-\Omega(\sqrt{n})}$.
\end{claim}

\begin{proof}
Note that the probability
\begin{align}
	\lefteqn{\Pr_{\vX, \vZ_b}\left( \vX \in \bobtypxgybrv, \vY_b \in \bobtypyz \setminus \bobtypyo | \tranrv = 0 \right)} \nonumber \\
	&= \sum_{\vy_b \in \bobtypyz \setminus \bobtypyo} p(\vy_b) \Pr_{\vX}\left( \vX \in \bobtypxgyb | \tranrv = 0 \right) \nonumber \\
	&\leq \max_{\vy_b \in \bobtypyz \setminus \bobtypyo} \left( \Pr_{\vX}\left( \vX \in \bobtypxgyb | \tranrv = 0 \right) \right) \label{eq:avg_pr_over_x_z_t0}
\end{align}
Note that for any message $\vM$ and private randomness $\vM_s$, the corresponding codeword $\vX(\vM, \vM_s)$ is generated {\it i.i.d.} according to $Bern(\rho)$. So, the probability that a randomly generated codeword is in the decoding ball $\bobtypxgyb$ (where $\vy_b \in \bobtypxgyb$) is,
\begin{align}
	\lefteqn{ \Pr_{\vX}\left( \vX \in \bobtypxgyb \left| \tranrv = 0 \right. \right) } \nonumber \\
		&= \sum_{\vx \in \bobtypxgyb} p(\vx) \nonumber \\
		&= \sum_{(\fozb, \foob) \in \typFb} \sum_{\vx \in \bobtypexgyb} p(\vx) \nonumber \\
		& \leq (\xi n^{1/4 + \delta/2}) {n(1 - \fyb) \choose n\fozb} {n\fyb \choose n\foob} \nonumber \\
		& \eqspacing \times \rho^{n(\fozb + \foob)} (1 - \rho)^{n(1 - \fozb - \foob)} \nonumber \\
		&= (\xi n^{1/4 + \delta/2}) 2^{-n(\mutual{\vx}{\vy_b} + \KL{\vx}{\rho})} \label{eq:avg_pr_over_x_t0}
\end{align}
for some constant $\xi$. Therefore, combining equation \eqref{eq:avg_pr_over_x_z_t0} and \eqref{eq:avg_pr_over_x_t0} we have
\begin{multline*}
	\Pr_{\vX, \vZ_b}\left( \vX \in \bobtypxgybrv, \vY_b \in \bobtypyz \setminus \bobtypyo \left| \tranrv = 0 \right. \right) \\
		\leq (\xi n^{1/4 + \delta/2}) 2^{-n(\mutual{\vx}{\vy_b} + \KL{\vx}{\rho})}. \nonumber
\end{multline*}
	
Note that $n(\mutual{\vx}{\vy_b} + \KL{\vx}{\rho}) = \thr_u \sqrt{n} + \Oh{n^{1/4}} $ by Claim~\ref{clm:pinsker} and Claim~\ref{clm:emp_mut_info} in the Appendix, and for a randomly generated codebook $\bm{\code}$, there are $2^{\thr \sqrt{n}}$ codewords. Hence, by union bound, the expected probability that there exists a codeword in $\code$ such that the codeword in the decoding ball $\bobtypxgyb$ equals
\begin{align}
	\lefteqn{ \Exp_{\bm{\code}} \left[ \Pr_{\vZ_b} \left( \exists \vx \in \code \mbox{ s.t. } \vx \in \bobtypxgybrv, \right. \right. } \nonumber \\
		& \eqspacing \left. \left. \vY_b \in \bobtypyz \setminus \bobtypyo \left| \tranrv = 0 \right. \right) \right] \nonumber \\
		& \leq \Pr_{\vX, \vZ_b} \left( \vX \in \bobtypxgybrv, \vY_b \in \bobtypyz \setminus \bobtypyo \left| \tranrv = 0 \right. \right) |\code| \nonumber \\
		&\leq 	(\xi n^{1/4 + \delta/2}) 2^{-n(\mutual{\vx}{\vy_b} + \KL{\vx}{\rho})} 2^{\thr \sqrt{n}} \nonumber \\
		&= 2^{-\Omega(\sqrt{n})}. \label{eq:exp_pr_t0}
\end{align}
So, apply Markov inequality on equation \eqref{eq:exp_pr_t0}, we have the probability that
\begin{multline*}
\Pr_{\vZ_b} \left( \exists \vx \in \code \mbox{ s.t. } \vx \in \bobtypxgybrv, \vY_b \in \bobtypyz \setminus \bobtypyo \left| \tranrv = 0 \right. \right) \\
	\geq 2^{-\tilde{c}' \sqrt{n}}
\end{multline*}
is at most $2^{-\Omega(\sqrt{n})}$ for some carefully chosen constant $\tilde{c}'$.
\end{proof}

\noindent \underline{\it Analysis of $\Pr(\hat{\vM} \neq \msg | \vX = \vx(\msg, \msg_s), \tranrv = 1)$:} \\
	When Alice does transmit ($\tranrv = 1$), we assume that a specific codeword $\vx(\msg, \msg_s)$ is transmitted. Since Alice's messages are equiprobabe, and each codeword is generated i.i.d., the probability of error analysis below is symmetric regardless of which codeword ends up being transmitted.
\begin{align}
	\lefteqn{\Pr_{\vZ_b}(\hat{\vM} \neq \msg | \vX = \vx(\msg, \msg_s), \tranrv = 1)} \nonumber \\
		& \leq \Pr_{\vZ_b}(\vY_b \notin \bobtypyo | \vX = \vx(\msg, \msg_s), \tranrv = 1) \nonumber \\
		& \eqspacing + \Pr_{\vZ_b}\Big(\bigcap_{\msg_s} \{\vx(\msg, \msg_s) \notin \bobtypxgybrv, \vY_b \in \bobtypyo \} \nonumber \\
		& \eqspacing \eqspacing | \vX = \vx(\msg, \msg_s), \tranrv = 1 \Big) \nonumber \\
		& \eqspacing + \Pr_{\vX' \neq \vx(\msg, \cdot), \vZ_b}\Big(\exists \vX' \in \bobtypxgybrv, \vY_b \in \bobtypyo \nonumber \\
		& \eqspacing \eqspacing | \vX = \vx(\msg, \msg_s), \tranrv = 1 \Big). \nonumber
\end{align}

To show $\Pr(\hat{\vM} \neq \msg | \tranrv = 1, \vX = \vx(\msg, \msg_s))$ is small, we have the following three lemmas.

\begin{claim}
\label{claim:delybo_val}
When Alice does transmit a codeword $\vx(\msg, \msg_s)$,
$$ \Pr_{\vZ_b}(\vY_b \notin \bobtypyo | \vX = \vx(\msg, \msg_s), \tranrv = 1) < 2^{-\hat{c} n^\delta} $$
with probability over $\code$ at least $1 - 2^{-\Oh{n^\delta}}$ if $\delybo = \hat{k} n^{-1/2 + \delta/2}$.
\end{claim}

\begin{proof}
Applying Chernoff bound, we can upper bound
\begin{align}
	\lefteqn{\Exp_{\bm{\code}} \left( \Pr_{\vZ_b}(\vY_b \notin \bobtypyo | \vX = \vx(\msg, \msg_s), \tranrv = 1) \right)} \nonumber \\
		&= \Pr_{\vX, \vZ_b}(\vY_b \notin \bobtypyo | \tranrv = 1) \nonumber \\
		&< 2 \exp\left( -\frac{1}{3} \left(\delybo)\right)^2 \rho \ast \pb n \right) . \nonumber
\end{align}

Since $\delybo = \hat{k} n^{-1/2 + \delta/2}$, we have that
\begin{equation}
	\Exp_{\bm{\code}} \left( \Pr_{\vZ_b}(\vY_b \notin \bobtypyo | \tranrv = 1) \right) < 2^{-\hat{c} n^\delta} \nonumber
\end{equation}
for some constant $\hat{c}$.

Hence, applying Markov's inequality, we have
\begin{equation*}
	\Pr_{\bm{\code}} \left( \Pr_{\vZ_b}(\vY_b \notin \bobtypyo | \vX = \vx(\msg, \msg_s), \tranrv = 1) \geq 2^{-\hat{c}' n^\delta} \right) \leq 2^{-\Omega(n^\delta)}
\end{equation*}
for some constant $\hat{c}'$.
\end{proof}

\begin{claim}
\label{claim:decerr:twocw}
When Alice does transmit the codeword $\vx(\msg, \msg_s)$,
\begin{multline*}
\Pr_{\vZ_b}\Big(\exists \vx'(\msg', \msg_s') \in \code \mbox{ s.t. } \vx' \in \bobtypxgybrv, \vY_b \in \bobtypyo \\
\Big| \vX = \vx(\msg, \msg_s), \tranrv = 1 \Big) < 2^{-\Omega(\sqrt{n})}
\end{multline*}
with probability over $\code$ at least $1 - 2^{-\Oh{\sqrt{n}}}$.
\end{claim}

\begin{proof}
This proof is similar to Claim~\ref{claim:exist_code_err_no_tran}. The following is the sketch of the proof. We first note that
\begin{multline*}
	\Pr_{\vX', \vZ_b}\left( \vX' \in \bobtypxgybrv, \vY_b \in \bobtypyo \left| \vX = \vx(\msg, \msg_s), \tranrv = 1 \right. \right) \nonumber \\
		\leq \max_{\vy_b \in \bobtypyo} \left(\Pr_{\vX'}\left( \vX' \in \bobtypxgyb \left| \vX = \vx(\msg, \msg_s), \tranrv = 1 \right. \right)\right). \nonumber
\end{multline*}
We then have
\begin{multline*}
	\Pr_{\vX'}\left( \vX' \in \bobtypxgyb \left| \vX = \vx(\msg, \msg_s), \tranrv = 1 \right. \right) \nonumber \\
		\leq (\xi' n^{1/4+\delta/2}) 2^{-n(\mutual{\vx}{\vy_b}+\KL{\vx}{\rho})}
\end{multline*}
for some constant $\xi'$. Therefore, we have
\begin{multline*}
\Pr_{\vX', \vZ_b}\left( \vX' \in \bobtypxgybrv, \vY_b \in \bobtypyo \left| \vX = \vx(\msg, \msg_s), \tranrv = 1 \right. \right) \\
	\leq (\xi' n^{1/4+\delta/2} ) 2^{-n(\mutual{\vx}{\vy_b}+\KL{\vx}{\rho})}.
\end{multline*}
So,
\begin{multline*}
	\Exp_{\bm{\code}} \Big[ \Pr_{\vZ_b}\Big(\exists \vx' \in \bm{\code} \mbox{ s.t. } \vx' \in \bobtypxgybrv, \vY_b \in \bobtypyo \\
	\Big| \vX = \vx(\msg, \msg_s), \tranrv = 1 \Big) \Big] \leq 2^{-\Omega(\sqrt{n})}.
\end{multline*}
Hence, apply Markov's inequality, we have that
\begin{multline*}
\Pr_{\vZ_b}\Big(\exists \vx' \in \bm{\code} \mbox{ s.t. } \vx' \in \bobtypxgybrv, \vY_b \in \bobtypyo \Big| \\
	\vX = \vx(\msg, \msg_s), \tranrv = 1 \Big) \geq 2^{-\hat{c}'' \sqrt{n}}
\end{multline*}
is at most $2^{-\Omega(\sqrt{n})}$ for some constant $\hat{c}''$.
\end{proof}

\begin{claim}
\label{claim:delozb_deloob}
When Alice does transmit the codeword $\vx(\msg, \msg_s)$,
\begin{multline*}
\Pr_{\vZ_b}\Big( \bigcap_{\msg_s} \{ \vx(\msg, \msg_s) \notin \bobtypxgyb \}, \vY_b \in \bobtypyo \\
\Big| \vX = \vx(\msg, \msg_s), \tranrv = 1 \Big) < 2^{- \Omega(n^\delta)}.
\end{multline*}
\end{claim}

\begin{proof}
Denote the event $\mathcal{E}_{\msg, \msg_s} = \{ \vx(\msg, \msg_s) \notin \bobtypxgybrv \}$ and the event $\mathcal{E}_{\msg, \msg_s'} = \{ \vX'(\msg, \msg_s') \notin \bobtypxgybrv \}$ for $\msg_s' \neq \msg_s$. We then have the error probability
\begin{align}
	\lefteqn{ \Pr_{\vX', \vZ_b} \left( \mathcal{E}_{\msg, \msg_s}, \bigcap_{\msg_s' \neq \msg_s} \mathcal{E}_{\msg, \msg_s'}, \vY_b \in \bobtypyo \Big| \vX = \vx(\msg, \msg_s), \tranrv = 1 \right) } \nonumber \\
		& \leq \Pr_{\vZ_b} \left(\mathcal{E}_{\msg, \msg_s}, \vY_b \in \bobtypyo \Big| \vX = \vx(\msg, \msg_s), \tranrv = 1 \right) \nonumber \\
		&= \Pr_{\vZ_b} \left(\vx(\msg, \msg_s) \notin \bobtypxgybrv, \vY_b \in \bobtypyo \Big| \vX = \vx(\msg, \msg_s), \tranrv = 1 \right) \nonumber \\
		& \leq \Pr_{\vZ_b} \left(\vx(\msg, \msg_s) \notin \bobtypxgybrv \Big| \vY_b \in \bobtypyo, \tranrv = 1 \right) \nonumber
\end{align}
Note that for any $\vy_b \in \bobtypyo$, if the codeword $\vx(\msg, \msg_s) \notin \typx$, this codeword $\vx(\msg, \msg_s) \notin \bobtypxgyb$. Hence, the only case that remains is that when $\vx(\msg, \msg_s) \in \typx$. So, $\mathcal{E}_{\msg, \msg_s}$ happens only when the bit-flips in the support of $\vx(\msg, \msg_s)$ is outside the range $(n \rho \pb (1 - \delozb), n \rho \pb (1 + \delozb))$. That is, the fraction of bit-flips in the support of $\vx(\msg, \msg_s)$ is outside the range $(\pb(1 - \hat{k}' n^{-1/4+\delta/2}), \pb(1 + \hat{k}' n^{-1/4+\delta/2}))$ for some constant $\hat{k}'$ and $\delta \ll 1$. Therefore,
\begin{multline*}
	\Pr_{\vZ_b} (\vx(\msg, \msg_s) \notin \bobtypxgybrv | \vY_b \in \bobtypyo, \tranrv = 1) \\
		< 2 \exp\left(-\hat{c}''' n^\delta\right)
\end{multline*}
for some constant $\hat{c}'''$.
\end{proof}


Combining the above claims, we obtain
\begin{multline*}
	\Pr(\hat{\vM} \neq 0 | \tranrv = 0) \\ 
		+ \max_\msg \sum_{\msg_s \in \code} \Pr(\hat{\vM} \neq \msg | \vX = \vx(\msg, \msg_s), \tranrv = 1) \Pr(\vM_s = \msg_s) < 2^{-\Omega(n^\delta)}
\end{multline*}
with probability greater than $1 - \exp\left( -\Oh{n^{\delta}} \right)$ over the random code $\code$.

\subsection{Achievability of Fixed Channel Model (Theorem~\ref{thm:model:known:achi})}

\subsubsection{Achievability: Deniability}
\label{subsec:achi:deni}


Recall that in equation~\eqref{eq:var_dist_tri_ineq} the deniability part of the achievability is equivalent to show that $\variation{\Pz}{\Po} < \epsdeni$. This can be further divided into that showing that a code is, with high probability, $(1-\epsdeni)$-deniable is implied by showing that $\variation{\Pz}{\Exp(\Po)} < \epsdeni$ and $\variation{\Exp(\Po)}{\Po} < 2^{-\Omega(n^\delta)}$ are true, {\it i.e.}, the variational distance between $\Pz$ and $\Exp_\code(\Po)$ is less than $\epsdeni$, and the variational distance between $\Exp_\code(\Po)$ and $\Po$ is exponentially small with (super-exponentially) high probability. In fact, the bulk of the proof of deniability focuses on the latter inequality, {\it i.e.}, Lemma~\ref{lemma:achi_deni_EcP1_P1}, since it is the most technically challenging part of the proof.

\begin{lemma}
\label{lemma:achi_deni_P0_EcP1}
If the codebook $\code$ is drawn from the $\rho n$-weight random ensemble with $\rho < \frac{2 \sqrt{\pw(1 - \pw)}}{1 - 2\pw} \frac{\epsdeni}{\sqrt{n}}$, then $\variation{\Pz}{\Exp(\Po)} < \epsdeni$.
\end{lemma}

While Lemma~\ref{lemma:achi_deni_P0_EcP1} is relatively straightforward, following broadly from ``standard techniques'' in information theory (such as Pinsker's inequality -- indeed, such an approach was followed in the work of Bash {\it et al}, the major novelty in this proof is in Lemma~\ref{lemma:achi_deni_EcP1_P1}, which occupies the bulk of the remainder of this Section.)

\begin{lemma}
\label{lemma:achi_deni_EcP1_P1}
If the codebook $\code$ is drawn from the $\rho n$-weight random ensemble with $\rho < \frac{2 \sqrt{\pw(1 - \pw)}}{1 - 2\pw} \frac{\epsdeni}{\sqrt{n}}$ and the throughput $\thr_L < \thr < \thr_U$, $\variation{\Exp(\Po)}{\Po} < 2^{-\Omega(n^\delta)}$ with probability greater than $1 - \exp\left(-2^{\Oh{n^\delta}}\right)$ over the code $\code$.
\end{lemma}

So, combining Lemma~\ref{lemma:achi_deni_P0_EcP1} and Lemma~\ref{lemma:achi_deni_EcP1_P1}, we conclude that $\alpha + \beta \geq 1 - \epsdeni$ with probability greater than $1 - \exp\left(-2^{\Oh{n^\delta}}\right)$ over the code $\code$.





\noindent{\bf Proof of Lemma~\ref{lemma:achi_deni_P0_EcP1}} \\
Note that $\Pz$ corresponds to the $n$-letter distribution (over a support of size $2^n$) induced by the codebook $\code$ and $n$ Bernoulli-$(\pw)$ random variables corresponding to entries of $\vZ_w$. Similarly, the ``smoothed'' distribution $\Exp_\code(\Po)$  corresponds to the $n$-letter distribution (also over a support of size $2^n$) induced by $n$ Bernoulli-$(\rho \ast \pw)$ random variables corresponding to entries of $\vX \oplus \vZ_w$. 

Hence by further ``standard statistical arguments'',
\begin{eqnarray}
	\variation{\Pz}{\Exp_\code(\Po)} & \leq & \sqrt{\frac{\ln 2}{2}\KL{\Pz}{\Exp_\code(\Po)}}   \label{eq:pins} \\
	&=& \sqrt{\frac{n \ln 2}{2}\KL{\pw}{\rho \ast \pw}} \label{eq:prod} 
\end{eqnarray}
where \eqref{eq:pins} follows from Pinsker's inequality (\cite[Lemma 11.6.1]{cover2012book}, reprised in~\cite{BasGT:12J} as Fact~$2$), and~\eqref{eq:prod} from the chain rule for relative entropy (\cite[Equation (2.67)]{cover2012book}, reprised in~\cite{BasGT:12J} as Fact~$3$).

Using the Taylor series bound on the Kullback-Leibler divergence (See Claim~\ref{clm:pinsker} in the Appendix, we have $\KL{\pw}{\rho \ast \pw} \leq \frac{\rho^2 (1-2\pw)^2}{ 2\pw (1-\pw)\ln 2 }$. Therefore, substituting in \eqref{eq:prod},
\begin{equation}
	\variation{\Pz}{\Exp_\code(\Po)} \leq \frac{1-2\pw}{\sqrt{\pw(1-\pw)}} \frac{\rho \sqrt{n}}{2}.
\label{eq:1st_term}
\end{equation}
Finally, using the fact that $\rho < \frac{2 \sqrt{\pw(1 - \pw)}}{1 - 2\pw} \frac{\epsdeni}{\sqrt{n}}$, we have $\variation{\Pz}{\Exp_\code(\Po)} < \epsdeni$. $\hfill\square$

\noindent{\bf Proof of Lemma~\ref{lemma:achi_deni_EcP1_P1}} \\
Using the definitions in Section~\ref{sec:notation_def}, $\variation{\Exp_\code(\Po)}{\Po}$  from \eqref{eq:var_dist_tri_ineq} may be further expanded as the following
\begin{align}
	\lefteqn {\variation{\Exp_\code(\Po)}{\Po}} \nonumber\\ 
	&= {\frac{1}{2} \sum_{\vy_w \in \{0,1\}^n} | \Exp_\code(\po(\vy_w)) -  \po(\vy_w) |}  \nonumber\\ 
	& \leq \frac{1}{2} \sum_{\vy_w \in \typyw} \left| \sum_{\code} \Pr(\code) \sum_{\vx \in \code \cap \typxgyw} \po(\vy_w|\vx) p(\vx)) \right. \nonumber \\
	& \eqspacing \eqspacing \left. - \sum_{\vx \in \code \cap \typxgyw} \po(\vy_w|\vx) p(\vx)) \right| \nonumber \\
	& \eqspacing + \frac{1}{2} \sum_{\vy_w \in \typyw} \left| \sum_\code \Pr(\code) \sum_{\vx \in \code \setminus \typxgyw} \po(\vy_w|\vx) p(\vx)) \right. \nonumber \\
	& \eqspacing \eqspacing \left. - \sum_{\vx \in \code \setminus \typxgyw} \po(\vy_w|\vx) p(\vx)) \right| \nonumber \\
	& \eqspacing + \frac{1}{2} \sum_{\vy_w \notin \typyw} \left| \sum_{\code} \Pr(\code) \sum_{\vx \in \code} \po(\vy_w|\vx) p(\vx)) \right. \nonumber \\
	& \eqspacing \eqspacing \left. - \sum_{\vx \in \code} \po(\vy_w|\vx) p(\vx)) \right| \label{eq:expan_var_dist} \displaybreak\\
	& \leq \frac{1}{2} \sum_{\vy_w \in \typyw} \left| \sum_{\code} \Pr(\code) \sum_{\vx \in \code \cap \typxgyw} \po(\vy_w|\vx) p(\vx)) \right. \nonumber \\
	& \eqspacing \eqspacing \left. - \sum_{\vx \in \code \cap \typxgyw} \po(\vy_w|\vx) p(\vx)) \right| \nonumber \\
	& \eqspacing + \frac{1}{2} \sum_{\vy_w \in \typyw} \Exp_\code \left( \sum_{\vx \in \code \setminus \typxgyw} \po(\vy_w|\vx) p(\vx)) \right) \nonumber \\
	& \eqspacing + \frac{1}{2} \Exp_\code \left( \sum_{\vy_w \notin \typyw} \po(\vy_w) \right) \nonumber \\
	& \eqspacing + \frac{1}{2} \sum_{\vy_w \in \typyw} \sum_{\vx \in \code \setminus \typxgyw} \po(\vy_w|\vx) p(\vx)) \nonumber \\
	& \eqspacing + \frac{1}{2} \sum_{\vy_w \notin \typyw} \sum_{\vx \in \code} \po(\vy_w|\vx) p(\vx)). \label{eq:var_further_tri}
\end{align}
The terms in equation~\eqref{eq:expan_var_dist} above correspond to the difference between the distribution on $\vy_w$ observed by Willie due to the actual code $\code$ used, and the distribution on $\vy_w$ if the ensemble distribution (over random codebooks) had been used. The first two terms in equation~\eqref{eq:expan_var_dist} deal with the difference between these distributions for ``typical'' $\vy_w$, and the last one for ``atypical'' $\vy_w$ -- bounding different terms require different techniques, as outlined later.

The rest of the proof focuses on showing that with (super-exponentially high probability over code design) each of the terms in \eqref{eq:var_further_tri} is small.

The key tool used in proving that with probability super-exponentially close to $1$ the first term is small (which is perhaps the ``trickiest'' art of the proof) is in showing that for typical $\vY_w$ and conditionally typical $\vX$, with super-exponentially high probability over code design the expected number of codewords in these typical types is very close to the expected number of codewords (over the randomness in code design). This is possible since the number of codewords is super-polynomially large ($2^{\thr\sqrt{n}}$) and the codebook $\code$ is designed {\it i.i.d.}, hence the Chernoff bound gives us the desired fact. Once we prove this statement, we use the fact that once you have eliminated the highly probable, whatever remains is highly improbable,
\footnote{A slight twist on Sherlock Holmes' dictum ``Once you eliminate the impossible, whatever remains, no matter how improbable, must be the truth.''~\cite{doyle2010sign}} carefully defined triangle inequalities do the trick here.

\begin{figure}
	\centering
	\includegraphics[width=0.8\columnwidth]{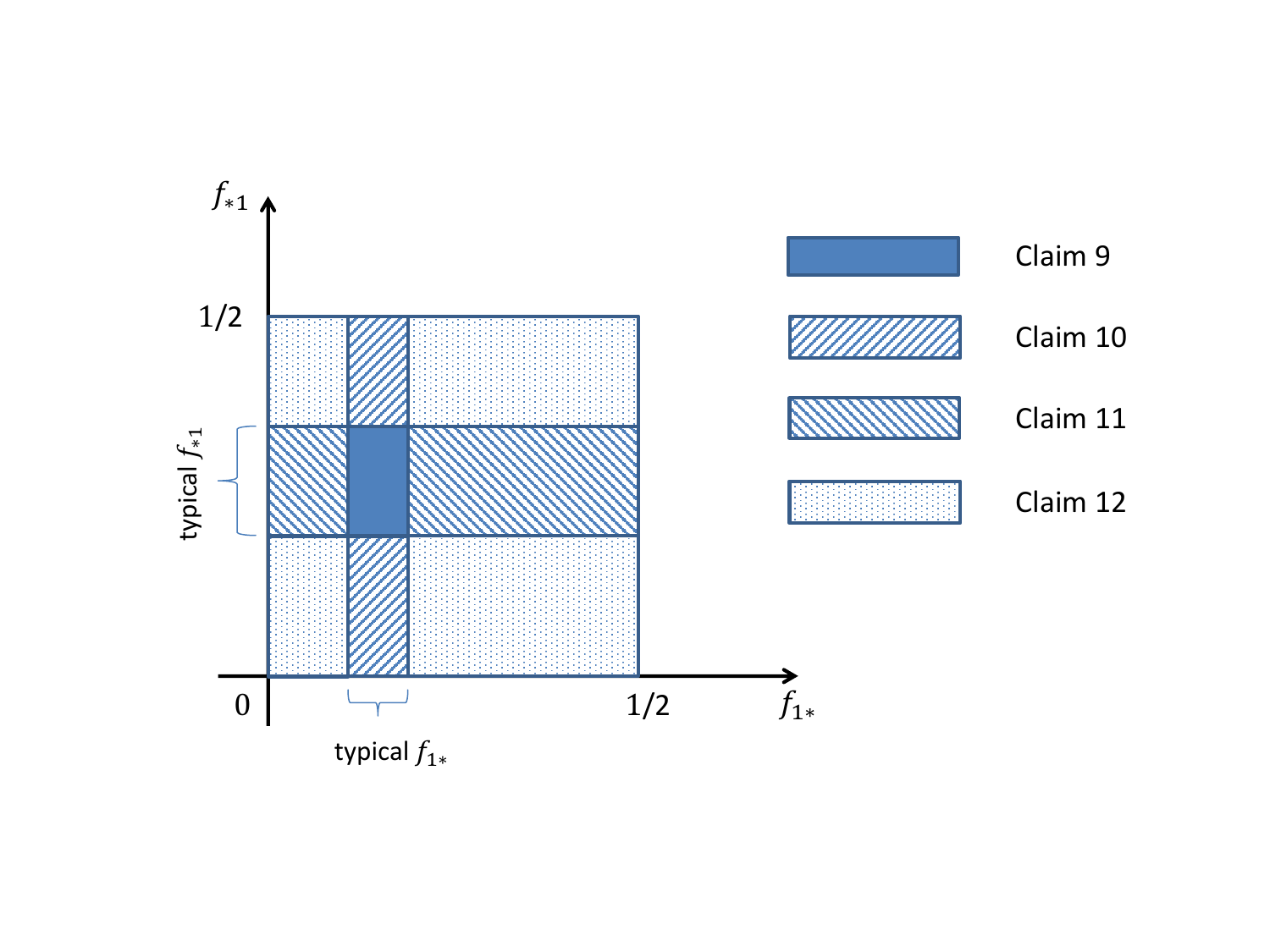}
	\caption[Regions of the Claims in $f_{1*} f_{*1}$ -- plane]{As shown in this figure, The following claims fall into the corresponding typicality condition of $\vx$ and $\vy_w$.}
	\label{fig:achi_deni}
\end{figure}

\begin{claim} \label{clm:typ_yw_typ_x}
For any $\delta \in (0, 1/2)$,
\begin{multline}
	\frac{1}{2} \sum_{\vy_w \in \typyw} \Bigg| \sum_{\code} \Pr(\code) \sum_{\vx \in \code \cap \typxgyw} \po(\vy_w|\vx) p(\vx)) \\
	- \sum_{\vx \in \code \cap \typxgyw} \po(\vy_w|\vx) p(\vx)) \Bigg| < 2^{-n^\delta} \nonumber
\end{multline}
with probability greater than $1 - \exp\left( -\Oh{ 2^{\sqrt{n}} } \right)$ over the code $\code$, where $\delta \ll 1/2$.
\end{claim}

\begin{proof}
Note that the first part of the first term in \eqref{eq:expan_var_dist} can be expanded as,
\begin{align}
	\lefteqn{\sum_\code \Pr(\code) \sum_{\vx \in \code \cap \typxgyw} \po(\vy_w|\vx) p(\vx)} \nonumber \\
	&= \sum_\code \Pr(\code) \sum_{\fozw, \foow} \sum_{\vx \in \code \cap \typexgyw} \po(\vy_w|\vx) p(\vx) \nonumber \\
	&= \sum_\code \Pr(\code) \sum_{\fozw, \foow} | \code \cap \typexgyw | \po(\vy_w|\vx) p(\vx) \label{eq:mo1_prob_are_const} \\
	&= \sum_{\fozw, \foow} \Exp_\code(| \code \cap \typexgyw |) \po(\vy_w|\vx) p(\vx). \label{eq:mo1_prob_are_const2}
\end{align}
In words, Equation~\eqref{eq:mo1_prob_are_const2} says that the expected probability (over the ensemble average over all codes and over the codewords $\vx$ in the specific code $\code$) of observing a $\vy_w$, conditioned on it being ``caused'' by a conditionally typical codeword $\vx$, equals the weighted average (over ``typical type-classes'' $\typexgyw$, weighted by the appropriate probabilities of codewords $p(\vx)$ and transition probability $p_1(\vy_w|\vx)$ of the expected number of codewords in each conditionallly typical type-class $\typxgyw$. To obtain \eqref{eq:mo1_prob_are_const}, we note that $\po(\vy_w|\vx)$ is a constant for particular type $\typexgyw$, and $p(\vx)$ is always constant and equals $1/2^{(\thr + \thr_s)\sqrt{n}}$, since each codeword is generated {\it i.i.d.}. To obtain \eqref{eq:mo1_prob_are_const2}, we exchange the order of the summations. We also note that the expected number of codewords that fall into a high probability type-class $\Exp_{\bm \code}(| \code \cap \typexgyw |) = \Pr_{\vM, \vM_s}(\vX \in \typexgyw) |\code|$. By standard counting arguments, we have
\begin{align}
	\lefteqn{\Pr_{\vM, \vM_s}(\vX \in \code \cap \typexgyw)} \label{eq:pr_xintype_start} \\
		&= {n(\fzow + \foow) \choose n\foow} \rho^{n\foow} (1-\rho)^{n\fzow} \nonumber \\
		& \eqspacing \times {n(\fzzw + \fozw) \choose n\fozw} \rho^{n\fozw} (1-\rho)^{n\fzzw} \label{eq:w_type_0} \\
		& \geq \frac{1}{(n+1)^2} 2^{n(\fzow + \foow)\entropy{\frac{\foow}{\fzow + \foow}}} 2^{n(\fzzw + \fozw)\entropy{\frac{\fozw}{\fzzw + \fozw}}} \nonumber \\
		& \eqspacing \times \rho^{n(\fozw + \foow)} (1-\rho)^{n(\fzzw + \fzow)} \label{eq:w_type_1} \\
		&= \frac{1}{(n+1)^2} 2^{n\condentropy{\vx}{\vy_w}} 2^{n[(\fozw + \foow) \log \rho + (1-\fozw-\foow) \log(1-\rho)]} \\
		&= \frac{1}{(n+1)^2} 2^{n\condentropy{\vx}{\vy_w}} 2^{-n\left[\entropy{\vx} + \KL{\vx}{\rho}\right]} \\
		&= \frac{1}{(n+1)^2} 2^{-n\left[\mutual{\vx}{\vy_w} + \KL{\vx}{\rho}\right]} \label{eq:pr_xintype_end}
\end{align}
Here, \eqref{eq:w_type_0} is the probability that $\vX$ satisfies the constraints that define the type $\typexgyw$ given the received vector $\vy_w$. The term ${\fyw n \choose \foow n} {(1 - \fyw) n \choose \fozw n}$ in \eqref{eq:w_type_1} counts the total number of binary sequences $\vx$ that satisfy the constraint that the fraction of $1$'s in the support of the given $\vy_w$ equals $\foow$, and a similar constraint for $\fozw$. Recall that $\rho$ is the codebook generation probability, and hence $\rho^{\foow n} (1 - \rho)^{\fzow n} \rho^{\fozw n} (1 - \rho)^{\fzzw n}$ is the probability that such $\vx$ are generated. In \eqref{eq:w_type_1}, we also use the fact that ${n \choose k} \geq \frac{1}{n+1} 2^{n \entropy{k/n}}$ \cite{cover2012book}.

We show that for all $(\vx,\vy_w)$ such that $\vy_w$ in $\typyw$, and $\vx \in \typxgyw$, $\mutual{\vx}{\vy_w} = \rho (1 - 2\pw) \log \frac{1 - \pw}{\pw} + \Oh{n^{-\frac{3}{4}}}$ and $\KL{\vx}{\rho} = \Oh{n^{-1}}$ in the Appendix. We can choose the throughput $\thr > \thr_L$ so that $\Exp_\code(| \code \cap \typexgyw |) = 2^{c \sqrt{n}}$ for a constant $c = \thr - \thr_L > 0$. Thus, we can concentrate the value of $| \code \cap \typexgyw |$ around its expectation by using the Chernoff bound,
\begin{multline} \label{eq:type_chernoff_bd}
	\Pr_{{\bm \code}, \vM, \vM_s} \Bigg( \Big| | \code \cap \typexgyw | \\
		- \Exp_\code(| \code \cap \typexgyw |) \Big| \\
		> \epstyp \Exp_\code(| \code \cap \typexgyw |) \Bigg) \\
		< 2 \exp\left(-\frac{1}{3} \epstyp^2 2^{c\sqrt{n}} \right).
\end{multline}
Here $\epstyp$ is a code-design parameter. By choosing $\epstyp = 2^{-n^{\delta}}$ and $\delta \in (0, 1/2)$, we have with probability at least $1 - 2 \exp\left(- \Oh{2^{\sqrt{n}}} \right)$ over the code $\code$.
\begin{align}
	\lefteqn{\sum_{\vy_w \in \typyw} \left| \sum_\code \Pr(\code) \sum_{\vx \in \code \cap \typxgyw} \po(\vy_w|\vx) p(\vx) \right.} \nonumber \\
		& \eqspacing \left. - \sum_{\vx \in \code \cap \typxgyw} \po(\vy_w|\vx) p(\vx) \right| \nonumber \\
		&= \sum_{\vy_w \in \typyw} \nonumber \\
		& \eqspacing \left| \sum_{(\fozw, \foow) \in \typFw} \Exp_{\bm \code}(| \code \cap \typexgyw |) \po(\vy_w|\vx) p(\vx) \right. \nonumber \\
		& \eqspacing \left. - \sum_{(\fozw, \foow) \in \typFw} | \code \cap \typexgyw | \po(\vy_w|\vx) p(\vx) \right| \label{eq:sum_x_to_f} \\
		& \leq \sum_{\vy_w \in \typyw} \sum_{(\fozw, \foow) \in \typFw} \left| \Exp_{\bm \code}(| \code \cap \typexgyw |) \right. \nonumber \\
		& \eqspacing \eqspacing \left. - | \code \cap \typexgyw | \right| \po(\vy_w|\vx) p(\vx) \label{eq:type_hold_super} \\
		&< \sum_{\vy_w \in \typyw} \sum_{\fozw, \foow} \epstyp \Exp_{\bm \code}(| \code \cap \typexgyw |) \po(\vy_w|\vx) p(\vx) \nonumber \\
		&< \epstyp = 2^{-n^{\delta}}, \label{eq:mo1_ep_var_typ_x}
\end{align}
where we change the summation $\sum_{\vx \in \code \cap \typxgyw}$ over the typical set into the summation $\sum_{(\fozw, \foow) \in \typFw} | \code \cap \typexgyw |$ over typical type-classes in \eqref{eq:sum_x_to_f} as defined in \ref{def:typexgyw}. Since \eqref{eq:type_chernoff_bd} for each typical type-class with super-exponential high probability close to $1$, and there are only super-polynomially many typical type-classes. Taking a union bound, we obtain inequality~\eqref{eq:type_hold_super}.


Hence, in summary, with probability at least $1 - \exp\left(- \Oh{2^{\sqrt{n}}} \right)$ over code design, the contribution to the variational distance between $\Po$ and $\Exp_\code(\Po)$ by the first term in \eqref{eq:var_further_tri} is at most $2^{-n^\delta}$. Recall that this first term corresponds to only those $\vy_w$ which are typical, and those codewords $\vx$ that are conditionally typical with respect to such typical $\vy_w$. This proves Claim~\ref{clm:typ_yw_typ_x}.
\end{proof}

Next, we focus on bounding the remaining terms in equation~\eqref{eq:var_further_tri}.

\begin{claim} \label{clm:atyp_yw}
	We have $$ \sum_{\vy_w \notin \typyw} \Exp_{\bm \code}(\po(\vy_w)) < 2^{-c' n^\delta} $$ for some constant $c'$ by choosing $\delyw = n^{- 1/2 + \delta/2}$.
\end{claim}

\begin{proof}
By the Chernoff bound, we have
\begin{align}
	\lefteqn{ \sum_{\vy_w \notin \typyw} \Exp_{\bm \code}(\po(\vy_w)) } \nonumber \\
	 	&= \Pr_{{\bm \code}, \vM, \vM_s, \vZ_w}\left( \vY_w \notin \typyw \right) \nonumber \\
		&= \Pr_{{\bm \code}, \vM, \vM_s, \vZ_w}\left( \fyw \notin ( (1 - \delyw) \rho \ast \pw, (1 + \delyw) \rho \ast \pw ) \right) \nonumber \\
		&< 2 \exp\left(-\frac{1}{3} \delyw^2 (\rho \ast \pw) n\right). \nonumber
\end{align}

Choosing
\begin{equation}
	\delyw = n^{- 1/2 + \delta/2},
	\label{eq:mo1_delyw}
\end{equation}
we have
\begin{align}
	\sum_{\vy_w \notin \typyw} \Exp_{\bm \code}(\po(\vy_w)) &< 2 \exp\left(-\frac{1}{3} \delyw^2 (\rho \ast \pw) n\right) \nonumber \\
		&= 2 \exp\left( -\frac{1}{3} (\rho \ast \pw) n^\delta \right) \nonumber \\
		&= 2^{-c' n^\delta}, \label{eq:mo1_ep_atyp_yw}
\end{align}
for some constant $c'$.
\end{proof}

\begin{claim} \label{clm:typ_yw_atyp_x}
	We have $$ \Exp_{\bm \code}\left(\sum_{\vy_w \in \typyw} \sum_{\vx \in \code \setminus \typxgyw} \po(\vy_w|\vx) p(\vx) \right) < 2^{-c'' n^\delta} $$ for some constant $c''$ by choosing $ \delozw = \deloow = n^{-1/4 + \delta/2} $.
\end{claim}

\begin{proof}
Similarly to Claim~\ref{clm:atyp_yw}, we have
\begin{align}
	\lefteqn{\Exp_{\bm \code} \left(\sum_{\vy_w \in \typyw} \sum_{\vx \in \code \setminus \typxgyw} \po(\vy_w|\vx) p(\vx) \right)} \nonumber \\
		&= \Exp_{\bm \code} \left( \sum_{\vy_w \in \typyw} \sum_{\vx \in \code \setminus \typxgyw} \po(\vy_w) p(\vx|\vy_w) \right) \nonumber \\
		&= \sum_{\vy_w \in \typyw} \po(\vy_w) \Exp_{\bm \code} \left( \sum_{\vx \in \code \setminus \typxgyw} p(\vx|\vy_w) \right) \nonumber \\
		&< \left( \sum_{\vy_w \in \typyw} \po(\vy_w) \right) \nonumber \\
		& \eqspacing \times \left[ 2 \exp\left( -\frac{1}{3} \delozw \rho \pw n \right) + 2 \exp\left( -\frac{1}{3} \deloow \rho (1 - \pw) n \right) \right] \label{eq:known:achi:deni:actypx} \\
		&< 2 \exp\left( -\frac{1}{3} \delozw \rho \pw n \right) + 2 \exp\left( -\frac{1}{3} \deloow \rho (1 - \pw) n \right), \nonumber
\end{align}
where equation~\eqref{eq:known:achi:deni:actypx} holds since the Chernoff bound is applied to the following equation,
\begin{multline*}
	\Exp_{\bm \code} \left( \sum_{\vx \in \code \setminus \typxgyw} p(\vx|\vy_w) \right) \\
	= \Pr_{{\bm \code}, \vM, \vM_s, \vZ_w}\Bigg( \{ \fozw \notin ( (1 - \delozw) \rho \pw, (1 + \delozw) \rho \pw ) \} \\
						\bigcup \{ \foow \notin (1 - \deloow) \rho (1 - \pw), (1 + \deloow) \rho (1 - \pw) ) \} \Bigg).
\end{multline*}

Choosing
\begin{equation}
	\delozw = \deloow = n^{-1/4 + \delta/2},
	\label{eq:mo1_delozw}
\end{equation}
we have
\begin{equation}
	\Exp_{\bm \code} \left(\sum_{\vy_w \in \typyw} \sum_{\vx \in \code \setminus \typxgyw} \po(\vy_w|\vx) p(\vx) \right) < 2^{- c'' n^\delta}.
	\label{eq:mo1_ep_atyp_x}
\end{equation}
\end{proof}

\begin{claim} \label{clm:atyp_yw_atyp_x}
$$ \sum_{\vy_w \notin \typyw} \po(\vy_w) + \sum_{\vy_w \in \typyw} \sum_{\vx \in \code \setminus \typxgyw} \po(\vy_w|\vx) p(\vx) < 2^{- \Omega(n^\delta)} $$ with probability greater than $1 - \exp\left( -\Oh{2^{\sqrt{n}}} \right)$ over code design.
\end{claim}

\begin{proof}
By Claim~\ref{clm:typ_yw_typ_x} and Claim~\ref{clm:typ_yw_atyp_x}, with probability greater than $1 - \exp\left( -\Oh{2^{\sqrt{n}}} \right)$ over code design, we have
\begin{align}
	\lefteqn{2^{-n^\delta} + 2^{- c'' n^\delta}} \nonumber \\
		&> \sum_{\vy_w \in \typyw} \left| \sum_\code \Pr(\code) \sum_{\vx \in \code \cap \typxgyw} \po(\vy_w|\vx) p(\vx) \right. \nonumber \\
		& \eqspacing \left. - \sum_{\vx \in \code \cap \typxgyw} \po(\vy_w|\vx) p(\vx) \right| \nonumber \\
		& \eqspacing \eqspacing + \Exp_\code\left(\sum_{\vy_w \in \typyw} \sum_{\vx \in \code \setminus \typxgyw} \po(\vy_w|\vx) p(\vx) \right) \label{eq:abs_eps} \\
		& \geq \left[ \sum_{\vy_w \in \typyw} \sum_\code \Pr(\code) \sum_{\vx \in \code \cap \typxgyw} \po(\vy_w|\vx) p(\vx) \right. \nonumber \\
		& \eqspacing \left. + \Exp_\code\left(\sum_{\vy_w \in \typyw} \sum_{\vx \in \code \setminus \typxgyw} \po(\vy_w|\vx) p(\vx) \right) \right] \nonumber \\
		& \eqspacing \eqspacing - \sum_{\vy_w \in \typyw} \sum_{\vx \in \code \cap \typxgyw} \po(\vy_w|\vx) p(\vx) \label{eq:abs_to_noabs} \\
		&= \sum_{\vy_w \in \typyw} \Exp_\code(\po(\vy_w)) - \sum_{\vy_w \in \typyw} \sum_{\vx \in \code \cap \typxgyw} \po(\vy_w|\vx) p(\vx) \nonumber \\
		&> 1 - 2^{- c' n^\delta} \nonumber \\
		& \eqspacing - \left[ 1 - \sum_{\vy_w \notin \typyw} \po(\vy_w) - \sum_{\vy_w \in \typyw} \sum_{\vx \in \code \setminus \typxgyw} \po(\vy_w|\vx) p(\vx) \right] \label{eq:typ_yw} \\
		&= \sum_{\vy_w \notin \typyw} \po(\vy_w) + \sum_{\vy_w \in \typyw} \sum_{\vx \in \code \setminus \typxgyw} \po(\vy_w|\vx) p(\vx) - 2^{- c' n^\delta} \nonumber
\end{align}

We obtain \eqref{eq:abs_to_noabs} by using the triangle inequality $|a - b| \geq a - b$ in the first term in \eqref{eq:abs_eps}. Equation~\eqref{eq:typ_yw} holds directly from Claim~\ref{clm:atyp_yw}. Therefore, we have
\begin{align}
	& \sum_{\vy_w \notin \typyw} \po(\vy_w) + \sum_{\vy_w \in \typyw} \sum_{\vx \in \code \setminus \typxgyw} \po(\vy_w|\vx) p(\vx)  \nonumber \\
	&< 2^{- n^\delta} + 2^{- c' n^\delta} + 2^{- c'' n^\delta} \nonumber \\
	&= 2^{- \Omega(n^\delta)}. \label{eq:mo1_atyp_sum}
\end{align}
\end{proof}

Hence, combining Claim~\ref{clm:typ_yw_typ_x}, Claim~\ref{clm:atyp_yw}, Claim~\ref{clm:typ_yw_atyp_x} and Claim~\ref{clm:atyp_yw_atyp_x}, with probability greater than $1 - \exp\left( -\Oh{2^{\sqrt{n}}} \right)$ over code design, the variational distance $\variation{\Exp_\code(\Po)}{\Po}$ can be bounded from above as follows,
\begin{align}
	\lefteqn{\variation{\Exp_\code(\Po)}{\Po}} \nonumber \\
		&= \frac{1}{2} \sum_{\vy_w \in \{0,1\}^n} \left|\Exp_\code(\po(\vy_w)) - \po(\vy_w)\right| \nonumber \\
		&< \frac{1}{2} \sum_{\vy_w \in \typyw} \left| \sum_\code \Pr(\code) \sum_{\vx \in \code \cap \typxgyw} \po(\vy_w|\vx) p(\vx) \right. \nonumber \\
		& \eqspacing \eqspacing \left. - \sum_{\vx \in \code \cap \typxgyw} \po(\vy_w|\vx) p(\vx) \right| \nonumber \\
		& \eqspacing + \frac{1}{2} \sum_{\vy_w \in \typyw} \sum_\code \Pr(\code) \sum_{\vx \in \code \setminus \typxgyw} \po(\vy_w|\vx) p(\vx) \nonumber \\
		& \eqspacing \eqspacing + \frac{1}{2} \sum_{\vy_w \in \typyw} \sum_{\vx \in \code \setminus \typxgyw} \po(\vy_w|\vx) p(\vx) \nonumber \\
		& + \frac{1}{2} \sum_{\vy_w \notin \typyw} \Exp_\code(\po(\vy_w)) + \frac{1}{2} \sum_{\vy_w \notin \typyw} \po(\vy_w) \label{eq:mo1_var_full_tri_ineq} \\
		&< 2^{-\Omega(n^\delta)}. \nonumber 
\end{align} $\hfill\square$

This (finally!) concludes the proof of Lemma~\ref{lemma:achi_deni_EcP1_P1}, which first started on page 70!

This Lemma shows that the variational distance between $\Exp_\code(\Po)$ and $\Po$ is exponentially small with probability super-exponentially close to $1$. This says that the bulk of the contribution of the variational distance between $\Pz$ and $\Po$ is due to the variational distance between the two smooth distributions $\Pz$ and $\Exp_\code(\Po)$ -- the ``lumpiness'' (as shown in Figures~\ref{fig:ps} and \ref{fig:pmfs}) corresponding to the variational distance between the ensemble average distribution of $\vY_w$ and the actual distribution on $\vY_w$ due to a randomly chosen codebook.


\subsubsection{Achievability: Reliability}

The reliability follows from Proposition~\ref{prop:ran_sto_code} by choosing the parameter $\code(\thr\sqrt{n},1)$.
\subsection{Hidability of Fixed Channel Model (Theorem~\ref{thm:model:known:hide})}
Note that the reliability follows from Proposition~\ref{prop:ran_sto_code} by choosing the parameter $\code(\thr\sqrt{n},\thr_s\sqrt{n})$.

And, the hidability follows directly from the standard secrecy arguments.


\newpage

\subsection{Converse for Slow Fading Channel Model (Theorem~\ref{thm:model:unknown:conv})}

We now move to analyzing a channel model in which the throughput that Alice can both reliably and deniably get through to Bob scales {\it linearly} in the block-length $n$, rather than as the {\it square-root} of $n$. The reason for this behaviour is because the value of the noise parameter (in the case of the Binary Symmetric Channels we consider in this work, the noise parameter we consider corresponds to the probability of bit-flips) itself has some uncertainty. Since the ``SNR'' of the channel itself has some uncertainty, from Willie's perspective, so the standard deviation of the noise itself increases substantially, from $\Oh{\sqrt{n}}$, to being linear in $n$ in some cases of interest (as in the specific case we analyze, where the noise parameter is uniformly distributed in an interval).\footnote{This insightful way of viewing matters is due to a conversation with Gerhard Kramer.} Hence in the following discussion we discuss the {\it rates} of codes (which scale linearly with block-length $n$), rather than their {\it throughputs} (which scale linearly with $\sqrt{n}$).

\subsubsection{Upper bound on $\alpha + \beta$}

In the outer bound for the Fixed Channel model discussed in Section~\ref{sec:fixed_channel:conv}, the estimator discussed was a threshold estimator, which used the Hamming weight of the received transmission to estimate the``energy'' injected into the transmitted codeword by Alice. This resulted in an outer bound on the throughput which did not match (up to small constant factors explicitly calculated) in that model.

However, in the Slow Fading Channel Model considered in this section, the noise parameters $\pb$ and $\pw$ themselves are drawn (independently) from uniform distributions on pre-specified sub-intervals $(\lb, \ub)$ and $(\lw, \uw)$ of $(0,1/2)$ respectively. For this model we demonstrate that using essentially the same estimator as in the Fixed Channel Model results in outer bounds on the $(1-\epsreli)$-reliable $(1-\epsdeni)$-deniable rate achievable by Alice that are asymptotically essentially optimal, by showing an achievability scheme in the next section (Section~\ref{sec:slow_fading:achi}) that has performance essentially meeting those outer bounds.\footnote{The difference in behaviour between these two channel models stems perhaps from the fact that our outer bounding techniques for the Fixed Channel models aren't yet fully optimized -- indeed, we are currently studying the techniques of Wang {\it et al}~\cite{Wang2015}, and Bloch~\cite{Bloch2015} to understand their techniques, since they have matching achievabilities and converses for the Fixed Channel models as well.}


Let $\fyw$ denote the fractional Hamming weight of $\vy_w$, $\wt{\vy_w}/n$, Willie chooses as threshold $\thre$ (a parameter whose value we specify later in this section). When Willie receives the vector $\vy_w$, he generates his estimate $\hat{\tranrv}$ of Alice's transmission status $\tranrv$ as follows:
\begin{itemize}
	\item $\hat{\tranrv} = 1$, if $\fyw > \thre$;
	\item $\hat{\tranrv} = 0$, otherwise.
\end{itemize}

In Lemma~\ref{lemma:up_bd_alpha_beta} below, we show that among the class of such threshold-based estimators, there is a specific optimal choice of the threshold $\thre$ that Willie can choose so as to minimize the deniability of Alice's communication scheme. Then, we show that if Willie's deniability is at most $1 - \epsdeni$, the fractional weight of Alice's codewords is at least $\frac{1}{\gamma(\rho)} \frac{\uw - \lw}{1 - 2\lw} \epsdeni$. Here, $\gamma(\rho)$ denotes the probability mass of codewords whose fractional weight is larger than $\rho$.

\begin{lemma} \label{lemma:up_bd_alpha_beta}
The sum of probability of false alarm $\alpha$ and probability of missed detection $\beta$ satisfies
\[\alpha + \beta \leq 1 - \rho \gamma(\rho) \frac{1 - 2\lw}{\uw - \lw},\]
if Alice's codebook has a $\gamma(\rho)$ fraction of codewords with fractional weight at least $\rho$.
\end{lemma}

\begin{proof}

This proof consist of three parts. In the first part, we express the probability of false alarm $\alpha$ as an explicit function of the threshold $\thre$. Similarly, in the second part, for each possible codeword we derive an explicit expression for the probability of missed detection $\beta_{\rho}$ as a function of the threshold $\thre$. In the third part of this proof, we combine the results from the first two parts, and suggest an optimal choice of threshold $\thre$ for Willie to choose so as to minimize the deniability $1-\epsdeni$ among the class of threshold-based estimators.

\begin{enumerate}
\item {\bf Calculation of the probability of false alarm $\alpha$}
	We break our anaysis of the probability of false alarm into three cases. Specifically, if Willie chooses a threshold based estimator, we bound from above the probability of false alarm in $3$ parts summarized in equation~\eqref{eq:falarm_up_bd_simp} below, derived by considering three possible ranges into which the threshold $\thre$ may fall. And $\delta$ below is a proof-technique parameter to be specified in equation~\eqref{eq:falarm_up_bd}.
	\begin{enumerate}
	
	\item If Willie's threshold $\thre \leq \lw + \delta$, we have
	
	\begin{equation}
		\prob_{\Pw, \vZ_w}(\hat{\tranrv} = 1 | \tranrv = 0) \leq 1.
		\label{eq:falarm_t_less_Lw}
	\end{equation}

	\item If $\thre \in (\lw + \delta, \uw + \delta)$, we have that Willie's probability of false alarm is bounded from above as
	\begin{align}
		\lefteqn{\Pr_{\Pw, \vZ_w}(\hat{\tranrv} = 1 | \tranrv = 0)} \nonumber \\
			&= \Pr_{\Pw, \vZ_w}(\hat{\tranrv} = 1 | \tranrv = 0, \Pw \in (\lw, \thre - \delta])\frac{\thre - \delta - \lw}{\uw - \lw} \nonumber \\
			& \eqspacing + \Pr_{\Pw, \vZ_w}(\hat{\tranrv} = 1 | \tranrv = 0, \Pw \in (\thre - \delta, \uw))\frac{\uw - \thre + \delta}{\uw - \lw} \label{eq:falarm_brk} \\
			& \leq \Pr_{\vZ_w}\left( \fyw > \thre | \Pw = \thre - \delta \right) \frac{\thre - \delta - \lw}{\uw - \lw} + \frac{\uw - \thre + \delta}{\uw - \lw} \label{eq:falarm_brk2} \\
			&< \exp\left( -\frac{\delta^2}{3(\thre-\delta)} n \right) \frac{\thre - \delta - \lw}{\uw - \lw} + \frac{\uw - \thre + \delta}{\uw - \lw}, \label{eq:falarm_brk3}
	\end{align}
	In \eqref{eq:falarm_brk}, we divide the range of $\Pw$ into two parts, $(\lw, \thre - \delta]$ and $(\thre - \delta, \uw)$, where the value of $\delta$ will be determined later. The inequality \eqref{eq:falarm_brk2} holds since $\Pr_{\Pw, \vZ_w}(\hat{\tranrv} = 1 | \tranrv = 0, \Pw \in (\thre - \delta, \uw)) \leq 1$ and
	\begin{align}
		\lefteqn{\Pr_{\Pw, \vZ_w}(\hat{\tranrv} = 1 | \tranrv = 0, \Pw \in (\lw, \thre - \delta])} \nonumber \\
		&= \frac{1}{\thre - \delta - \lw} \int_{\lw}^{\thre - \delta} \Pr_{\vZ_w}(\hat{\tranrv} = 1 | \tranrv = 0, \Pw = p) \di p \nonumber \\
		& \leq \frac{1}{\thre - \delta - \lw} \int_{\lw}^{\thre - \delta} \max_{p \in [\lw, \thre - \delta]}\Pr_{\vZ_w}(\hat{\tranrv} = 1 | \tranrv = 0, \Pw = p) \di p \nonumber \\
		&= \frac{1}{\thre - \delta - \lw} \int_{\lw}^{\thre - \delta} \Pr_{\vZ_w}(\hat{\tranrv} = 1 | \tranrv = 0, \Pw = \thre - \delta) \di p \nonumber \\
		&= \Pr_{\vZ_w}(\hat{\tranrv} = 1 | \tranrv = 0, \Pw = \thre - \delta). \nonumber
	\end{align}
	We obtain \eqref{eq:falarm_brk3} by applying the Chernoff bound.

	\item If $\thre \geq \uw + \delta$, we note that
	\begin{align}
		\lefteqn{\Pr_{\Pw, \vZ_w}(\hat{\tranrv} = 1 | \tranrv = 0)} \nonumber \\
			&= \Pr_{\Pw, \vZ_w}(\fyw > \thre | \tranrv = 0) \nonumber \\
			& \leq \Pr_{\vZ_w}(\fyw > \uw + \delta | \tranrv = 0, \Pw = \uw) \nonumber \\
			&< \exp\left(-\frac{\delta^2}{3 \uw}n\right), \label{eq:falarm_t_large_Uw}
	\end{align}
	where \eqref{eq:falarm_t_large_Uw} holds by the Chernoff bound.

	\end{enumerate}
	
Combining \eqref{eq:falarm_t_less_Lw}, \eqref{eq:falarm_brk3} and \eqref{eq:falarm_t_large_Uw}, we have the probability of false alarm is bounded from above as
\begin{multline}
	\Pr_{\Pw, \vZ_w}(\hat{\tranrv} = 1 | \tranrv = 0) \\ \left\{ \begin{array}{ll}
		\leq 1, &\mbox{if } \thre \leq \lw + \delta, \\
		< \exp\left( -\frac{\delta^2}{3(\thre-\delta)} n \right) \frac{\thre - \delta - \lw}{\uw - \lw} + \frac{\uw - \thre + \delta}{\uw - \lw}, &\mbox{if } \thre \in (\lw + \delta, \uw + \delta), \\
		< \exp\left(-\frac{\delta^2}{3 \uw}n\right), &\mbox{if } \thre \geq \uw + \delta.
	\end{array} \right.
\label{eq:falarm_up_bd}
\end{multline}

Choosing $\delta = n^{-1/4}$, we have that for $\lambda \triangleq \exp\left( - \frac{n^{1/2}}{3 \uw} \right)$,
$$\exp\left( -\frac{\delta^2}{3(\thre-\delta)} n \right) \frac{\thre - \delta - \lw}{\uw - \lw} + \frac{\uw - \thre + \delta}{\uw - \lw} < \lambda + \frac{\uw - \thre + \delta}{\uw - \lw}.$$
So, the upper bound on \eqref{eq:falarm_up_bd} can be rewritten as 
\begin{equation}
	\Pr_{\Pw, \vZ_w}(\hat{\tranrv} = 1 | \tranrv = 0) \left\{ \begin{array}{ll}
		\leq 1, &\mbox{if } \thre \leq \lw + \delta, \\
		\leq \lambda + \frac{\uw - \thre + \delta}{\uw - \lw}, &\mbox{if } \thre \in (\lw + \delta, \uw + \delta), \\
		\leq \lambda, &\mbox{if } \thre \geq \uw + \delta.
	\end{array} \right.
\label{eq:falarm_up_bd_simp}
\end{equation}
where $\lambda$ can be made arbitrarily small as $n$ increases. This curve is visually depicted in Figure~\ref{fig:converse1} and \ref{fig:converse2} below.

\item {\bf Calculation of the probability of missed detection $\beta_{\zeta, \thre}$}

We first do a calculation on an outer bound on the probability of missed detection for a single codeword $\vx$ of fixed fractional weight $\zeta \triangleq \wt{\vx}/n$. Note that the calculation of $\beta_{\zeta, \thre}$ depends on the fractional weight of a codeword to be $\zeta$.

As for the bound on the probability of false alarm derived in equation~\eqref{eq:falarm_up_bd_simp}, we break our analysis of the probability of missed detection into three cases. Specifically, if Willie chooses a threshold-based estimator, we bound from above the probability of missed detection in $3$ parts summarized in equation~\eqref{eq:mdetect_up_bd_simp} below, derived by considering three possible ranges into which the threshold $\thre$ may fall. And $\bar{\delta}$ below is a proof-technique parameter to be determined in equation~\eqref{eq:mdetect_up_bd}.

\begin{enumerate}

	\item If Willie chooses his threshold $\thre \leq \zeta \ast \lw - \bar{\delta}$, then
	\begin{align}
		\lefteqn{\Pr_{\Pw, \vZ_w}(\hat{\tranrv} = 0 | \tranrv = 1, \vX = \vx)} \nonumber \\ &= \Pr_{\Pw, \vZ_w}(\fyw < t | \vX = \vx) \nonumber \\
			& \leq \Pr_{\vZ_w}(\fyw < \zeta \ast \lw - \bar{\delta} | \vX = \vx, \Pw = \lw) \nonumber \\
			&< \exp\left( -\frac{\bar{\delta}^2}{2 (\zeta \ast \lw)} n \right), \label{eq:mdetect_t_less_rhoLw}
	\end{align}
note that the codeword $\vx$ has fractional weight $\zeta$ and the channel parameter $\Pw = \lw$, in expectation, the fraction of $1$'s in the received vector $\fyw$ is $\zeta \ast \lw$. Therefore, equation~\eqref{eq:mdetect_t_less_rhoLw} can be obtained by the Chernoff bound.

	\item If the threshold $\thre \in \zeta \ast \lw - \bar{\delta}, \zeta \ast \uw - \bar{\delta})$, then
	\begin{align}
		\lefteqn{\Pr_{\Pw, \vZ_w}(\hat{\tranrv} = 0 | \tranrv = 1, \vX = \vx)} \nonumber \\
			&= \Pr_{\Pw, \vZ_w}(\hat{\tranrv} = 0 | \vX = \vx, \zeta \ast \Pw \in (\zeta \ast \lw, \thre + \bar{\delta})) \nonumber \\
			& \eqspacing \times \frac{\thre + \bar{\delta} - \zeta \ast \lw}{\zeta \ast \uw - \zeta \ast \lw} \nonumber \\
			& \eqspacing + \Pr_{\Pw, \vZ_w}(\hat{\tranrv} = 0 | \vX = \vx, \zeta \ast \Pw \in [\thre + \bar{\delta}, \zeta \ast \uw)) \nonumber \\
			& \eqspacing \eqspacing \times \frac{\zeta \ast \uw - \thre - \bar{\delta}}{\zeta \ast \uw - \zeta \ast \lw} \label{eq:mdetect_brk} \\
			& \leq  \frac{\thre + \bar{\delta} - \zeta \ast \lw}{\zeta \ast \uw - \zeta \ast \lw} \nonumber \\
			& \eqspacing + \Pr_{\vZ_w}\left( \fyw < \thre | \vX = \vx, \zeta \ast \Pw = \thre + \bar{\delta} \right) \nonumber \\
			& \eqspacing \eqspacing \times \frac{\zeta \ast \uw - \thre - \bar{\delta}}{\zeta \ast \uw - \zeta \ast \lw} \label{eq:mdetect_brk2} \\
			&< \frac{\thre + \bar{\delta} - \zeta \ast \lw}{\zeta \ast \uw - \zeta \ast \lw} \nonumber \\
			& \eqspacing + \exp\left( -\frac{\bar{\delta}^2}{2 (t + \bar{\delta})} n \right) \frac{\zeta \ast \uw - \thre - \bar{\delta}}{\zeta \ast \uw - \zeta \ast \lw}, \label{eq:mdetect_brk3}
	\end{align}
	In \eqref{eq:mdetect_brk}, we break the range of $\Pw$ into two parts $(\zeta \ast \lw, \thre+\bar{\delta})$ and $[\thre + \bar{\delta}, \zeta \ast \uw)$, where $\bar{\delta}$ will be determined later. Equation~\eqref{eq:mdetect_brk2} follows from arguments similar to those used to justify equation~\eqref{eq:falarm_brk2}. By the Chernoff bound, we obtain \eqref{eq:mdetect_brk3}.

	\item If the threshold $\thre \geq \zeta \ast \uw - \bar{\delta}$, we have
	\begin{equation}
		\Pr_{\Pw, \vZ_w}(\hat{\tranrv} = 0 | \tranrv = 1, \vX = \vx) \leq 1.
	\label{eq:mdetect_t_large_rhoUw}
	\end{equation}

	\end{enumerate}

Therefore, combining \eqref{eq:mdetect_t_less_rhoLw}, \eqref{eq:mdetect_brk3} and \eqref{eq:mdetect_t_large_rhoUw}, we have that the probability of missed detection given a fractional weight of a codeword to be $\zeta$ is bounded from above as
{\small \begin{multline}
	\Pr(\hat{\tranrv} = 0 | \tranrv = 1, \vX = \vx) \\ \left\{ \begin{array}{ll}
		< \exp\left( -\frac{\bar{\delta}^2}{2 (\zeta \ast \lw)} n \right), &\mbox{if } \thre \leq \zeta \ast \lw - \bar{\delta}, \\
		< \frac{\thre + \bar{\delta} - \zeta \ast \lw}{\zeta \ast \uw - \zeta \ast \lw} & \\
		\eqspacing + \exp\left( -\frac{\bar{\delta}^2}{2 (t + \bar{\delta})} n \right) \frac{\zeta \ast \uw - \thre - \bar{\delta}}{\zeta \ast \uw - \zeta \ast \lw}, &\mbox{if } \thre \in (\zeta \ast \lw - \bar{\delta}, \zeta \ast \uw - \bar{\delta}), \\
		\leq 1, &\mbox{if } \thre \geq \zeta \ast \uw - \bar{\delta}.
	\end{array} \right.
\label{eq:mdetect_up_bd}
\end{multline} }

Choosing $\bar{\delta} = n^{-1/4}$, we have that for $\bar{\lambda} \triangleq \exp\left( -\frac{n^{1/2}}{2 (\zeta \ast \uw)}\right)$,
and
\begin{multline}
	\frac{\thre + \bar{\delta} - \zeta \ast \lw}{\zeta \ast \uw - \zeta \ast \lw} + \exp\left( -\frac{\bar{\delta}^2}{2 (t + \bar{\delta})} n \right) \frac{\zeta \ast \uw - \thre - \bar{\delta}}{\zeta \ast \uw - \zeta \ast \lw} \\
	< \frac{\thre + \bar{\delta} - \zeta \ast \lw}{\zeta \ast \uw - \zeta \ast \lw} + \bar{\lambda}.
\end{multline}
Therefore, \eqref{eq:mdetect_up_bd} can be simplified to the following for large $n$,
{\small \begin{equation}
	\prob(\hat{\tranrv} = 0 | \tranrv = 1) \left\{ \begin{array}{ll}
		\leq \bar{\lambda}, &\mbox{if } \thre \leq \zeta \ast \lw - \bar{\delta}, \\
		< \frac{\thre + \bar{\delta} - \zeta \ast \lw}{\zeta \ast \uw - \zeta \ast \lw} + \bar{\lambda}, &\mbox{if } \thre \in (\zeta \ast \lw - \bar{\delta}, \zeta \ast \uw - \bar{\delta}), \\
		\leq 1, &\mbox{if } \thre \geq \zeta \ast \uw - \bar{\delta}.
	\end{array} \right.
\label{eq:mdetect_up_bd_simp}
\end{equation} }
This curve is visually depicted in Figure~\ref{fig:converse1} and \ref{fig:converse2} below.
	
\item {\bf Optimizing $\alpha + \beta$}

	\begin{enumerate}

	\item {\bf The case of ``small'' fractional weight --  [$\rho \ast \lw - \bar{\delta} < \uw + \delta$]}
	
	\begin{figure}
		\centering
		\includegraphics[width=0.8\columnwidth]{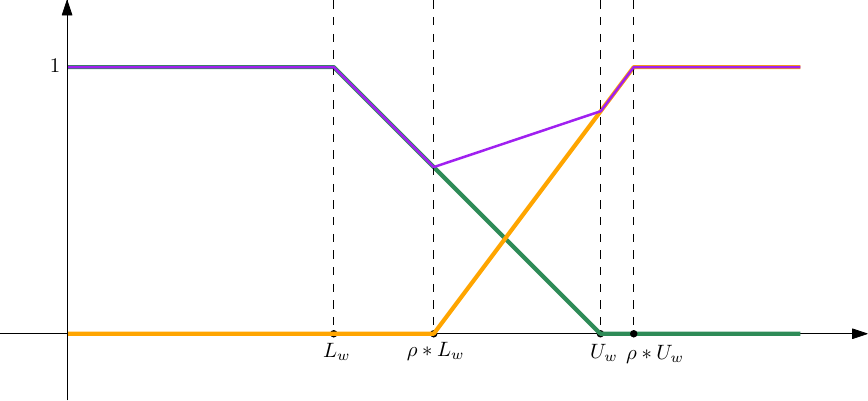}
		\caption[The case of ``small'' fractional weight : $\zeta \ast \lw - \bar{\delta} < \uw + \delta$]{{\bf The case of ``small'' fractional weight : $\zeta \ast \lw - \bar{\delta} < \uw + \delta$.} The $x$-axis represents the choice of the threshold $\thre$, and the $y$-axis corresponds to the probability of various events. This figure shows the asymptotic behaviour of the probability of false alarm $\alpha$, missed detection $\beta_\rho$ and the deniability $\alpha + \beta_\zeta$. The green line shows the probability of false alarm  $\alpha$, the orange line shows the probability of missed detection $\beta_\zeta$ given a codeword with fractional weight $\zeta$. The deniability is shown via a purple line, which is equal to the sum of false alarm and missed detection probabilities $\alpha + \beta_\zeta$.}
		\label{fig:converse1}
	\end{figure}

	When $\zeta \ast \lw - \bar{\delta} < \uw + \delta$, we obtain the curves depicted in Figure~\ref{fig:converse1}. Here, we can see that the optimal choice of threshold $\thre$ is $\zeta \ast \lw - \bar{\delta}$. For notational convenience, we use $\alpha$ to denote the probability false alarm, $\beta$ to denote the probability of missed detection, and $\beta_\zeta$ denote the probability of missed detection given that the transmitted codeword has fractional weight $\zeta$. So, combining \eqref{eq:falarm_up_bd_simp} and \eqref{eq:mdetect_up_bd_simp}, since the fact that $\bar{\lambda} > \lambda$, we obtain
	{\small \begin{equation}
		\alpha + \beta_{\zeta,\thre} \leq \left\{ \begin{array}{ll}
			1 + \bar{\lambda}, &\mbox{if } \thre \leq \lw + \delta, \\
		\frac{\uw - \thre + \delta}{\uw - \lw} + 2\bar{\lambda}, &\mbox{if } \thre \in (\lw + \delta, \zeta \ast \lw - \bar{\delta}], \\
		\frac{\uw - \thre + \delta}{\uw - \lw} + \frac{\thre + \bar{\delta} - \zeta \ast \lw}{\zeta \ast \uw - \zeta \ast \lw} + 2\bar{\lambda}, &\mbox{if } \thre \in (\zeta \ast \lw - \bar{\delta}, \uw + \delta), \\
		\frac{\thre + \bar{\delta} - \zeta \ast \lw}{\zeta \ast \uw - \zeta \ast \lw} + 2\bar{\lambda}, &\mbox{if } 
\thre \in [\uw + \delta, \zeta \ast \uw - \bar{\delta}), \\
		1 + \bar{\lambda}, &\mbox{if } \thre \geq \zeta \ast \uw - \bar{\delta}.
		\end{array} \right.
	\label{eq:alpha_beta_rho_thre}
	\end{equation} }
	We choose the optimal threshold $\thre^* = \zeta \ast \lw - \bar{\delta}$ in different regions and simplify \eqref{eq:alpha_beta_rho_thre}, we obtain
	\begin{align}
		\lefteqn{\alpha + \beta_\zeta} \nonumber \\
		  	& \leq \frac{\uw - \zeta \ast \lw}{\uw - \lw} + 2\bar{\lambda} \nonumber \\
			&= 1 - \frac{\zeta(1 - 2\lw)}{\uw - \lw} + 2\bar{\lambda}.
	\label{eq:alpha_beta_rho}
	\end{align}

	Let $\gamma(\zeta)$ be the probability mass of codewords in the codebook $\code$ having fractional weight greater than $\zeta$. Then, we have
	\begin{align}
		\lefteqn{\alpha + \beta} \nonumber \\
			& \leq (1 - \gamma(\zeta)) + \gamma(\zeta) \left( 1 - \frac{\zeta(1 - 2\lw)}{\uw - \lw} \right) + 2\bar{\lambda} \label{eq:upbd_alpha+beta} \\
			&= 1 - \zeta \cdot \gamma(\zeta) \cdot \frac{1 - 2\lw}{\uw - \lw} + 2\bar{\lambda} \label{eq:upbd_deni}
	\end{align}
	
	Here, equation~\eqref{eq:upbd_alpha+beta} holds since for the $(1 - \gamma(\zeta))$ probability mass of codewords with fractional weight less than $\zeta$, $\alpha + \beta$ can be upper bounded by $1$. For codewords with fractional weight greater than $\zeta$, $\alpha + \beta$ can be upper bounded by $1 - \zeta \frac{1 - 2\lw}{\uw - \lw}$ from equation~\eqref{eq:alpha_beta_rho}.

	\item {\bf The case on ``large'' fractional weight --  [$\uw + \delta \leq \zeta \ast \lw - \bar{\delta}$]}

	On the other hand, if $\uw + \delta \leq \zeta \ast \lw - \bar{\delta}$, (see Figure~\ref{fig:converse2}) Willie could choose his threshold $t \in [\uw + \delta, \zeta \ast \lw - \bar{\delta}]$ such that $\alpha + \beta \leq 0$. Thus, Alice's transmission cannot be deniable from Willie.
	\begin{figure}
		\centering
		\includegraphics[width=0.8\columnwidth]{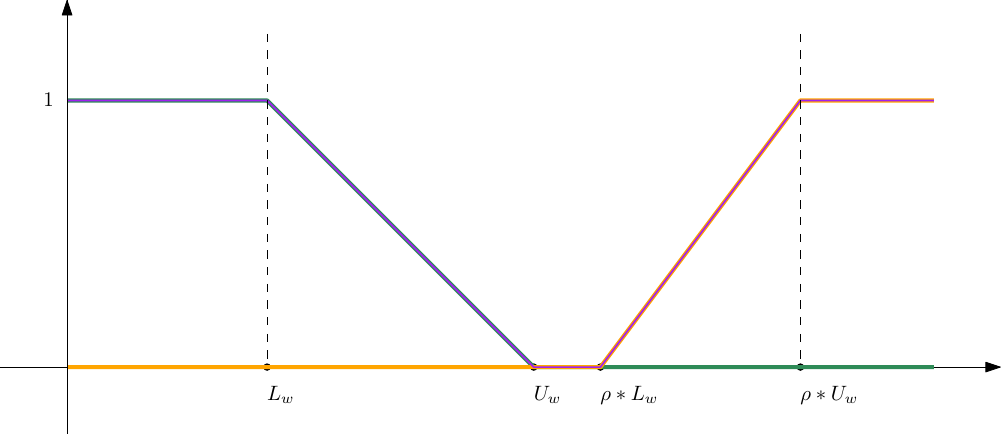}
		\caption[The case of ``large'' fractional weight -- $\rho \ast \lw - \bar{\delta} < \uw + \delta$]{{\bf The case of ``large'' fractional weight -- $\rho \ast \lw - \bar{\delta} < \uw + \delta$.} The green line shows the probability of false alarm, the orange line shows the probability of missed detection given a codeword with fractional weight $\rho$. The sum of false alarm probability and missed detection probability is shown in purple line.}
		\label{fig:converse2}
	\end{figure}

	\end{enumerate}

\end{enumerate}

\end{proof}


\subsubsection{Lower bound on the deniability parameter $\epsdeni$}

Given the observations in the prior section, we are interested in the case when the fractional weight of most of the codewords in the codebook $\code$ is reasonably small.

\begin{lemma}
Let $\epsreli^*(\epsdeni, \thr, n, \pb)$ be the smallest error probability among all codes with throughput $\thr$, deniability parameter $\epsdeni$, and block-length $n$ over a channel with transition probability $\pb$. Then, $\epsreli^*(\epsdeni, \thr, n, \pb)$ is an increasing function of $\pb$ for $\pb \leq 1/2$.
\end{lemma}

\begin{proof}
Consider $p_1 < p_2$. when $\pb = p_1$, decoder can simply flip each received bit i.i.d. with probability $q = \frac{p_2 - p_1}{1-2p_1}$. Thus, the effective channel is $BSC(p_2)$ since $p_2 = p_1 \ast q$.
\end{proof}

\begin{figure}
	\centering
	\includegraphics[width=0.8\columnwidth]{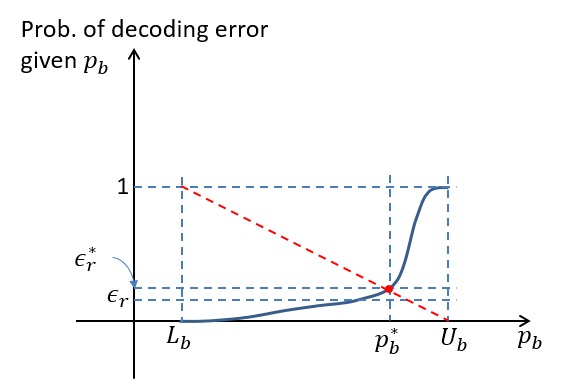}
	\caption[Decoding error analysis of the converse for slow fading]{Since the probability of error is an increasing function in $\pb$, there is a unique intersection (red dot) between this curve and the line segment $\pb = \ub - (\ub - \lb) \epsilon_{r | \pb}$.}
	\label{fig:err_slow_fading}
\end{figure}

From the above lemma, we first note that the probability of decoding error given the channel parameter $\pb$ is an increasing function in $\pb$. As shown in Figure~\ref{fig:err_slow_fading}, since the probability of decoding error given the channel parameter $\pb$ is an increasing function in $\pb$, there is a unique intersection between this function and the line segment $\pb = \ub - (\ub - \lb) \epsilon_{r | \pb}$, where $\epsilon_{r | \pb} \triangleq \Pr(\hat{\vM} \neq \vM | \Pb = \pb)$ denotes the probability of decoding error given the channel parameter $\pb$. Denote the intersection point as $(\pb^*, \epsreli^*)$, we have $\frac{\ub - \pb^*}{\ub - \lb} = \epsreli^*$.

Note that the probability of decoding error is
\begin{align}
	\epsreli &= \frac{1}{\ub - \lb} \int_{\lb}^{\ub} \Pr(\hat{\vM} \neq \vM | \Pb = \pb) \di \pb \nonumber \\
		& \geq \frac{1}{\ub - \lb} \epsreli^* (\ub - \pb^*) \nonumber \\
		&= (\epsreli^*)^2. \nonumber
\end{align}
\begin{figure}
	\centering
	\includegraphics[width=0.8\columnwidth]{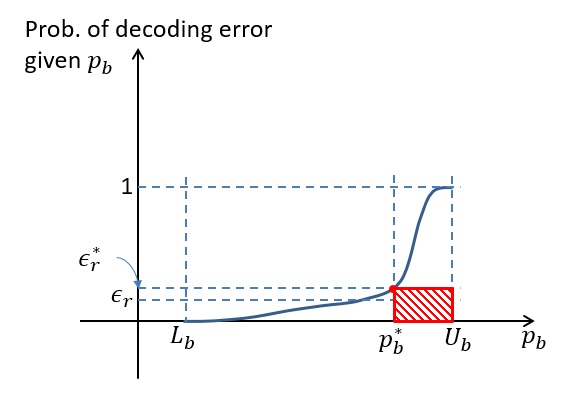}
	\caption[Lower bound of the decoding error for the slow fading converse]{In this figure, we see that the red shadowed area has a smaller area than the area between the error function $\epsilon_{r | \pb}$ and the $\pb$-axis.}
	\label{fig:err_slow_fading_2}
\end{figure}
Therefore, we have $\epsreli^* \leq \sqrt{\epsreli}$. As shown in Figure~\ref{fig:err_slow_fading_2}, the red shadowed rectangle represents $(\epsreli^*)^2$. 

\begin{figure}
	\centering
	\includegraphics[width=0.8\columnwidth]{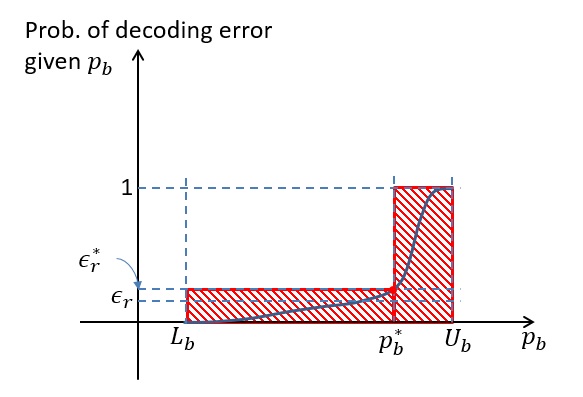}
	\caption[Upper bound of the decoding error for the slow fading converse]{In this figure, we see that the red shadowed area has a larger area than the area between the error function $\epsilon_{r | \pb}$ and the $\pb$-axis.}
	\label{fig:err_slow_fading_3}
\end{figure}
Similarly, we can obtain an upper bound of the decoding error $\epsreli$ as shown in Figure~\ref{fig:err_slow_fading_3},
\begin{align}
	\epsreli &= \frac{1}{\ub - \lb} \int_{\lb}^{\ub} \Pr(\hat{\vM} \neq \vM | \Pb = \pb) \di \pb \nonumber \\
		& \leq \frac{1}{\ub - \lb} \left[ \epsreli^* (\pb^* - \lb) + (\ub - \pb^*) \right] \nonumber \\
		&= \epsreli^* \frac{\pb^* - \lb}{\ub - \lb} + \frac{\ub - \pb^*}{\ub - \lb} \nonumber \\
		&= \epsreli^* (1 - \epsreli^*) + \epsreli^*
\end{align}
Therefore, we have $\epsreli^* \geq 1 - \sqrt{1 - \epsreli}$.

Let the input distribution of the codebook $\code$ is $\rho^*$, and recall that $\gamma:[0,1] \rightarrow [0,1]$ be a function of $\zeta$, $\gamma(\zeta)$ is the fraction of codewords with fractional weight at least $\zeta$. Then, we obtain the value of $\gamma(\rho^*)$ and simplify it as $\gamma^*$.

\begin{remark}
The fractional weight $\zeta$ of a codeword takes a value in $\left\{ 0, \frac{1}{n}, \ldots, 1 \right\}$.
\end{remark}

\begin{remark}
If $\gamma^*$ is a decreasing function in $n$, this means most codewords is the code $\code$ have small fractional weight, which means $\epsreli$ is large. Therefore, we focus on the case when $\gamma^*$ is a constant.
\end{remark}

Then, we have
\begin{align}
	\lefteqn{\epsreli} \nonumber \\
		&= \Pr_\code(\hat{\messrv} \neq \messrv) \nonumber \\
		&= \gamma^* \Pr_\code(\hat{\messrv} \neq \messrv | \wt{\vX} \geq \rho^* n) \nonumber \\
		& \eqspacing + (1 - \gamma^*) \Pr_\code(\hat{\messrv} \neq \messrv | \wt{\vX} < \rho^* n) \nonumber \\
		& \geq \gamma^* \cdot 0 + (1 - \gamma^*) \Pr_{\code^*}(\hat{\messrv} \neq \messrv) \label{eq:conv:separate_code_lb} \\
		&= (1 - \gamma^*) \Pr_{\code^*}(\hat{\messrv} \neq \messrv) \nonumber \\
		& \geq (1 - \gamma^*) \left[ \Pr_{\code^*}(\hat{\messrv} \neq \messrv | \Pb < \pb^*) \Pr(\Pb < \pb^*) \right. \nonumber \\
		& \eqspacing \left. + \Pr_{\code^*}(\hat{\messrv} \neq \messrv | \Pb \geq \pb^*) \Pr(\Pb \geq \pb^*) \right] \nonumber \\
		& \geq \epsreli^* (1 - \gamma^*) \Pr_{\code^*}(\hat{\messrv} \neq \messrv | \Pb \geq \pb^*) \nonumber \\
		& \geq (1 - \sqrt{1 - \epsreli}) ( 1 - \gamma^*) \Pr_{\code^*}(\hat{\messrv} \neq \messrv | \Pb = \pb^*) \label{eq:unknown:conv:reli}
\end{align}
In equation~\eqref{eq:conv:separate_code_lb}, we lower bounds of $\Pr_\code(\hat{\messrv} \neq \messrv | \wt{\vX} \geq \rho^* n)$ by $0$, and $\Pr_\code(\hat{\messrv} \neq \messrv | \wt{\vX} < \rho^* n)$ by $\Pr_{\code^*}(\hat{\messrv} \neq \messrv)$, where $\code^*$ is the sub-code of $\code$ containing all the codewords with fractional weight less than $\rho^*$ in $\code$.

Using \cite[Theorem 4]{kostina2015}, we have
\begin{align}
	\lefteqn{\Pr_{\code^*}(\hat{\messrv} \neq \messrv | \Pb = \pb^*)} \nonumber \\
	 	& \geq Q\left( \frac{c}{\sqrt{n}} (n\capacity(\code^*) - n\rate(\code^*) + c' \log n ) \right) \nonumber \\
		&= 1 - Q \left( c \sqrt{n} \left( \rate(\code^*) - \capacity(\code^*) - c' \frac{\log n}{n} \right) \right) \nonumber \\
		& \geq 1 - \exp\left( - \frac{1}{2} c^2 n \left( \rate(\code^*) - \capacity(\code^*) - c' \frac{\log n}{n} \right)^2 \right) \label{eq:unknown:conv:strong_conv}
\end{align}
Here, $c$ and $c'$ are both positive constants. $\rate(\code^*)$ is the rate for the sub-code $\code^*$, and $\rate(\code^*) = \rate + \frac{\log(1 - \gamma^*)}{n}$. $\capacity(\code^*)$ is the capacity of the sub-code $\code^*$, and $\capacity(\code^*) \leq I(\rho^*, \pb^*)$.

Therefore, using inequality~\eqref{eq:unknown:conv:strong_conv} into inequality~\eqref{eq:unknown:conv:reli}, we have
\begin{multline} \label{eq:unknown:conv:simp}
	\epsreli \geq (1 - \sqrt{1 - \epsreli})(1 - \gamma^*) \\
		\times \left[ 1 - \exp\left( - \frac{1}{2} c^2 n \left( \rate(\code^*) - \capacity(\code^*) - c' \frac{\log n}{n} \right)^2 \right) \right].
\end{multline}
Since $\sqrt{\epsreli} \geq 1 - \sqrt{1 - \epsreli}$, we simplify the above inequality~\eqref{eq:unknown:conv:simp} into the following,
\begin{equation}
	\sqrt{\epsreli} \geq (1 - \gamma^*)\left[ 1 - \exp\left( - \frac{1}{2} c^2 n \left( \rate(\code^*) - \capacity(\code^*) - c' \frac{\log n}{n} \right)^2 \right) \right].
\end{equation}
Then, we obtain
\begin{equation}
	\rate(\code^*) \leq C(\code^*) + c' \frac{\log n}{n} + \sqrt{\frac{2}{c^2 n} \log \left( \frac{1 - \gamma^*}{1 - \gamma^* - \sqrt{\epsreli}} \right)}.
\end{equation}

Replacing $\rate(\code^*)$ and $C(\code^*)$ by $\rate$ and $\mutual{\rho^*}{\pb^*}$ respectively, we have
\begin{align}
	\lefteqn{\rate} \nonumber \\
	&\leq \mutual{\rho^*}{\pb^*} + c' \frac{\log n}{n} + \sqrt{\frac{2}{c^2 n} \log \left( \frac{1 - \gamma^*}{1 - \gamma^* - \sqrt{\epsreli}} \right)} - \frac{\log ( 1 - \gamma^* )}{n} \nonumber \\
	&= \entropy{\rho^* \ast \pb^*} - \entropy{\pb^*} + c' \frac{\log n}{n} \nonumber \\
	& \eqspacing + \sqrt{\frac{2}{c^2 n} \log \left( \frac{1 - \gamma^*}{1 - \gamma^* - \sqrt{\epsreli}} \right)} - \frac{\log ( 1 - \gamma^* )}{n} \label{eq:unknown_reli_conv}
\end{align}

On the other hand, note that from equation~\eqref{eq:upbd_deni}, we have $\rho^* \leq \epsdeni \frac{\uw - \lw}{\gamma^*(1 - 2\lw)}$. So, replacing $\rho^*$ in equation~\eqref{eq:unknown_reli_conv}. And as $n$ large enough, we have
\begin{align}
	\rate &\leq \entropy{\rho^* \ast \pb^*} - \entropy{\pb^*} \nonumber \\
		& \leq \entropy{\left[\epsdeni \frac{\uw - \lw}{\gamma^*(1 - 2\lw)}\right] \ast \pw} - \entropy{\pw}. \nonumber
\end{align}
Therefore,
\begin{align}
	\rate &\leq \min_{\gamma^*} \entropy{\left[\epsdeni \frac{\uw - \lw}{\gamma^*(1 - 2\lw)}\right] \ast \pw} - \entropy{\pw} \nonumber \\
		&= \entropy{\left[\epsdeni \frac{\uw - \lw}{(1 - 2\lw)}\right] \ast \pw} - \entropy{\pw}. \nonumber
\end{align}




\subsection{Achievability of Slow Fading Channel Model (Theorem~\ref{thm:model:unknown:achi})}
\label{sec:slow_fading:achi}

\subsubsection{Achievability: Deniability}
In this part, we would like to show that the variational distance between $\Pz$ and $\Po$ is small where $\Pz$ and $\Po$ are the distributions of Willie's vector when Alice's transmission status are $0$ and $1$ respectively. That is, we would like to show $\variation{\Pz}{\Po} < \epsdeni$. We also note that $\Po$ depends on a particular codebook, it is hard to obtain in general. By the triangle inequality, we have $\variation{\Pz}{\Po} \leq \variation{\Pz}{\Exp(\Po)} + \variation{\Exp(\Po)}{\Po}$. So, similar to the Fixed Channel Model, we show $\variation{\Pz}{\Po} < \epsdeni$ holds with probability greater than $1 - \exp\left(-\Omega(n^\delta)\right)$ over the codebook $\code$. To show this, we first show $\variation{\Pz}{\Exp(\Po)} < \epsdeni$, we then show $\variation{\Exp(\Po)}{\Po} < 2^{-\Omega(n^\delta)}$ with probability greater than $1 - \exp\left(-\Omega(n^\delta)\right)$ over the codebook $\code$.

\begin{lemma}
\label{lemma:P0_vs_EP1}
Let $\rho$ to be the codebook generation parameter. If $\rho < \frac{\uw - \lw}{(1 - 2\lw)} \epsdeni$, then the variational distance $\variation{\Pz}{\Exp(\Po)} < \epsdeni$.
\end{lemma}

\begin{proof}
Let Willie receives $\vy_w$, and let the fractional weight of $\vy_w$ to be $\fyw = \frac{\wt{\vy_w}}{n}$. When Alice's transmission status $\tranrv = 0$, if $\fyw \in (\lw, \uw)$, then
\begin{align}
	\pz(\vy_w) &= \frac{1}{\uw - \lw} \int_{\lw}^{\uw} \pw^{\wt{\vy_w}}(1-\pw)^{n - \wt{\vy_w}} \di \pw \nonumber \\
		&= \frac{1}{\uw - \lw} \int_{\lw}^{\uw} \pw^{n \fyw} (1 - \pw)^{n (1 - \fyw)} \di \pw \nonumber \\
		&= \frac{1}{\uw - \lw} \int_{\lw}^{\uw} 2^{-n \left( \entropy{\fyw} + \KL{\fyw}{\pw} \right)} \di \pw \nonumber \\
		&= \frac{2^{-n \entropy{\fyw}}}{\uw - \lw} \int_{\lw}^{\uw} 2^{-n \KL{\fyw}{\pw}} \di \pw \nonumber \\
		&= \frac{2^{-n \entropy{\fyw}}}{\uw - \lw} \sqrt{\frac{2 \pi \fyw (1 - \fyw)}{n}}\left(1 + \Oh{\frac{1}{n}} \right). \label{eq:laplace_p0}
\end{align}
Equation \eqref{eq:laplace_p0} is obtained by the direct application of Laplace's method (see Appendix I Theorem~\ref{thm:laplace}).

Let
\begin{equation}
	\qz(\vy_w) = \left\{ \begin{array}{ll}
	\frac{1}{\lfloor n\uw \rfloor - \lfloor n\lw \rfloor} \frac{1}{{n \choose n \fyw}}, & \quad \mbox{if } \fyw \in (\lw, \uw) \\
	0, & \quad \mbox{otherwise.}
\end{array} \right. \label{eq:def_q0}
\end{equation}

\begin{remark}
Note that for this distribution $\Qz$, we have $\qz(\wt{\vy_w}) = \frac{1}{\lfloor n\uw \rfloor - \lfloor n\lw \rfloor}$. This means, the weight distribution on $\qz$ is uniform for $\fyw \in (\lw, \uw)$. Also note that,
$$ \left| \frac{1}{\lfloor n\uw \rfloor - \lfloor n\lw \rfloor} - \frac{1}{n(\uw - \lw)} \right| = \Oh{\frac{1}{n^2}}. $$
Therefore, for ease of computation, we neglect the floor functions $\lfloor \cdot \rfloor$ in the later computations.
\end{remark}

Note that $\sum_{\vy_w} \qz(\vy_w) = 1$, and therefore $\qz(\vy_w)$ is a probability mass function. We would like to show the variational distance between $\Pz$ and $\Qz$ is small. That is,
$$ \variation{\Pz}{\Qz} = \frac{1}{2} \sum_{\vy_w \in \{ 0,1 \}^n} \left|\pz(\vy_w) - \qz(\vy_w) \right| $$

Also note that when $\fyw \in (\lw, \uw)$, $p_0$ is closed to $\qz$. More precisely, by Theorem~\ref{thm:var_dist_small} in Appendix, we have
\begin{equation}
	\frac{1}{2} \sum_{\vy_w: \fyw \in (\lw, \uw)} \left|\pz(\vy_w) - \qz(\vy_w) \right| < \epn, \label{eq:epn}
\end{equation}
where $\epn$ is decreasing when $n$ is increasing. Then,
\begin{align}
	\lefteqn{\variation{\Pz}{\Qz}} \nonumber \\
	&= \frac{1}{2} \sum_{\vy_w: \fyw \in (\lw, \uw)} \left|\pz(\vy_w) - \qz(\vy_w) \right| \nonumber \\
	& \eqspacing + \frac{1}{2} \sum_{\vy_w: \fyw \notin (\lw, \uw)} \left|\pz(\vy_w) - \qz(\vy_w) \right| \nonumber \\
	&< \epn + \frac{1}{2} \sum_{\vy_w: \fyw \notin (\lw, \uw)} \left|\pz(\vy_w) - \qz(\vy_w) \right| \label{eq:epn_ref} \\
	& \leq \epn + \frac{1}{2} \sum_{\vy_w: \fyw \notin (\lw, \uw)} \left( \pz(\vy_w) + \qz(\vy_w) \right) \label{eq:trieq} \\
	&= \epn + \frac{1}{2} \sum_{\vy_w: \fyw \notin (\lw, \uw)} \pz(\vy_w) + \frac{1}{2} \sum_{\vy_w: \fyw \notin (\lw, \uw)} \qz(\vy_w) \nonumber \\
	&= \epn + \frac{1}{2} \left(1 - \sum_{\vy_w: \fyw \in (\lw, \uw)}\pz(\vy_w)\right) \nonumber \\
	& \eqspacing \eqspacing + \frac{1}{2} \left(1 - \sum_{\vy_w: \fyw \in (\lw, \uw)}\qz(\vy_w)\right) \nonumber \\
	&< \epn + \left(1 - \sum_{\vy_w: \fyw \in (\lw, \uw)}\qz(\vy_w)\right) + \frac{\epn}{2} \label{eq:abs_epn} \\
	&< \frac{3}{2} \epn. \label{eq:var_p0_q0}
\end{align}
Here, \eqref{eq:epn_ref} holds from \eqref{eq:epn}. \eqref{eq:trieq} is true since the triangle inequality $|a-b| \leq |a| + |b|$. \eqref{eq:abs_epn} holds from \eqref{eq:epn} and
\begin{multline}
\frac{1}{2} \sum_{\vy_w: \fyw \in [\lw + \deln, \uw - \deln]} \left( \qz(\vy_w) - \pz(\vy_w) \right) \\
	\leq \frac{1}{2} \sum_{\vy_w: \fyw \in [\lw + \deln, \uw - \deln]} \left|\pz(\vy_w) - \qz(\vy_w) \right| \leq \epn.
\end{multline}

Similar to the case $\pz$, $\Exp_\code \left( \po(\vy_w) \right)$ can be calculated as follows,
\begin{align}
	\lefteqn{\Exp_\code \left( \po(\vy_w) \right)} \nonumber \\
		&= \frac{1}{\uw - \lw} \int_{\lw}^{\uw} (\rho \ast\pw)^{\wt{\vy_w}}(1-\rho \ast \pw)^{n - \wt{\vy_w}} \di \pw \nonumber \\
		&= \frac{1}{\uw - \lw} \int_{\lw}^{\uw} (\rho \ast\pw)^{n \fyw}(1-\rho \ast \pw)^{n (1 - \fyw)} \di \pw \nonumber \\
		&= \frac{1}{\uw - \lw} \int_{\lw}^{\uw} 2^{-n \left( \entropy{\fyw} + \KL{\fyw}{\rho \ast \pw} \right)} \di \pw \nonumber \\
		&= \frac{2^{-n \entropy{\fyw}}}{\uw - \lw} \int_{\lw}^{\uw} 2^{-n \KL{\fyw}{\rho \ast \pw}} \di \pw \nonumber \\
		&= \frac{2^{-n \entropy{\fyw}}}{(\uw - \lw)(1 - 2 \rho)} \sqrt{\frac{2 \pi \fyw (1 - \fyw)}{n}}\left(1 + \Oh{\frac{1}{n}} \right), \label{eq:laplace_Ep1}
\end{align}
if $\fyw \in (\rho \ast \lw, \rho \ast \uw)$. Then, we define a probability mass function $\qo$,
\begin{equation}
	\qo(\vy_w) = \left\{ \begin{array}{ll}
	\frac{1}{\lfloor n (\rho \ast \uw) \rfloor - \lfloor n (\rho \ast \lw) \rfloor} \frac{1}{{n \choose n \fyw}}, & \quad \mbox{if } \fyw \in (\rho \ast \lw, \rho \ast \uw) \\
	0, & \quad \mbox{otherwise.}
\end{array} \right. \label{eq:def_q1}
\end{equation}

Similar to equation~\eqref{eq:epn}, for $\fyw \in (\rho \ast \lw, \rho \ast \uw)$,
\begin{equation}
	\frac{1}{2} \sum_{\vy_w: \fyw \in (\rho \ast \lw, \rho \ast \uw)} \left|\Exp_\code (\po(\vy_w)) - \qo(\vy_w) \right| < \epn', \label{eq:epn_p}
\end{equation}

Therefore, similar to \eqref{eq:var_p0_q0}, the variational distance between $\Exp_\code (\Po)$ and $\Qo$ can be upper bounded. That is,
\begin{equation}
	\variation{\Exp_\code (\Po)}{\Qo} < \frac{3}{2} \epn'. \label{eq:var_Ep1_q1}
\end{equation}

Note that,
\begin{equation}
	\variation{\Pz}{\Exp_\code (\Po)} \leq \variation{\Pz}{\Qz} + \variation{\Exp_\code (\Po)}{\Qo} + \variation{\Qz}{\Qo}. \label{eq:trieq_variation}
\end{equation}
From \eqref{eq:var_p0_q0} and \eqref{eq:var_Ep1_q1}, we know that the first two terms of RHS of \eqref{eq:trieq_variation} tend to $0$ as $n$ increasing. So, the variational distance $\variation{\Pz}{\Exp_\code (\Po)}$ is upper bounded by $\variation{\Qz}{\Qo}$.

Then, we compute $\variation{\Qz}{\Qo}$ as follows.
\begin{align}
	\lefteqn{\variation{\Qz}{\Qo}} \nonumber \\
		&= \frac{1}{2} \sum_{\vy_w \in \{0,1\}^n} \left| \qz(\vy_w) - \qo(\vy_w) \right| \nonumber \\
		&= \frac{1}{2} \sum_{\vy_w: \fyw \in [\lw, \rho \ast \lw)} \qz(\vy_w) + \frac{1}{2} \sum_{\vy_w: \fyw \in [\rho \ast \lw, \uw]} \left| \qz(\vy_w) - \qo(\vy_w) \right| \nonumber \\
		& \eqspacing + \frac{1}{2} \sum_{\vy_w: \fyw \in (\uw, \rho \ast \uw]} \qo(\vy_w) \label{eq:q0_q1_diff} \\
		&= \frac{1}{2} \frac{\rho \ast \lw - \lw}{\uw - \lw} + \frac{1}{2} \left( \frac{1}{1 - 2 \rho} - 1 \right) \frac{\uw - \rho \ast \lw}{\uw - \lw} \nonumber \\
		& \eqspacing + \frac{1}{2} \frac{\rho \ast \uw - \uw}{\rho \ast \uw - \rho \ast \lw} \label{eq:q0_q1_diff_value} \\
		&= \frac{1}{2}\frac{(1 - 2\rho)(\rho \ast \lw - \lw) + 2\rho(\uw - \rho \ast \lw) + (\rho \ast \uw - \uw)}{(1-2\rho)(\uw - \lw)} \nonumber \\
		&= \frac{\rho - 2\rho(\rho \ast \lw)}{(1-2\rho)(\uw - \lw)} \nonumber \\
		&= \frac{\rho(1 - 2\lw)(1 - 2\rho)}{(1 - 2\rho)(\uw -\lw)} \nonumber \\
		&= \frac{\rho(1 - 2\lw)}{\uw - \lw} \nonumber
\end{align}
Here, \eqref{eq:q0_q1_diff} is true since $\qz$ takes non-zero values when $\fyw \in (\lw, \uw)$ and $\qo$ takes non-zero values when $\fyw \in (\rho \ast \lw, \rho \ast \uw)$. The probability of $\wt{\vy_w} = n \fyw$ for $\qz$, equals ${n \choose n \fyw} \frac{1}{\lfloor n\uw \rfloor - \lfloor n\lw \rfloor} \frac{1}{{n \choose n \fyw}} = \frac{1}{\lfloor n\uw \rfloor - \lfloor n\lw \rfloor}$. As we specified, we neglect the floor functions for simplicity, so the probability $\qz$ of $\wt{\vy_w} = n \fyw$ equals $\frac{1}{n(\uw - \lw)}$. Similarly, the probability $\qo$ of $\wt{\vy_w} = n \fyw$ equals $\frac{1}{n(\rho \ast \uw - \rho \ast \lw)} = \frac{1}{n(1-2\rho)(\uw - \lw)}$. Therefore, in \eqref{eq:q0_q1_diff_value}, $\sum_{\vy_w: \fyw \in (\rho \ast \lw, \uw)} |\qz(\vy_w) - \qo(\vy_w)| = n(\uw - \rho \ast \lw)\left(\frac{1}{1-2\rho} - 1\right)\frac{1}{n(\uw - \lw)} = \frac{2\rho}{1 - 2\rho}\frac{\uw - \rho \ast \lw}{\uw - \lw}$. Therefore, given $\epsdeni$-deniability, we can take $\variation{\Pz(\vy_w)}{\Exp_\code (\Po(\vy_w))} \leq \frac{\rho(1 - 2\lw)}{\uw - \lw} < \epsdeni$, where $\rho < \frac{\uw - \lw}{(1 - 2\lw)} \epsdeni$. 
\end{proof}

\begin{remark}
If there is shared secret between Alice and Bob, (Bob knows the codebook that Alice uses while Willie only knows the codebook generation parameter), then $\variation{\Pz}{\Po} = \variation{\Pz}{\Exp(\Po)}$. Therefore, under this assumption, Lemma~\ref{lemma:P0_vs_EP1} completes the deniability part of our achievability. Also, Willie's channel does not need to be noisier than Bob's channel under this assumption.
\end{remark}

\begin{lemma}
\label{lemma:EP1_vs_P1}
If the codebook $\code$ is drawn from the $\rho n$-weight random ensemble with $\rho < \frac{\uw - \lw}{(1 - 2\lw)} \epsdeni$, the variational distance $\variation{\Exp(\Po)}{\Po} < 2^{-\Omega(n^\delta)}$ with probability greater than $1 - \exp\left(- 2^{\Oh{n^\delta}}\right)$ over the code $\code$.
\end{lemma}





\begin{proof}
It remains us to show that the variational distance between $\Exp_\code(\Po)$ and $\Po$ is small. Suppose a particular codebook $\code$ is used. We first define ``typicality'' for $\vy_w$. (In this setting, we define a {\it high probability set} for $\vy_w$ instead of typical set since the channel noise $\pw$ is not known). Recall that $\fyw = \frac{\wt{\vy_w}}{n}$ is the fractional weight of $\vy_w$. Since Willie's channel noise takes value from $\lw$ to $\uw$, the fractional weight of Willie's received vector is in the range around $\rho \ast \lw$ to $\rho \ast \uw$.
\begin{equation}
	\hpsyw \triangleq \left\{ \vy_w: \fyw \in (\rho \ast \lw (1 - \delyw), \rho \ast \uw (1 + \delyw)) \right\}
\label{def:typAyw}
\end{equation}

Also, we denote $\fzzw$, $\fzow$, $\fozw$ and $\foow$ be the fractions of $(0,0)$, $(0,1)$, $(1,0)$ and $(1,1)$ in $(\vx, \vy_w)$ respectively. Then, we define a {\it conditionally high probability set} of $\vx$ given a particular $\vy_w$. That is,
\begin{small}
\begin{equation}
	\hpsxgyw \triangleq \left \{ \vx: 
	\begin{array}{ll} 
		\fozw & \in \left( \rho \lw (1 - \delozw), \right. \\
		& \eqspacing \left. \rho \uw (1 + \delozw) \right), \\ 
		\foow & \in \left( \rho (1 - \uw) (1 - \deloow), \right. \\
		& \eqspacing \left. \rho (1 - \lw) (1 +  \deloow) \right), \\
		\fx = \fozw + \foow & \in ( \rho (1 - \delx), \rho(1 + \delx) )
\end{array}	   
            \right \}.
\label{def:typAxgyw}
\end{equation}
\end{small}

Note that given the tuple $(\fyw, \fozw, \foow)$, the values of $\fzzw$ and $\fzow$ can be computed as $1 - \fyw - \fozw$ and $\fyw - \foow$ respectively. Hence, the joint type of $(\vx, \vy_w)$ can be determined by this tuple $(\fyw, \fozw, \foow)$. Further, we define a conditionally type of $\vx$ given $\vy_w$,
\begin{small}
\begin{equation}
	\typeBxgyw \triangleq \left \{ \vx: 
		\begin{array}{ll} 
			\mbox{the fraction of $(1,0)$ in $(\vx, \vy_w)$ equals $\fozw$} ,\\ 
            \mbox{the fraction of $(1,1)$ in $(\vx, \vy_w)$ equals $\foow$}
		\end{array}	   
    \right \}.
\label{def:typeBxgyw}
\end{equation}
\end{small}

Then, $\hpsxgyw$ can be written as the union of the $\typeBxgyw$'s. Mathematically,
\begin{small}
\[ \hpsxgyw = \bigcup_{\substack{\fozw, \foow: \\
	\fozw \in ( \rho \lw (1 - \delozw),  \rho \uw (1 + \delozw) ), \\
	\foow \in ( \rho (1 - \uw) (1 - \deloow),  \rho (1 - \lw) (1 +  \deloow) ), \\
	\fozw + \foow \in ( \rho (1 - \delx), \rho (1 + \delx) ), \\
	n \fozw, n \foow \in \mathbb{Z} }} \typeBxgyw \]
\end{small}

Note that,
\begin{small}
\begin{align}
	\lefteqn{\variation{\Exp_\code(\po(\vy_w))}{\po(\vy_w)}} \nonumber \\
	 	&= \frac{1}{2} \sum_{\vy_w \in \{0,1\}^n} \left|\Exp_\code(\po(\vy_w)) - \po(\vy_w)\right| \nonumber \\
		&= \frac{1}{2} \sum_{\vy_w \in \hpsyw} \left| \Exp_\code(\po(\vy_w)) - \po(\vy_w) \right| + \frac{1}{2} \sum_{\vy_w \notin \hpsyw} \left| \Exp_\code(\po(\vy_w)) - \po(\vy_w) \right| \label{eq:variation_break_y} \\
		&= \frac{1}{2} \sum_{\vy_w \in \hpsyw} \left| \sum_\code \Pr(\code) \sum_{\vx \in \code \cap \hpsxgyw} \po(\vy_w|\vx)p(\vx) - \sum_{\vx \in \code_0 \cap \hpsxgyw} \po(\vy_w|\vx)p(\vx) \right| \nonumber \\
		& \eqspacing + \frac{1}{2} \sum_{\vy_w \in \hpsyw} \left| \sum_\code \Pr(\code) \sum_{\vx \in \code \setminus \hpsxgyw} \po(\vy_w|\vx)p(\vx) - \sum_{\vx \in \code_0 \setminus \hpsxgyw} \po(\vy_w|\vx)p(\vx) \right| \nonumber \\
		& \eqspacing \eqspacing + \frac{1}{2} \sum_{\vy_w \notin \hpsyw} \left| \Exp_\code(\po(\vy_w)) - \po(\vy_w) \right| \label{eq:variation_break_x}
\end{align}
\end{small}

Here, in \eqref{eq:variation_break_y}, we break the variational distance between $\Po$ and $\Exp_\code(\Po)$ into two summations, one corresponds to the high probability set $\vy_w \in \hpsyw$, the other one corresponds to the small probability set $\vy_w \notin \hpsyw$. We further break the summation of the high probability set $\vy_w$ in \eqref{eq:variation_break_x} into the conditional high probability set $\vx \in \hpsxgyw$, and $\vx \notin \hpsxgyw$. We first focus on the $\vy_w$ and $\vx$ which are likely to see. For the ``atypical'' part, that is, the $\vy_w$ and $\vx$ which are unlikely to see, it can be bounded by tail inequalities. Using the similar techniques by changing $\typyw$ and $\typxgyw$ by $\hpsyw$ and $\hpsxgyw$ respectively in Model~\ref{model:known}, $\variation{\Exp_\code(\Po)}{\Po} < \epsdeni$. 
\end{proof}

\subsubsection{Achievability: Reliability}
Consider the legitimate receiver Bob uses the following decoding procedure:
\begin{enumerate}
	\item Bob uses the threshold estimator to determine Alice's transmission status $\tranrv$;
	\item If Alice's transmission status $\tranrv = 0$, Bob decodes $\hat{\vx} = \vec{0}$;
	\item If Alice's transmission status $\tranrv = 1$, Bob uses the decoding rule in \cite{}.
\end{enumerate}

Consider Bob's probability of decoding error,
\begin{align}
	\lefteqn{\Pr(\errb | \tranrv = 0) + \Pr(\errb | \tranrv = 1)} \nonumber \\
		&= \Pr(\hat{\bX} \neq \vec{0} | \tranrv = 0) \nonumber \\
		& \eqspacing+ \sum_{\vx \in \code} \left[\Pr(\hat{\bX} \neq \vx | \tranrv = 1, \vX = \vx) \Pr(\vX = \vx) \right. \nonumber \\
		& \eqspacing \eqspacing + \left. \Pr(\hat{\bX} = \vec{0} | \tranrv = 1, \vX = \vx) \Pr(\vX = \vx) \right] \nonumber \\
		&= \left[ \Pr(\hat{\bX} \neq \vec{0} | \tranrv = 0) \right. \nonumber \\
		& \eqspacing + \left. \Pr(\hat{\bX} = \vec{0} | \tranrv = 1) \right] + \Pr(\hat{\bX} \neq \vX | \tranrv = 1). \nonumber
\end{align}
Note that
\begin{equation}
	\Pr(\hat{\bX} \neq \vec{0} | \tranrv = 0) + \Pr(\hat{\bX} = \vec{0} | \tranrv = 1)
\end{equation}
is the same as the deniability for Willie. So, from the converse proof, we have
\begin{equation}
	\Pr(\hat{\bX} \neq \vec{0} | \tranrv = 0) + \Pr(\hat{\bX} = \vec{0} | \tranrv = 1) \leq 1 - \gamma(\zeta) \zeta \frac{1 - 2\lb}{\ub - \lb},
\end{equation}
where $\gamma$ is the fraction of codewords in the codebook $\code$ having fractional weight $\zeta$.

Since the codebook ensemble
\begin{equation}
	\rho = \frac{\uw - \lw}{1 - 2\lw} \epsdeni, \nonumber
\end{equation}
by Chernoff's bound, we have
\begin{equation}
	\Pr(\zeta < (1 - \xi) \rho) < \exp\left( -\frac{1}{2} \xi^2 \rho n \right), \nonumber
\end{equation}
which is equivalent to say,
\begin{equation}
	\Pr(\zeta > (1 - \xi) \rho) > 1 - \exp\left( -\frac{1}{2} \xi^2 \rho n \right). \nonumber
\end{equation}

On the other hand, we have
\begin{align}
	& & \Pr(\hat{\bX} \neq \vec{0} | \tranrv = 0) + \Pr(\hat{\bX} = \vec{0} | \tranrv = 1) &< \epsreli \nonumber \\
	&\Leftarrow & 1 - \gamma \zeta \frac{1 - 2\lb}{\ub - \lb} &< \epsreli \nonumber \\
	&\Leftrightarrow & \gamma \zeta \frac{1 - 2\lb}{\ub - \lb} &> 1 - \epsreli \nonumber \\
	&\Leftrightarrow & \zeta &> \frac{1 - \epsreli}{\gamma} \frac{\ub - \lb}{1 - 2\lb}. \label{eq:unknown:equiv}
\end{align}

Also note that
\begin{equation}
	\zeta \geq (1 - \xi)\rho = \frac{\uw - \lw}{1 - 2\lw} (1 - \xi)\epsdeni \nonumber
\end{equation}
with high probability. So, equation~\eqref{eq:unknown:equiv} can be obtained from
\begin{align}
	& & \frac{\uw - \lw}{1 - 2\lw} (1 - \xi) \epsdeni &> \frac{1 - \epsreli}{\gamma} \frac{\ub - \lb}{1 - 2\lb} \nonumber \\
	&\Leftrightarrow & \frac{\uw - \lw}{\ub - \lb} \frac{1 - 2\lb}{1 - 2\lw} &> \frac{1 - \epsreli}{(1 - \xi) \gamma \epsdeni}. \label{eq:unknown_cond}
\end{align}
Therefore, if
\begin{equation}
	 \frac{\uw - \lw}{\ub - \lb} \frac{1 - 2\lb}{1 - 2\lw} > \frac{1 - \epsreli}{(1 - \xi) \epsdeni} \nonumber
\end{equation}
by taking $gamma = 1$ in equation~\eqref{eq:unknown_cond}, we have
\begin{equation}
	\Pr(\hat{\bX} \neq \vec{0} | \tranrv = 0) + \Pr(\hat{\bX} = \vec{0} | \tranrv = 1) < \epsreli \nonumber
\end{equation}

On the other hand, we can obtain
\begin{equation}
	\Pr(\hat{\bX}\neq \vX | \tranrv = 1) < \epsreli
\end{equation}
with rate $\rate = \entropy{\left( \frac{\uw - \lw}{1 - 2\lw} \epsdeni \right) \ast \ub} - \entropy{\ub}$ from the result in \cite[Chapter 7, Problem 17 and 18]{yeung2008book} and \cite{Ahlswede70}.







\appendix
%

\section{Useful Tools}


\begin{claim}[Reverse Pinsker's inequality~\cite{BerK:12}]
\label{clm:pinsker} For any $p \in (0,1)$ and all sufficiently small $x>0$, the Kullback-Leibler divergence between two binary random variables can be bounded as
\begin{enumerate}
\item {\bf Additive:} $\KL{p}{p+x} \leq \frac{x^2}{2 p(1-p)\ln 2 }$ 
\item {\bf Convolutive:} $\KL{p}{p*x} \leq \frac{x^2 (1-2p)^2}{ 2p (1-p)\ln 2 }$ 
\end{enumerate}
\end{claim}

\begin{proof}
We prove the additive part (which provides a result that matches the corresponding result in~\cite{BerK:12}specialized to the case of binary random variables), and use it to prove the convolutive part by substituting $p*x=p+x(1-2p)$.
\begin{eqnarray}
	\lefteqn{\KL{p}{p+x}} \nonumber \\
		&=& p \log \left ( \frac{p}{p+x} \right ) + (1-p) \log \left ( \frac{1-p}{1-p-x} \right ) \nonumber \\
		&=& -p \log \left( 1+\frac{x}{p} \right) - (1-p) \log \left( 1 - \frac{x}{1-p} \right) \nonumber \\
		&=& -\frac{p}{\ln 2} \left( \frac{x}{p} - \frac{x^2}{2p^2} + \frac{x^3}{3p^3} \right) 
		 -\frac{1-p}{\ln 2} \left( -\frac{x}{1-p} - \frac{x^2}{2(1-p)^2} - \frac{x^3}{3(1-p)^3} \right) + \Oh(x^4) \label{eq:taylor} \\
		&=& \frac{x^2}{2 \ln 2} \left(\frac{1}{p} + \frac{1}{1-p}\right) - \frac{x^3}{3 \ln 2}\left(\frac{1}{p^2} - \frac{1}{(1-p)^2}\right)+ \Oh(x^4) \nonumber \\
		& \leq & \frac{x^2}{2 \ln 2} \left(\frac{1}{p} + \frac{1}{1-p}\right) = \frac{x^2}{2 p(1-p)\ln 2 }. \nonumber
\end{eqnarray}
Here (\ref{eq:taylor}) follows from the Taylor series expansion of the binary logarithm.
\end{proof}

\begin{claim} \label{clm:ent_diff}
The difference between two binary entropy functions equals
\begin{enumerate}
	\item {\bf Additive:} $\entropy{p+x} - \entropy{p} = \KL{p}{p+x} + x \log \left ( \frac{1-p-x}{p+x} \right )$
	\item {\bf Convolutive:} $\entropy{p \ast x} - \entropy{p} = \KL{p}{p \ast x} + x(1-2p) \log \left ( \frac{1 - p \ast x}{p \ast x} \right ) $
\end{enumerate}
\end{claim}

\begin{proof} The proof directly follows from algebraic manipulations of the definitions of the quantities involved. We prove the additive part, and use it to prove the convolutive part by substituting $p*x=p+x(1-2p)$.
\begin{eqnarray}
	\lefteqn{\entropy{p+x} - \entropy{p}} \nonumber \\
	&=& - (p+x) \log (p+x) - (1-p-x) \log (1-p-x) \nonumber 
	 + p \log p + (1-p) \log (1-p) \nonumber \\
	&=& p \log \left (\frac{p}{p+x} \right ) + (1-p) \log \left (\frac{1-p}{1-p-x} \right ) \nonumber 
	 + x \log \left ( \frac{1-p-x}{p+x} \right ) \nonumber \\
	&=& \KL{p}{p+x} + x \log \left (\frac{1-p-x}{p+x} \right ) \nonumber
\end{eqnarray}
\end{proof}

\begin{claim}
\label{clm:emp_mut_info}
Consider a binary symmetric channel with cross-over probability $p < 1/2$, suppose that a codeword $\vx$ with fractional Hamming weight $\rho \triangleq \crho/\sqrt{n}$ is sent, and $\vy$ is received. Assume that 
\[ \vy \in \mathcal{A}(\vY) \triangleq \left\{ \vy: \py \in (\rho \ast p - \Delta_{*1}, \rho \ast p + \Delta_{*1}) \right\} \]
and
\[ \vx \in \mathcal{A}(\vX|\vy) \triangleq \left \{ \vx: 
 \begin{array}{ll} 
  \poz & \in ( \rho (p - \Delta_{10}),  \rho (p + \Delta_{10})),\\ 
            \poo & \in ( \rho (1 - p - \Delta_{11}),  \rho (1 - p + \Delta_{11}))
\end{array}	   
            \right \}, \]
where $\Delta_{*1} = \Oh{n^{-1/2}}$, $\Delta_{10} = \Oh{n^{-1/4}}$ and $\Delta_{11} = \Oh{n^{-1/4}}$. Then, for sufficiently large $n$, the empirical mutual information between $\vx$ and $\vy$
\begin{equation}
	\mutual{\vx}{\vy} = \frac{\crho (1 - 2p)}{n^{1/2}} \log \left(\frac{1 - p}{p}\right) + \Oh{{n}^{-3/4}}.
\end{equation}
\end{claim}

\begin{proof}
Recall that we use $\py$ to denote $\poz + \poo$ and $\px$ to denote $\poz + \poo$. By the definition of the empirical mutual information for binary random variables,
\begin{equation}
\mutual{\vx}{\vy} = \pzz \log \left ( \frac{\pzz}{(1 - \px)(1 - \py)} \right ) + \pzo \log \left ( \frac{\pzo}{(1 - \px)\py} \right ) + \poz \log \left ( \frac{\poz}{\px(1 - \py)} \right ) + \poo \log \left ( \frac{\poo}{\px\py} \right ). \label{eq:emp_mut_info}
\end{equation}
Recall that given a tuple $(\py, \poz, \poo)$, the values of $\pzo$ and $\pzz$ can be computed (as $\py - \poo$ and $1-\py-\poz$ respectively). 
Hence, for fixed $(\poz, \poo)$, one may treat $\mutual{\vx}{\vy}$ in (\ref{eq:emp_mut_info}) as a function only of $\py$. It can then be shown by a sequence of algebraic manipulations (in which many terms cancel out) that
\begin{equation}
	\frac{\partial \mutual{\vx}{\vy}}{\partial \py} = \log \left ( \frac{(\py - \poo)(1 - \py)}{\py(1 - \py - \poz)} \right ).
\label{eq:didfy}
\end{equation}
Similarly, the partial derivative of $\mutual{\vx}{\vy_b}$ in $\poz$ and $\poo$ are
\begin{equation}
	\frac{\partial \mutual{\vx}{\vy}}{\partial \poz} = \log \left( \frac{\poz (1 - \poz - \poo)}{(\poz + \poo)(1 - \py - \poz)} \right)
\end{equation}
and
\begin{equation}
	\frac{\partial \mutual{\vx}{\vy}}{\partial \poo} = \log \left( \frac{\poo(1 - \poz - \poo)}{(\poz + \poo)(\py - \poo)} \right)
\end{equation}
respectively.

We are interested in the value of $\mutual{\vx}{\vy}$ where the triple $(\py, \poz, \poo)$ is near their expectation. That is, $\py \in (\rho \ast p (1 - \Delta_{*1}), \rho \ast p (1 + \Delta_{*1}))$, $\poz \in (\rho p(1  -  \Delta_{10}), \rho p(1 + \Delta_{10}))$ and $\poo \in (\rho (1 - p)(1 - \Delta_{11}), \rho (1 - p)(1 + \Delta_{11}))$. Note that the mutual information at the center point of the cube equals
\begin{eqnarray}
	\mutual{\vx}{\vy}|_{(\rho \ast p, \rho p, \rho (1 - p))} &=& \entropy{\rho \ast p} - \entropy{p} \\
	&=& \KL{p}{\rho \ast p} + \parc(1 - 2p) \log \left(\frac{1 - \rho \ast p}{\rho \ast p}\right) \\
	&=& \Oh{n^{-1}} + \frac{\crho (1 - 2p)}{n^{1/2}} \log \left(\frac{1 - p + \Oh{n^{-1/2}}}{p + \Oh{n^{-1/2}}}\right) \\
	&=& \Oh{n^{-1}} + \frac{\crho (1 - 2p)}{n^{1/2}} \log \left[ \frac{1 - p}{p} + \Oh{n^{-1/2}} \right] \\
	&=& \frac{\crho (1 - 2p)}{n^{1/2}} \log \left(\frac{1 - p}{p}\right) + \Oh{n^{-1}}
\end{eqnarray}

By Taylor's expansion with center at $(\rho \ast p, \rho p, \rho (1 - p))$, the empirical mutual information $\mutual{\vx}{\vy}$ can be obtained by the following equation,
\begin{eqnarray}
	\mutual{\vx}{\vy} &=& \mutual{\vx}{\vy}|_{(\rho \ast p, \rho p, \rho (1 - p))} \nonumber \\
	& & + (\py - \rho \ast p) \frac{\partial \mutual{\vx}{\vy}}{\partial \py}|_{(\rho \ast p, \rho p, \rho (1 - p))} \nonumber \\
	& & + (\poz - \rho p) \frac{\partial \mutual{\vx}{\vy}}{\partial \poz}|_{(\rho \ast p, \rho p, \rho (1 - p))} \nonumber \\
	& & + (\poz - \rho (1 - p)) \frac{\partial \mutual{\vx}{\vy}}{\partial \poo}|_{(\rho \ast p, \rho p, \rho (1 - p))} + \Oh{n^{-1}}.
\end{eqnarray}
Here, the second and higher derivative terms can be bounded by $\Oh{n^{-1}}$. So, the dominant terms is the center value of the empirical mutual information $\mutual{\vx}{\vy}$ plus the first derivative terms. And we have,
\begin{eqnarray}
	\frac{\partial \mutual{\vx}{\vy}}{\partial \py}|_{(\rho \ast p, \rho p, \rho (1 - p))} &=& \log \left( \frac{p (1 - \rho \ast p)}{(1 - p)(\rho \ast p)} \right) \nonumber \\
	&=& \log \left( \frac{p (1 - \rho \ast p)}{p (1 - \rho \ast p) + \rho(1 - 2p)} \right) \nonumber \\
	&=& \log \left( 1 - \frac{\rho(1 - 2p)}{p (1 - \rho \ast p) + \rho(1 - 2p)} \right) \nonumber \\
	&=& \Oh{\rho} \nonumber \\
	&=& \Oh{n^{-1/2}}.
\end{eqnarray}
So, $(\py-\rho \ast p) \frac{\partial \mutual{\vx}{\vy}}{\partial \py}|_{(\rho \ast p, \rho p, \rho (1 - p))} = \Oh{n^{-1}}$ since $\py \in (\rho \ast p - \Delta_{*1}, \rho \ast p + \Delta_{*1})$ where $\Delta_{*1} = \Oh{n^{-1/2}}$.

Similarly, it can be shown that
\begin{equation}
	\frac{\partial \mutual{\vx}{\vy_b}}{\partial \poz}|_{(\rho \ast p, \rho p, \rho (1 - p))} = \log\left( \frac{p}{1 - p} \right) = \Oh{1}
\end{equation}
and
\begin{equation}
	\frac{\partial \mutual{\vx}{\vy_b}}{\partial \poo}|_{(\rho \ast p, \rho p, \rho (1 - p))} = \log\left( \frac{1 - p}{p} \right) = \Oh{1}.
\end{equation}
Therefore, $(\poz - \rho p) \frac{\partial \mutual{\vx}{\vy}}{\partial \poz}|_{(\rho \ast p, \rho p, \rho (1 - p))} = \Oh{n^{-3/4}}$ and $(\poz - \rho (1 - p)) \frac{\partial \mutual{\vx}{\vy}}{\partial \poo}|_{(\rho \ast p, \rho p, \rho (1 - p))} = \Oh{n^{-3/4}}$ for $\poz \in (\rho (p  -  \Delta_{10}), \rho (p + \Delta_{10}))$ and $\poo \in (\rho (1 - p - \Delta_{11}), \rho (1 - p + \Delta_{11}))$ where $\Delta_{10} = \Oh{n^{-1/4}}$ and $\Delta_{11} = \Oh{n^{-1/4}}$. This means that any changes in $\py$, $\poz$ or $\poo$ in the cube will only contribute at most $\Oh{n^{-3/4}}$. Hence the empirical mutual information $\mutual{\vx}{\vy} =\frac{\crho (1 - 2p)}{n^{1/2}} \log \left(\frac{1 - p}{p}\right) + \Oh{n^{-3/4}}$ within the cube for sufficiently large $n$.
\end{proof}

\begin{proposition}[Laplace's method]
\label{thm:laplace}
Assume that $f(x)$ is smooth, $f(x)$ has a unique global minimum at $x_0 \in (a,b)$ and $f'(x_0) = 0$. Then,
	$$ \int_a^b e^{-n f(x)} \di x = \sqrt{\frac{2 \pi}{n f''(x_0)}} e^{-nf(x_0)}\left(1 + \frac{c}{n} + \Oh{\frac{1}{n^2}} \right), $$
where $c = - \frac{1}{8 (f''(x_0))^2} \frac{\di^4}{\di x^4} f(x_0) + \frac{5}{24 (f''(x_0))^3} \left(\frac{\di^3}{\di x^3} f(x_0)\right)^2$.
\end{proposition}

\begin{proposition}[Stirling's Approximation]
	$$ \sqrt{2 \pi n} \left(\frac{n}{e}\right)^n \leq n! \leq \sqrt{2 \pi n} \left(\frac{n}{e}\right)^n e^{\frac{1}{12 n}} $$
\end{proposition}

\begin{lemma}
\begin{multline}
	\sqrt{\frac{1}{2 \pi n \frac{k}{n}\left(1 - \frac{k}{n}\right)}} 2^{n \entropy{\frac{k}{n}}} \exp\left(- \frac{1}{12 n \left(\frac{k}{n}\left(1 - \frac{k}{n}\right)\right)}\right) \\
	 \leq {n \choose k} \leq \sqrt{\frac{1}{2 \pi n \frac{k}{n}\left(1 - \frac{k}{n}\right)}} 2^{n \entropy{\frac{k}{n}}} \exp\left(\frac{1}{12 n}\right).
\end{multline}
\end{lemma}

\begin{proof}
\begin{align}
	{n \choose k} &= \frac{n!}{k!(n-k)!} \nonumber \\
		& \leq \frac{\sqrt{2 \pi n} \left(\frac{n}{e}\right)^n e^{\frac{1}{12 n}}}{\sqrt{2 \pi k} \left(\frac{k}{e}\right)^k \sqrt{2 \pi (n-k)} \left(\frac{n-k}{e}\right)^{n-k}} \nonumber \\
		&= \sqrt{\frac{n}{2 \pi k (n-k)}} \frac{n^n}{k^k (n-k)^{n-k}} e^{\frac{1}{12 n}} \nonumber \\
		&= \sqrt{\frac{1}{2 \pi n \frac{k}{n}\left(1 - \frac{k}{n}\right)}} \left(\frac{k}{n}\right)^{-k} \left(1 - \frac{k}{n}\right)^{-(n-k)} e^{\frac{1}{12 n}} \nonumber \\
		&= \sqrt{\frac{1}{2 \pi n \frac{k}{n}\left(1 - \frac{k}{n}\right)}} 2^{n \entropy{\frac{k}{n}}} e^{\frac{1}{12 n}} \nonumber \\
		&= \sqrt{\frac{1}{2 \pi n \frac{k}{n}\left(1 - \frac{k}{n}\right)}} 2^{n \entropy{\frac{k}{n}}} \exp\left(\frac{1}{12 n}\right) \nonumber
\end{align}
Similarly,
\begin{align}
	{n \choose k} & \geq \sqrt{\frac{1}{2 \pi n \frac{k}{n}\left(1 - \frac{k}{n}\right)}} 2^{n \entropy{\frac{k}{n}}} e^{-\frac{1}{12k}} e^{-\frac{1}{12(n-k)}} \nonumber \\
		&= \sqrt{\frac{1}{2 \pi n \frac{k}{n}\left(1 - \frac{k}{n}\right)}} 2^{n \entropy{\frac{k}{n}}} \exp\left(-\frac{1}{12n}\left( \frac{n}{k} + \frac{n}{n-k} \right)\right) \nonumber \\
		&= \sqrt{\frac{1}{2 \pi n \frac{k}{n}\left(1 - \frac{k}{n}\right)}} 2^{n \entropy{\frac{k}{n}}} \exp\left(-\frac{1}{12n}\left( \frac{1}{\frac{k}{n}\left( 1 - \frac{k}{n} \right)} \right)\right) \nonumber
\end{align}
\end{proof}

\begin{lemma}
	$ \sum_{\vy_w: \fyw \in (\lw, \uw)} \left| p_0(\vy_w) - q_0(\vy_w) \right| \leq \epn \xrightarrow{n} 0$, where
	\[ \epn = \sqrt{\pi n} \left( \left( \frac{5}{12 n \lw (1 - \lw)} - \frac{29}{24} \right) \frac{1}{n} + \Oh{\frac{1}{n^2}} \right). \]
\label{thm:var_dist_small}
\end{lemma}

\begin{proof}
Note that for $\fyw \in (\lw, \uw)$
	\[ p_0(\vy_w) = \frac{2^{-n \entropy{\fyw}}}{\uw - \lw} \int_{\lw}^{\uw} 2^{-n \KL{\fyw}{\pw}} \di \pw, \]
and
	\[ \int_{\lw}^{\uw} 2^{-n \KL{\fyw}{\pw}} \di \pw = \sqrt{\frac{2 \pi \fyw (1 - \fyw)}{n}} \left( 1 + \frac{a}{n} + \Oh{\frac{1}{n^2}} \right), \]
where
\begin{eqnarray}
	a &=& \frac{1}{8} \fyw^2(1 - \fyw)^2 \frac{-\fyw^3 - (1 - \fyw)^3}{\fyw^3(1 - \fyw)^3} - \frac{5}{24} \fyw^3(1 - \fyw)^3 \frac{\left( (1 - \fyw)^2 - \fyw^2 \right)^2}{\fyw^4 (1 - \fyw)^4} \nonumber \\
		&=& - \frac{1}{8} \frac{\fyw^2 - \fyw (1 - \fyw) + (1 - \fyw)^2}{\fyw (1 - \fyw)} - \frac{5}{24} \frac{(1 - 2 \fyw)^2}{\fyw (1 - \fyw)} \nonumber \\
		&=& - \frac{3 \fyw^2 - 3 \fyw + 3 \fyw^2 + 3 - 6 \fyw + 3 \fyw^2 + 5 - 20 \fyw + 20 \fyw^2}{24 \fyw (1 - \fyw)} \nonumber \\
		&=& - \frac{29 \fyw^2 - 29 \fyw + 8}{24 \fyw (1 - \fyw)} \nonumber \\
		&=& \frac{29}{24} - \frac{1}{3 \fyw (1 - \fyw)} \nonumber \\
		& \geq & \frac{29}{24} - \frac{1}{3 \lw (1 - \lw)} \nonumber
\end{eqnarray}
\end{proof}

Therefore, for $\fyw \in (\lw, \uw)$,
\[ p_0(\vy_w) \geq \frac{2^{-n \entropy{\fyw}}}{\uw - \lw} \sqrt{\frac{2 \pi \fyw (1 - \fyw)}{n}} \left( 1 + \left( \frac{29}{24} - \frac{1}{3 \lw (1 - \lw)} \right) \frac{1}{n} + \Oh{\frac{1}{n^2}} \right). \]

On the other hand, for $\fyw \in (\lw, \uw)$
	\[ q_0(\vy_w) = \frac{1}{n(\uw - \lw)} \frac{1}{{n \choose n \fyw}}. \]

Also note that
\begin{eqnarray}
	\frac{1}{{n \choose n \fyw}} & \leq & \sqrt{2 \pi n \fyw (1 - \fyw)} 2^{- n \entropy{\fyw}} \exp\left(\frac{1}{12 n \fyw (1 - \fyw)}\right) \nonumber \\
		&=& \sqrt{2 \pi n \fyw (1 - \fyw)} 2^{- n \entropy{\fyw}} \left( 1 + \frac{1}{12 n \fyw (1 - \fyw)} + \Oh{\frac{1}{n^2}} \right) \nonumber \\
		& \leq & \sqrt{2 \pi n \fyw (1 - \fyw)} 2^{- n \entropy{\fyw}} \left( 1 + \frac{1}{12 n \lw (1 - \lw)} + \Oh{\frac{1}{n^2}} \right). \nonumber
\end{eqnarray}

So,
\[ q_0(\vy_w) \leq \frac{2^{-n \entropy{\fyw}}}{\uw - \lw} \sqrt{\frac{2 \pi \fyw (1 - \fyw)}{n}} \left( 1 + \frac{1}{12 n \lw (1 - \lw)} + \Oh{\frac{1}{n^2}} \right). \]

We then have,
\begin{eqnarray}
	\left| p_0(\vy_w) - q_0(\vy_w) \right| &=& q_0(\vy_w) - p_0(\vy_w) \nonumber \\
		& \leq & \frac{2^{-n \entropy{\fyw}}}{\uw - \lw} \sqrt{\frac{2 \pi \fyw (1 - \fyw)}{n}} \left[ \left( \frac{5}{12 n \lw (1 - \lw)} - \frac{29}{24} \right) \frac{1}{n} + \Oh{\frac{1}{n^2}} \right] \nonumber \\
		& \leq & \frac{1}{\uw - \lw} \sqrt{\frac{\pi}{2 n}} \left[ \left( \frac{5}{12 n \lw (1 - \lw)} - \frac{29}{24} \right) \frac{1}{n} + \Oh{\frac{1}{n^2}} \right]. \nonumber
\end{eqnarray}
Hence,
\begin{eqnarray}
	\sum_{\vy_w: \fyw \in (\lw, \uw)} \left| p_0(\vy_w) - q_0(\vy_w) \right| & \leq & n(\uw - \lw)\frac{1}{\uw - \lw} \sqrt{\frac{\pi}{2 n}} \left[ \left( \frac{5}{12 n \lw (1 - \lw)} - \frac{29}{24} \right) \frac{1}{n} + \Oh{\frac{1}{n^2}} \right] \nonumber \\
		&=& \sqrt{\pi n} \left[ \left( \frac{5}{12 n \lw (1 - \lw)} - \frac{29}{24} \right) \frac{1}{n} + \Oh{\frac{1}{n^2}} \right] \nonumber \\
		& \triangleq & \epn \nonumber \\
		& \rightarrow & 0. \nonumber
\end{eqnarray}




\bibliographystyle{IEEEtran}
\bibliography{stego_journal}

\end{document}